%% file: ms.tex
\documentclass[twocolumn]{aastex63}

\usepackage{lineno}

\input{mydef.tex}

\received{04-Jul-2021}
\revised{27-Feb-2022}
\accepted{28-Mar-2022}
\submitjournal{APJ}
\shorttitle{dust accretion}
\shortauthors{Arras et al.}

\newcommand{\Ms}{M_{\rm s}}

\newcommand{\Rsub}{R_{\rm sub}}
\newcommand{\Mp}{M_{\rm p}}
\newcommand{\Rp}{R_{\rm p}}
\newcommand{\md}{m_{\rm d}}
\renewcommand{\ap}{a_{\rm p}}
\newcommand{\vp}{v_{\rm p}}
\newcommand{\rhill}{r_{\rm Hill}}

\begin{document}

\title{Dust Accretion onto Exoplanets}

\correspondingauthor{Phil Arras}
\email{arras@virginia.edu}
\author[0000-0001-5611-1349]{Phil Arras}
\author[0000-0002-0310-8530]{Megan Wilson }
\author[0000-0001-5751-8239]{Matthew Pryal}
\author[0000-0003-0934-0805]{Jordan Baker }
\affiliation{Department of Astronomy, University of Virginia, Charlottesville, VA 22904, USA}

\begin{abstract}
Accretion of interplanetary dust onto gas giant exoplanets is considered. Poynting-Robertson drag causes dust particles from distant reservoirs to slowly inspiral toward the star. Orbital simulations for the three-body system of the star, planet, and dust particle show that a significant fraction of the dust may accrete onto massive planets in close orbits. The deceleration of the supersonic dust in the planet's atmosphere is modeled, including ablation by thermal evaporation and sputtering. The fraction of the accreted dust mass deposited as gas-phase atoms is found to be large for close-in orbits and massive planets. If mass outflow and vertical mixing are sufficiently weak, the accreted dust produces a constant mixing ratio of atoms and remnant dust grains below the stopping layer. When vertical mixing is included along with settling, the solutions interpolate between the mixing ratio due to the meteoric source above the homopause, and that of the well-mixed deeper atmosphere below the homopause. The line opacity from atoms and continuum opacity from remnant dust may be observable in transmission spectra for sufficiently large dust accretion rates, a grain size distribution tilted toward the blowout size, and sufficiently weak vertical mixing. If mixing is strong, the meteoric source may still act to augment heavy elements mixed up from the deep atmosphere as well as provide nucleation sites for the formation of larger particles. The possible role of the Lorentz drag force in limiting the flow speeds and mixing coefficient for pressures $\la 1\, \rm mbar$ is discussed.
\end{abstract}

\section{Introduction} \label{sec:intro}

In our solar system, dust is constantly produced by both asteroidal and cometary sources (see \citealt{2013pss3.book..431M} for a review).  Under the action of Poynting-Robertson (PR) drag, this dust will slowly spiral inward toward the Sun. When the dust approaches a planet, there are several possible outcomes: a physical collision with the planet; ejection from the solar system; or the dust will pass by the planet and again inspiral until it reaches another planet or eventually the Sun. Dust which approaches close enough to the Sun may sublimate completely or be ejected due to enhanced radiation pressure after partial sublimation (the $\beta$-meteoroids).

Interplanetary dust may be directly observed as {\it zodiacal light} through the scattering of sunlight, and also through thermal emission \citep{KRUGER2014657}. The inner part of the zodiacal light is called the F-corona \citep{2020ApJS..246...27S}, and it dominates the reflected light down to a few solar radii, inside of which the dust is  destroyed \citep{2004SSRv..110..269M, 2019Natur.576..232H}. Upon entering the Earth's atmosphere, the stopping of the fast-moving dust particles is observable through radar reflections off their ionized trails, or as visible meteoroids \citep{1950PNAS...36..687W,1976iecr.book.....O,2012ChSRv..41.6507P}. The gas-phase atoms released by ablation give rise to an enhancement of heavy atoms, which is utilized by astronomers to make laser guide-star observations with adaptive optics systems \citep{1994IAUS..158..283O}. Some gas-phase atoms condense back into small dust grains called {\it meteoric smoke particles} \citep{1961JAtS...18..736R,1980JAtS...37.1342H}. They may act as nucleation sites around which an ice mantle is accreted, generating significant scattering opacity. These high altitude {\it noctilucent clouds} are coincident with the dust-stopping layer, and their composition and influx rate have been constrained by extinction observations \citep{2017JGRD..12213495H}.

This paper presents a study of the accretion of dust and its possible observable consequences for exoplanets orbiting very near their star. Infrared excess due to dust thermal emission has been observed from other stars, with the far, mid, and near infrared arising from {\it cold}, {\it warm}, and {\it hot} dust at different distances from the star 
 \citep{2008ApJ...674.1086T, 2013A&A...555A.104A, 2013A&A...555A..11E, 2014A&A...570A.128E, 2020AJ....159..177E, 2021A&A...651A..45A}. It is the hot dust component that would directly interact with close-in planets. 
In our solar system, the inner solar system sources of dust may be dominated by Jupiter-family comets with a smaller component from the asteroid belt  \citep{2010ApJ...713..816N,2011ApJ...743..129N}. The dust mass input rate ($\dot{M}_{\rm d, iss}$) to the inner solar system, which is required to explain the zodiacal light and meteoroids impacting the Earth's atmosphere, is roughly constrained to be $\dot{M}_{\rm d, iss} \sim 10^7\,
\rm g\, s^{-1}$ \citep{2011ApJ...743..129N}.

There are two 
reasons why higher dust accretion rates may be possible for extrasolar planets as compared to the Earth. 
First, in order for a star to have a detectable excess from hot dust requires vastly more dust than found in our own solar system (the {\it zodi}; \citealt{2012PASP..124..799R}). The near-infrared interferometric detection of a {\it H-} or {\it K-}band excess requires $\ga \, 3600 $ zodi's \citep{2013A&A...555A.104A}. 
Detection is more favorable for warm dust at $\sim \rm au$ separations, e.g., LBTI can detect warm dust down to tens of zodi for some systems \citep{2020AJ....159..177E}. While $\sim 15-20\%$ of stars have detectable hot dust excess \citep{2013A&A...555A.104A,  2014A&A...570A.128E, 2021A&A...651A..45A}, undetected systems with far larger $\dot{M}_{\rm d} $ could be more common and perhaps ubiquitous.

Second, \citet{2018MNRAS.480.5560B} have shown that massive, close-in planets are surprisingly effective at accreting dust as compared to smaller and more distant planets. The Earth is estimated to accrete only a tiny fraction of the dust passing through its orbital radius ($\dot{M}_{\rm Earth} \sim 10^2-10^3\, \rm g\, s^{-1}$, \citealt{2011ApJ...743..129N}); hence, some exoplanets may experience much larger dust accretion rates much larger than Earth's, with correspondingly higher gas-phase atomic and dust densities in the atmosphere. 

The goal of this paper is to estimate, in the context of simplified models, if these densities can be large enough to affect the atmospheric abundances, or to give rise to significantly larger transit radii. It will be shown that this may be accomplished for a range of  rates  $\dot{M}_{\rm d} = 10^7-10^{9}\, \rm g\, s^{-1}$ for systems with favorable parameters. 

One motivation for this work came from the puzzling transmission spectra of extrasolar planets (see the detailed review in \citealt{2019ARA&A..57..617M}). For the case of a {\it clear} atmosphere, atomic lines such as the Na D doublet may be expected to dominate the opacity over a wide band in the optical, matching onto Rayleigh scattering by H$_2$ toward shorter wavelengths. While such a feature has been observed
 \citep{2018Natur.557..526N}, it is much more common that only a narrow range of wavelengths will show the expected Na D feature, and outside the center of the line the  transit radius varies slowly with wavelength. At longer wavelengths,  absorption features from  molecules such as H$_2$O and CH$_4$ are expected in a clear atmosphere, but again these features are often muted. An interpretation of these observations is that the continuum absorption corresponds to aerosols made of condensates or due to photochemical reactions high in the atmosphere. The estimated pressure range for these high altitude clouds must be $\rm \mu bar - mbar$, well above the $\sim 0.1\, \rm bar$ level where the optical continuum is absorbed. Lofting large particles to such high altitudes requires  strong vertical mixing (e.g. \citealt{2019ApJ...887..170P}).

In addition, 
species like Mg, Si, and Fe, which readily condense to form clouds near the $P=1\, \rm bar$ level of the atmosphere (e.g. \citealt{1999ApJ...512..843B}), are observed with transit radii well above that of the optical continuum (e.g. \citealt{2019ARA&A..57..617M}). Since the abundances of these elements should have been greatly reduced above the cloud deck \citep{2010ApJ...716.1060V},  the origin of these elements in atomic form high in the atmosphere is difficult to explain. While large amounts of vertical mixing are required to mix heavy atoms and dust grains to high in the atmosphere, the meteoric source discussed in this paper would provide an alternative explanation that does not require strong mixing.

Earlier work by \citet{2017ApJ...847...32L} has already raised the possibility that meteoroid ablation may seed the upper atmosphere with heavy elements, which may then form the aerosols and give rise to the observed continuum absorption in the transit spectrum. However, that study was mainly concerned with the formation of photochemical hazes through mixing of heavy molecular species upward from the deep atmosphere to the $\sim 1\, \rm \mu bar$ level, where they would undergo photolysis. The authors concluded that while meteoroid accretion would deposit heavy elements in the upper atmosphere, the rate and hence the importance of the meteoric source was uncertain. This was a motivation for the present work in which a plausible range for the dust accretion rate is discussed, even if the value for individual systems is still uncertain. The more detailed calculations in \citet{2017ApJ...847...32L} provide context for the present study in that once a dust accretion rate is assumed, the coagulation of the small particles and effect on the transit spectrum may be estimated.

A possible source of heavy elements that may be accreted into the planet's atmosphere, aside from the case of interplanetary dust studied here, is gas loss from hypothetical satellites orbiting the planet \citep{2006PASP..118.1136J, 2009ApJ...704.1341C, 2019ApJ...885..168O}, similar to the Io plasma torus around Jupiter.

A somewhat analogous situation to the exoplanets occurs for white dwarfs \citep{2006Sci...314.1908G,2009ApJ...694..805F, 2010ApJ...714.1386F}. As settling times on white dwarfs can be a small fraction of their age, one might expect pristine photospheres showing little evidence of heavy elements. However, evidence of heavy elements are indeed found in their spectra for a significant fraction of white dwarfs, implying recent accretion. The observed abundance in the atmosphere is then a competition between accretion and settling (e.g. \citealt{2019ApJ...872...96B}), similar to the case studied here for accretion of dust onto planets. The existence of accreted heavy elements is perhaps more surprising in the white dwarf context as the post-main-sequence evolution of the star would effectively destroy their inner solar systems, and the orbits of any remaining bodies would expand significantly as the star loses mass (e.g. \citealt{2012ApJ...761..121M,2018MNRAS.481.4077S}).

The outline of the paper is as follows. In Section \ref{sec:orbits}, the dust orbital dynamics calculation of \citet{2018MNRAS.480.5560B} is revisited and their central result affirmed. 
Section \ref{sec:stopping} discusses the entry of the dust into the planet's atmosphere, the deceleration to subsonic speeds, and subsequent ablation as a function of altitude. Section \ref{sec:sourcedensity} shows that the mixing ratios below the source, if gravitational settling is dominant, may approach solar abundance for a dust accretion rate comparable to $\dot{M}_{\rm d, iss}$.  Section \ref{sec:Kzz} contains a discussion of the expected amount of mixing by fluid circulation in the upper atmosphere. The dust source layer is matched onto the rest of the atmosphere in a more complete manner in Section \ref{sec:diffusion}, including the effects of turbulent mixing, gravitational settling, and upward escape of hydrogen. Section \ref{sec:lines} uses the profiles of mixing ratio to compute the dependence of the transit radius on vertical mixing coefficient and dust accretion rate.
The size distribution and number density of the remnant dust is discussed in Section \ref{sec:remnantdust}.
Discussion, comparison to previous work, and estimates for Lorentz drag are given in Section \ref{sec:discussion}. Section \ref{sec:summary} contains a summary of the calculations and main results.
The Appendix contains a calculation of the accretion fraction onto the planet using the \"{O}pik approximation, and comparison to the {\it Rebound} simulations.

Calculations throughout this paper will be presented for three different planet sizes, a small planet with $M_{\rm p}=0.1\, M_{\rm J}$ and $R_{\rm p}= 0.7\, R_{\rm J}$ , a medium-size planet with $M_{\rm p}= M_{\rm J}$ and $R_{\rm p}= R_{\rm J}$, and a large planet with with $M_{\rm p}=2.0\, M_{\rm J}$ and $R_{\rm p}= 2.0\, R_{\rm J}$. A Sun-like star has parameters $M_{\rm s} = M_\odot$, luminosity $L_{\rm s}=L_\odot$, radius $R_{\rm s}=R_\odot$ and dust sublimation radius $R_{\rm sub}=4\, R_\odot$. 

\section{ dust orbital dynamics}
\label{sec:orbits}

Solutions of the three-body problem, including the orbital motion of the star, planet, and dust particle are considered in this section. The goal is to calculate the fraction of the dust that is accreted onto the planet, sublimated near the star, or ejected from the system. For the dust that hits the planet, the impact speed and angle of incidence will be computed. 

Two solution methods are used: numerical integration of the equations of motion (Section \ref{sec:rebound}), which makes few approximations but is time-consuming; and the \"{O}pik approximation (Appendix \ref{sec:opik}), which greatly speeds up the calculation at the  expense of approximations for the orbital dynamics.

\begin{figure*}[ht!]
\epsscale{1.0}
\plotone{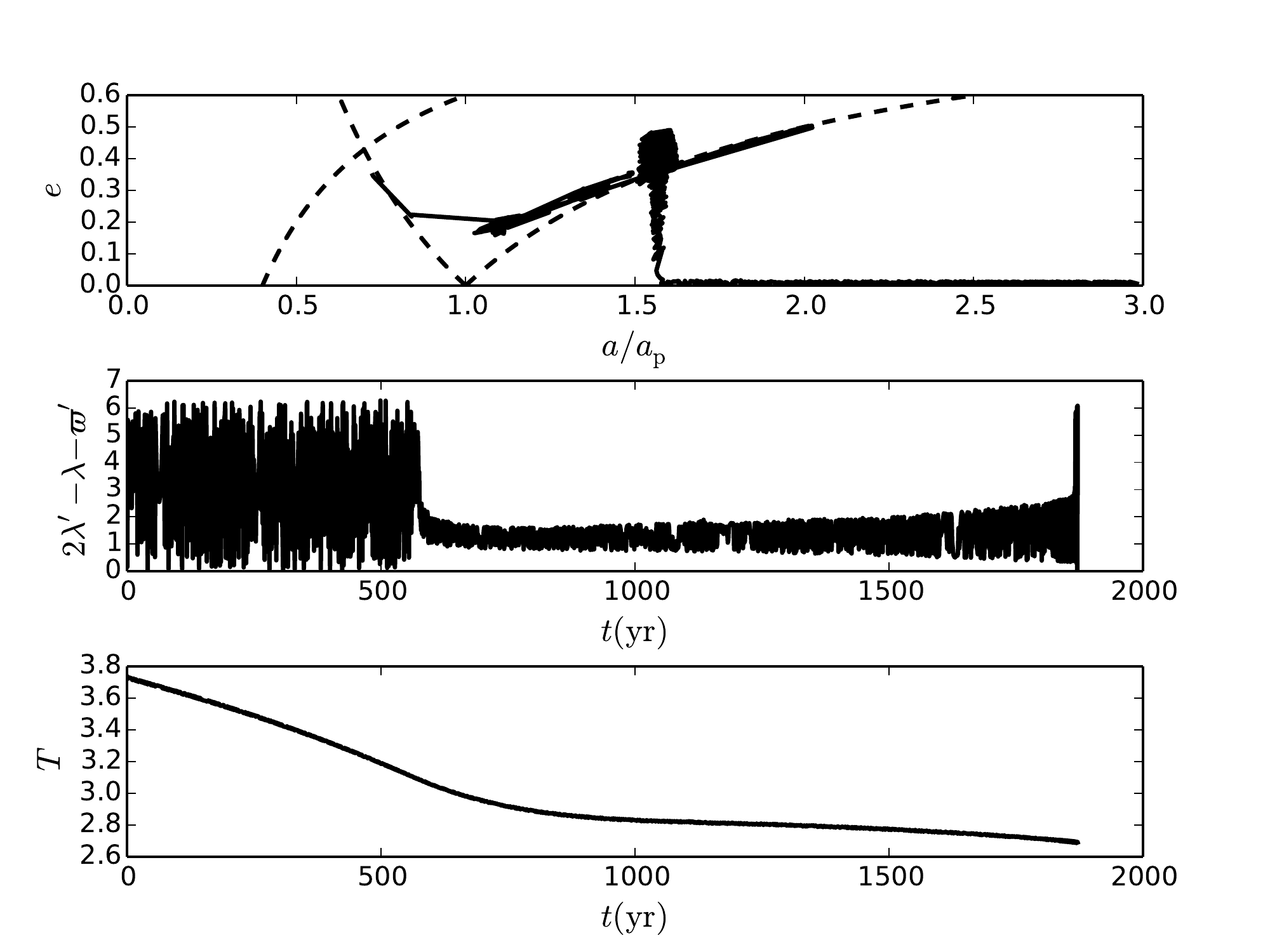} 
\caption{Dust particle inspiral for $\ap=10\,R_\odot$, $\beta=0.01$, and an inclination of $I=10^\circ$. (Top panel) The solid line shows the eccentricity versus semi-major axis of the orbit. The dashed lines correspond to $a(1\pm e)=\ap$ and $a(1-e)=\Rsub$. (Middle panel) Resonant angle versus time. (Bottom panel) Jacobi constant versus time. 
\label{fig:trajectory}}
\end{figure*}

The collisional dust source and any distant planets that scatter the dust are assumed to be distant so that PR drag has nearly circularized the dust orbit by the time it reaches the inner planet of interest. This planet is assumed to be on a circular orbit around the star. The dust feels the gravitational forces from both the star and planet, as well as radiation pressure and PR drag forces from the star. There are several phases for the orbital dynamics, as illustrated in Figure \ref{fig:trajectory}.

First, as PR drag removes energy and angular momentum from the orbit, the dust will slowly spiral inward on nearly circular orbits. Second, when it gets close enough to the planet, and for assumed small initial eccentricity and inclination, it will enter into a mean-motion resonance with the planet, temporarily halting the inspiral. 
Continued PR drag causes the eccentricity to rise to the point that the dust orbit begins to cross the planet's orbit. Due to the resonance, the planet is initially far away from the dust near its pericenter and close encounters are avoided. However, as PR drag pushes the orbit further into the resonance, the libration amplitude of the resonant angle increases and eventually the orbit leaves resonance, ending the second phase. 

In the third phase, the dust orbits are no longer in resonance, and begin with semi-major axis outside that of the planet, but pericenter inside the planet's orbit, allowing close encounters between the two. There are three fates for the dust particle in this phase: physical collision with the planet, orbits that intersect the dust sublimation zone of the star, or ejection of the dust to infinity. For planets that are sufficiently distant from the star, dust particles that avoid both ejection and collision with the planet may have time to circularize before they encounter the dust sublimation zone. For close orbits, the eccentricity may be raised to sufficiently large values to enter the sublimation zone directly. 

\subsection{ three-body simulations }
\label{sec:rebound}

Direct numerical integration of the equations of motion is performed with the {\it Rebound} code (version 2.20.5, \citealt{2012A&A...537A.128R}). The star, of mass $\Ms$ and dust sublimation radius $\Rsub$, and planet, of mass $\Mp$ and radius $\Rp$, are on a circular orbit of radius $a_p$.  The 15th order Gauss-Radau integrator with adaptive stepsize (IAS15) is used. The radiation pressure and PR drag forces due to the stellar flux are \citep{1979Icar...40....1B}
\be
\vec{F}_{\rm rad} & = & \beta \frac{G\Ms\md}{r_{\rm ds}^2} \left[ \left( 1 - \frac{\dot{r}_{\rm ds}}{c} \right) \vec{n}_{ds} - \frac{\vec{v}_{\rm ds}}{c} \right]
\ee
and are included as additional, velocity dependent forces. The dust position is $\vec{x}$ and the positions of the star and planet are $\vec{x}_{\rm s}$ and $\vec{x}_{\rm p}$, respectively. The star-dust separation vector is then $\vec{x}_{\rm ds} = \vec{x} - \vec{x}_{\rm s} = r_{\rm ds} \vec{n}_{\rm ds} $, and the relative velocity is $\vec{v}_{\rm ds} = \vec{v} - \vec{v}_{\rm s}$. The quantity 
\be
\beta & = & \frac{F_{\rm rad}}{F_{\rm grav}} = \frac{3L_{\rm s} Q_{\rm pr}}{16\pi GM_{\rm s} c\rho_{\rm d} s}
\label{eq:beta}
\ee
 is the ratio of radiation  to gravitational force on the dust, $L_{\rm s}$ is the stellar luminosity, $\rho_{\rm d}$ is the dust particle density, $s$ is the dust radius, and $m_{\rm d}=4\pi \rho_{\rm d} s^3/3$ is the dust mass. The quantity $Q_{\rm pr}$ is a dimensionless coefficient depending on the dust size, composition and the stellar spectrum \citep{1979Icar...40....1B}. For large dust, which obeys the geometrical optics limit, $Q_{\rm pr} \simeq 1$ and $\beta \simeq  2.9 \times 10^{-3} \, (2\, \rm g\, cm^{-3}/\rho_{\rm d})(s/100\mu \rm m)^{-1}(L_{\rm s}/L_\odot)(M_{\rm s}/M_\odot)^{-1}$. The radiation pressure term effectively decreases the mass of the star to $\Ms(1-\beta)$ so that the dust would orbit more slowly than the planet for a circular orbit at the same separation. The inspiral timescale, for a circular orbit of radius $a$, is
\be
t_{\rm pr}(a) & = & \frac{ca^2}{4\beta G \Ms}  
\nonumber \\ & = &  1.3 \times 10^3\, \rm yr\, \left( \frac{a}{0.1\, \rm au} \right)^2 \left( \frac{  3 \times 10^{-3} }{ \beta} \right) \left( \frac{ M_\odot }{ \Ms } \right),
\ee
much longer than the orbital timescale. 

The simulations are run until a physical collision or ejection occurs, or until a maximum run time of $90\, t_{\rm pr}(a_{\rm p})$, which is sufficient to ensure that most dust particles have been accreted or ejected. A by-product of using fixed integration times is that some runs in which particles have diffused to  $a \gg \ap$ orbits will not have had sufficient time to finish. However, these runs are small in number and do not affect the outcome probabilities significantly.

{\it Rebound's direct collision} scheme is used to detect collisions of the dust with either planet or the sublimation zone around the star.  
The relative velocity and angle of the velocity with respect to the radial direction are tabulated at the time when the dust particle hits the surface.

Calculations were carried out for a Sun-like star, each of the three planet sizes, and a range of semi-major axis $a_{\rm p}=(4-20) \times R_\odot$. Results are presented for a dust orbital inclination $I=10^\circ$ relative to the planet's orbit. The initial dust semi-major axis is $a=3 a_{\rm p}$, initial eccentricity is zero, and for each orbital separation, 1000 runs were carried out distributed uniformly over initial orbital phase. 

The dust does not evolve during the simulation in that the value of $\beta$ is fixed and the dust is assumed to be instantly destroyed if it enters the sublimation radius $R_{\rm sub}$ around the star. For a Sun-like star this corresponds to a radius where the dust equilibrium temperature would be higher than 2000 K. 
Depending on the sublimation timescale as compared to the orbital period, dust mass loss may lead to an increase in the importance of the radiation pressure force. In the case without a planet, this has been shown to lead to ejection from the system when $\beta \ga 1$ (the $\beta$-metereoroids; e.g. \citealt{2009Icar..201..395K}). However, when a planet is included collisions with the planet will compete with ejection. This effect is not included in the present study.

\begin{figure*}[htb!]
\epsscale{0.9}
\plotone{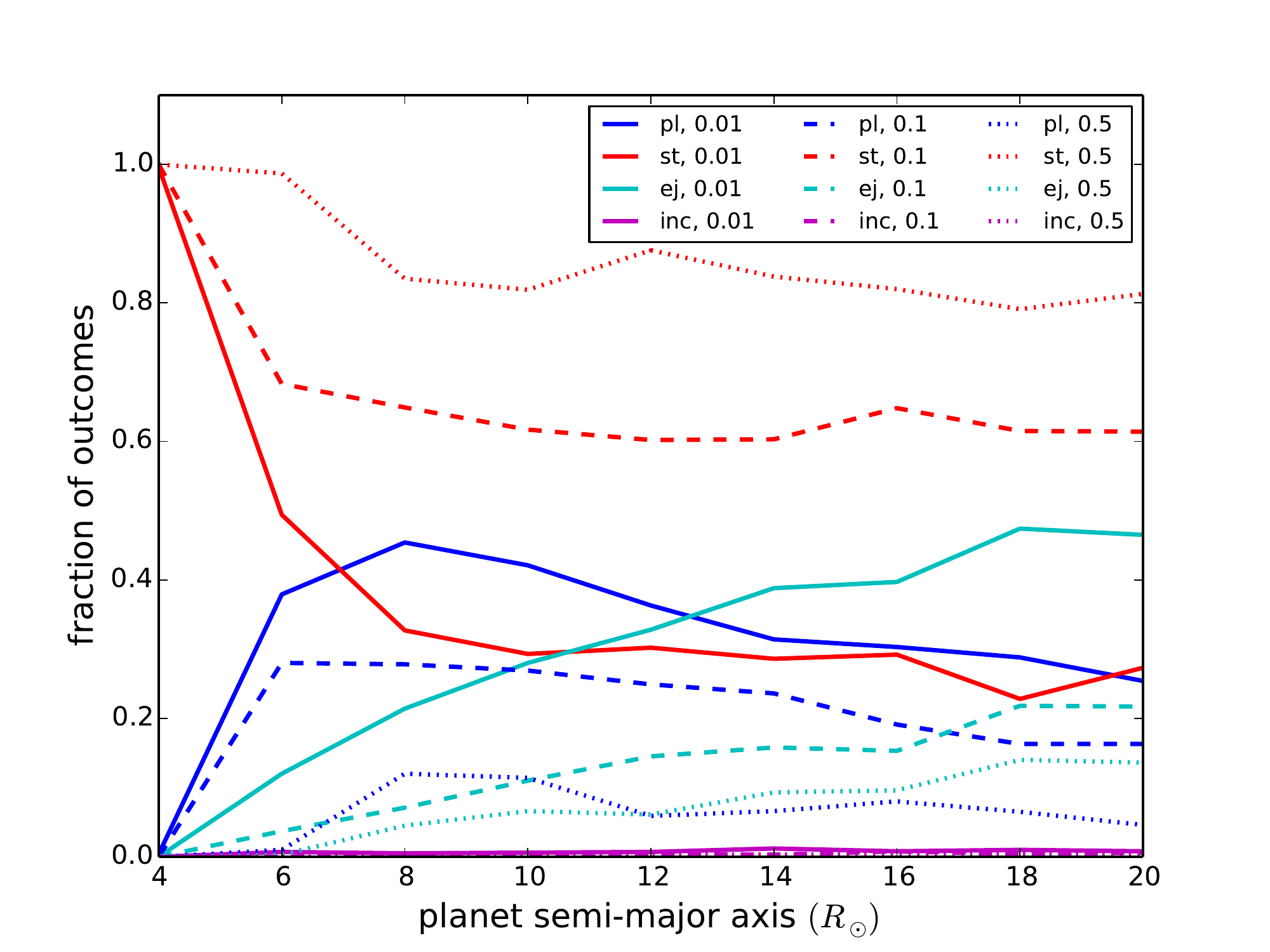} 
\caption{ Outcome probability versus planet semi-major axis $a_p$ for a Sun-like star and medium-size planet. The three physical outcomes are colliding with the planet (pl, blue lines), dust sublimation near the star (st, red lines), and ejection from the system (ej, cyan lines). Since the simulations are run for a fixed time, $90t_{\rm pr}(a_p)$, a small number had not yet achieved one of the three outcomes and are {\it incomplete} (inc, magenta lines). Each line is labeled by $\beta=0.01, 0.1$, and $ 0.5$. An initial inclination $I=10^\circ$ has been used for all runs. 
\label{fig:medfractions}}
\end{figure*}

\begin{figure*}[htb!]
\epsscale{0.9}
\plotone{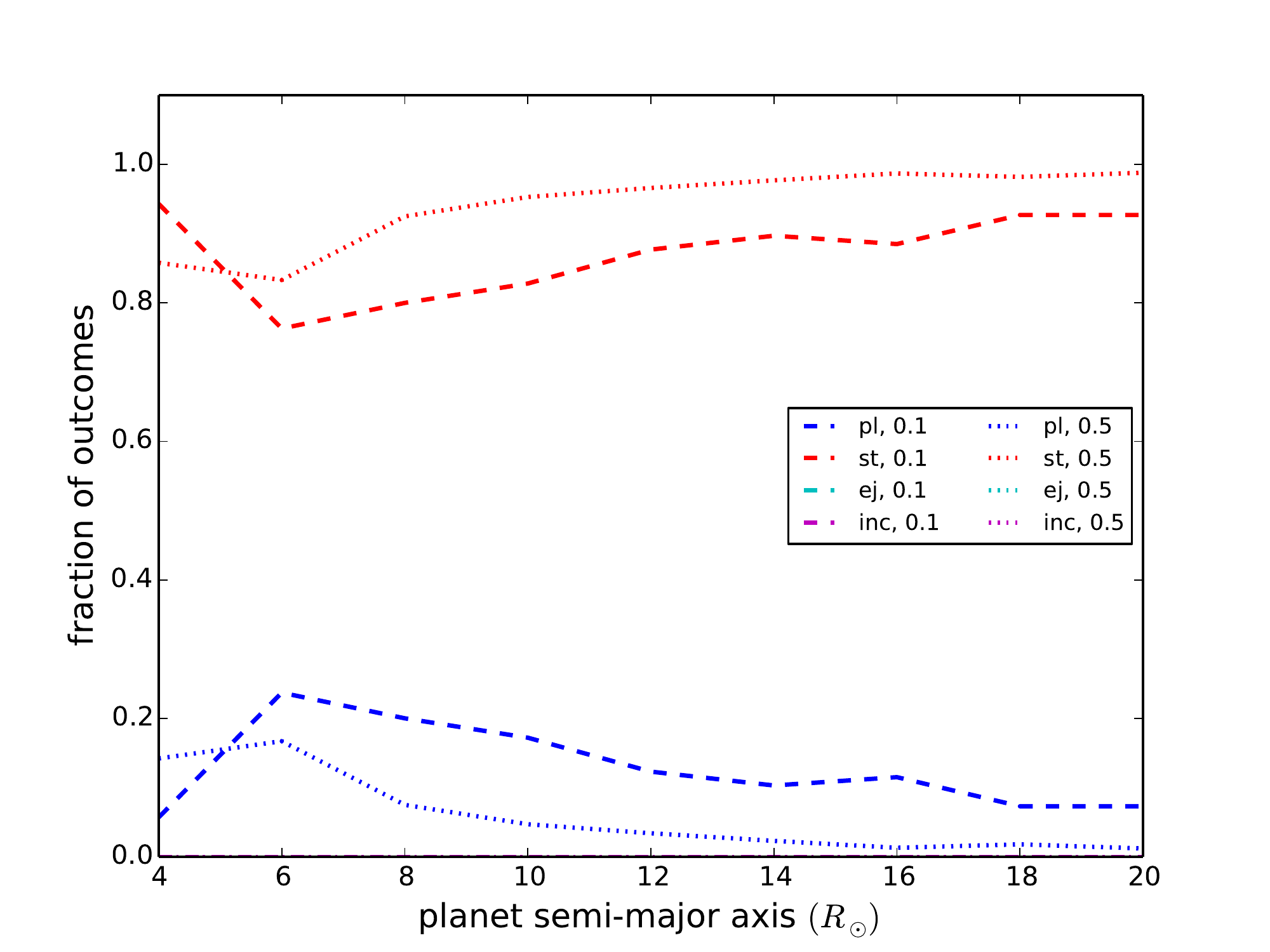} 
\caption{ The same as Figure \ref{fig:medfractions} for a small planet and $\beta=0.1,0.5$.
\label{fig:smallfractions}}
\end{figure*}

\begin{figure*}[htb!]
\epsscale{0.9}
\plotone{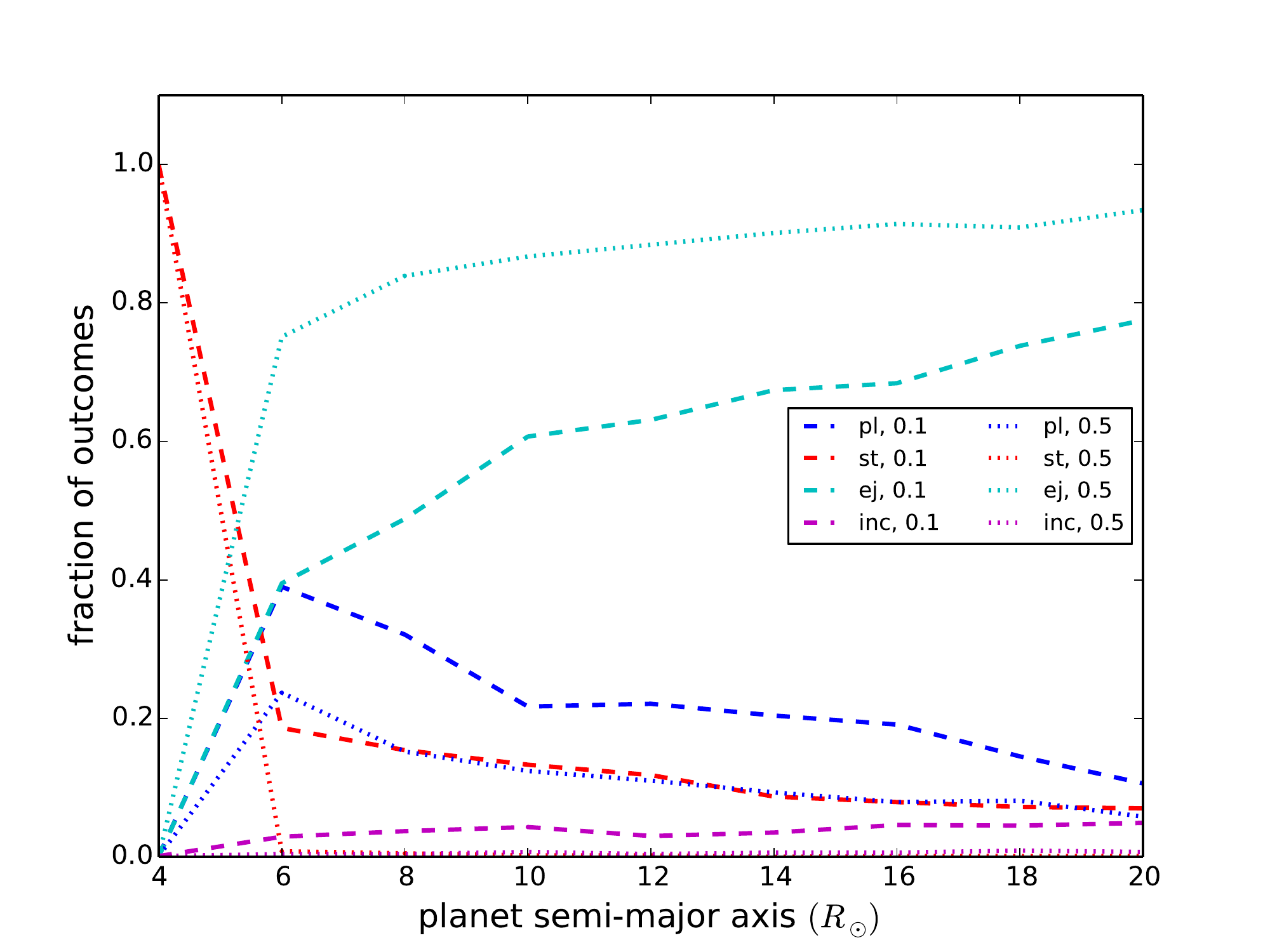} 
\caption{ The same as Figure \ref{fig:medfractions} for a large planet and $\beta=0.1,0.5$.
\label{fig:bigfractions}}
\end{figure*}

\begin{figure*}[htb!]
\epsscale{0.9}
\plotone{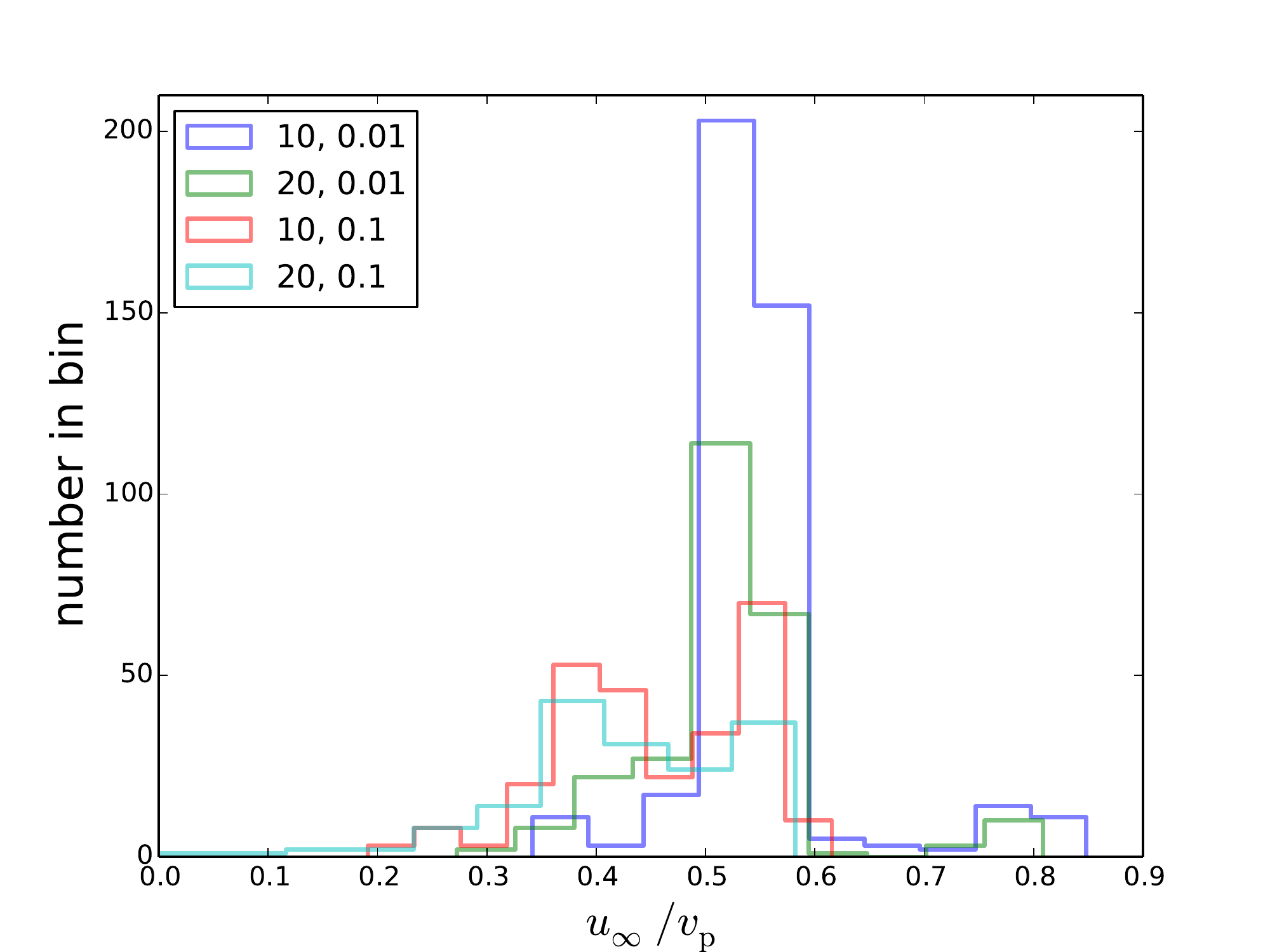}
\caption{ Histogram of $u_\infty = \sqrt{u^2 - 2G\Mp/\Rp}$, the planet-dust relative velocity at the time of the physical collision with the escape speed contribution subtracted off. Each line is labeled by $(\ap/R_\odot, \beta)$. The same star and planet parameters as in Figure \ref{fig:medfractions} are used.
\label{fig:velocities}}
\end{figure*}

\begin{figure*}[htb!]
\epsscale{0.9}
\plotone{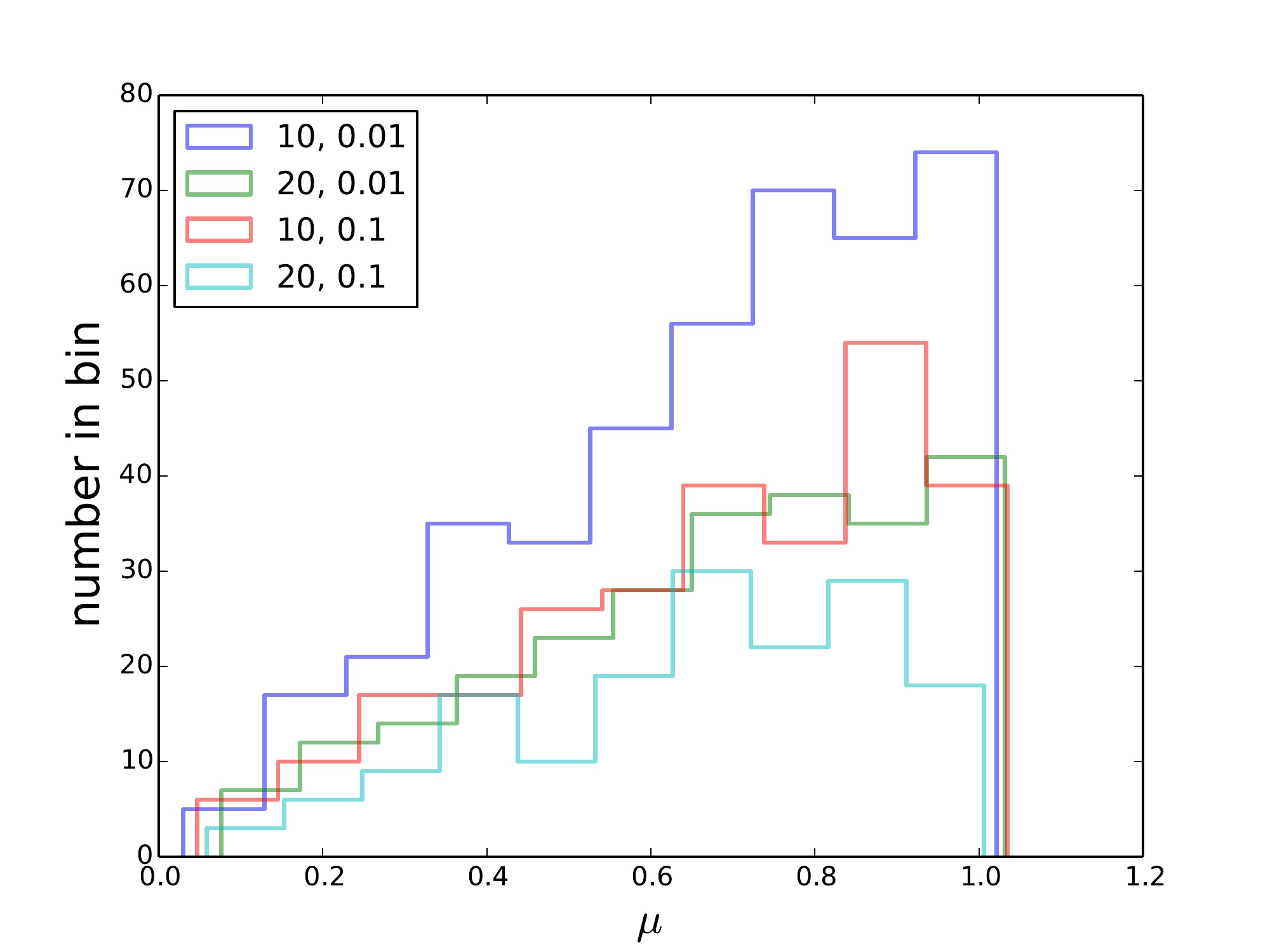} 
\caption{ Histogram of $\mu$, the cosine of angle of the dust with respect to the radial at the time of the physical collision, for the same runs as in Figure \ref{fig:velocities}. The value $\mu=0$ is for nearly horizontal entry and $\mu=1$ is for downward vertical entry. The same star and planet parameters as in Figure \ref{fig:medfractions} are used.
\label{fig:mus}}
\end{figure*}

Selected results are presented in Figures \ref{fig:trajectory} - \ref{fig:mus}. 
An example trajectory for $\beta=0.01$ is shown in Figure \ref{fig:trajectory}, and can be interpreted in terms of the analytic calculations in \citet{1993Icar..104..244W}. The top panel shows the semi-major axis versus eccentricity for the orbit of the dust around the star. Initially the orbit has small eccentricity and is decaying due to PR drag. It then enters a 2:1 mean motion resonance with the planet \citep{1993CeMDA..57..373S, 1993Icar..104..244W, 2006MNRAS.365.1367Q}, after which the semi-major axis becomes constant at $a=2^{2/3}(1-\beta)^{1/3}a_{\rm p}$ and the eccentricity rises. The trajectory crosses the dashed line for $a(1-e)=\ap$ and the dust and planet orbits now cross. The eccentricity rises to a maximum $e \simeq 0.5$ \citep{1993Icar..104..244W}, and then decreases again as oscillations in $a$ increase. When the resonance is broken, subsequent encounters with the planet's Hill sphere give rise to large changes in $a$ and $e$ ({\it cometary diffusion}, see e.g. \citealt{1980A&A....85...77Y}, and the Appendix). This particular trajectory ends with the dust particle hitting the planet.
 
The middle panel of Figure \ref{fig:trajectory} shows the resonant angle $\varphi = \lambda^\prime - \lambda - \varpi^\prime$ versus time, where $\lambda^\prime$ and $\varpi^\prime$ are the mean longitude and longitude of pericenter for the dust particle, and $\lambda$ is the mean longitude of the planet. Initially this angle is circulating, but begins to librate in resonance at $t \simeq 600\, \rm yr$. Thereafter, the libration amplitude slowly increases over $\sim 15\, t_{\rm pr}(\ap)$. Near $t \simeq 1800\, \rm yr$, the orbit falls out of resonance and $\varphi$ begins to circulate. At this point close encounters may occur, and a short time later the dust particle hits the planet.

The lower panel of Figure \ref{fig:trajectory} shows the Jacobi constant
\be
T & = & - \frac{2}{v_{\rm p}^2} \left[ \frac{v^2}{2} - \frac{G\Ms (1-\beta)}{r_{\rm ds}} - \frac{G\Mp}{r_{\rm dp}} - n_{\rm p} \left( x v_y - y v_x \right) \right]
\ee
as calculated in the inertial frame. In the absence of PR drag, $T$ is a constant of the motion. The value of $T$ as the resonance is exited is crucial for determining the fate of the dust and the speed at which it will hit the planet. 
Here $r_{\rm dp}$ is the dust-planet separation, 
$v_{\rm p}=(G\Ms/\ap)^{1/2}$ is the planet's orbital speed and  $n_{\rm p} = (G\Ms/\ap^3)^{1/2}$ is the planet's mean motion. The figure shows that $T$ initially decreases  due to PR drag during the inspiral phase. The planet enters the resonance at $T \simeq 3.1$ and continues to decrease, albeit more slowly, and exits the resonance with $T \simeq 2.7$. Subsequent decay of $T$ on a timescale $\sim t_{\rm pr}(\ap)$ can then occur due to PR drag outside of resonance, but it does not change due to encounters inside the Hill sphere. 

Figure \ref{fig:medfractions} shows the fraction for each of the four possible outcomes as a function of $\ap$ for the medium-size planet. These fractions will be called $f_{\rm pl}$, $f_{\rm st}$ and $f_{\rm ej}$ to hit the planet, hit the dust sublimation zone, or to be ejected, respectively. Each line is labeled with the dust fate as well as the value of $\beta$. 

Immediately apparent is that physical collisions with the planet (blue lines) are a significant fraction of the outcomes for massive planets near the star. This important result was first shown by \citet{2018MNRAS.480.5560B}. It motivates the following sections of this work on the effect of dust accretion on the planetary upper atmosphere.

For dust sublimation radius $\Rsub=4\, R_\odot$, the $\ap = 4.2\, R_\odot$ runs have $f_{\rm st} \simeq 100\%$. 
At  larger $\ap$, $f_{\rm st}$ is still significant as the random walk in $a$ and $e$ allows some particles to lower their pericenter while holding $T$ fixed. Interestingly, $f_{\rm st}$ is sensitive to PR drag since the  $\beta=0.01$ run has much smaller $f_{\rm st}$ than the $\beta=0.1$ and 0.5 runs. The ejection fraction $f_{\rm ej}$ is the opposite, with ejections more common for smaller $\beta$. The fraction hitting the planet is larger for smaller $\beta$ dust due to the longer inspiral timescale.
The ejected fraction grows with $\ap$ and is expected to dominate at sufficiently large $\ap$ (see the Appendix).

Figure \ref{fig:smallfractions} shows the results for the small planet with $\beta=0.1,0.5$. The main difference from Figure \ref{fig:medfractions} is now the complete absence of ejected particles. Impacts on the dust sublimation zone dominate, and the fraction hitting the planet has values of $f_{\rm pl} =10-20\%$, similar to the medium-planet case for the same $\beta$. Additional runs for an Earth-mass planet and $\beta=0.01$ and $0.1$ find maximum $f_{\rm pl}=1-10\%$ and a similar outward decrease.

The opposite happens for the large planet case shown in Figure \ref{fig:bigfractions}. Now ejections dominate and the sublimation zone only dominates at very small $\ap$. The fraction hitting the planet is larger than the medium-size case for the same $\beta$.

While not shown here, runs were also performed for inclination $I=0^\circ$ and they showed significant differences. In particular, hitting the planet was even more dominant, with $f_{\rm pl} \ga 0.9$ for the medium-size planet  over nearly the whole semi-major axis range, and $f_{\rm st}$ much smaller. A few runs were then done at $I=1^\circ-2^\circ$ and it was found that the $f_{\rm pl}$ rapidly decreased to near the $I=10^\circ $ value over that small range. This may be due to a critical inclination $\sim \Rp/\ap \sim 1^\circ$ for which $z \la \Rp$ and collisions are enhanced. Hence, unless the dust accretion disk is very thin, the more conservative $I=10^\circ$ results may be more appropriate, but in any event low inclination dust sources favor accretion onto the planet. The value $I=10^\circ$ is comparable to the {\it Hill inclination} $r_{\rm Hill}/\ap \simeq (\Mp/3\Ms)^{1/3} \sim 5^\circ$ for which $z \la r_{\rm Hill}$. For larger inclinations it is expected that the collision rate decreases. It is also expected that if the planet were allowed finite eccentricity that the results would be similar for $e_{\rm p} \la \rhill/\ap\simeq (\Mp/3\Ms)^{1/3} \sim 0.07$.

The impact speed at which the dust enters the planet's atmosphere is closely related to the Jacobi constant. Let $\vec{u} = \vec{v} - \vec{v}_{\rm p}$ be the relative velocity of the dust and planet. Using the rotating-frame expression for the Jacobi constant evaluated at the position of the planet, the impact speed can be written as
\be
u^2 & = & u_{\rm esc}^2 + u_\infty^2
\label{eq:impactspeed}
\ee
where $u_{\rm esc} = \sqrt{2G\Mp/\Rp}$ is the escape speed from the surface of the planet and the second term 
$u_\infty \simeq v_{\rm p} \sqrt{ 3 - 2\beta - T_\infty}$ is the {\it relative velocity at infinity} as the particle enters the planet's Hill radius. For close-in orbits the two terms in Equation \ref{eq:impactspeed} are comparable but for more distant orbits the escape speed term dominates. Since $\Mp$ varies much more than $\Rp$, the escape speed varies significantly between the small, medium, and large planets. In this context, a Keplerian orbit around the star may be used to approximate
\be
T_\infty & \simeq & (1-\beta)\frac{\ap}{a} + 2 \left[(1-\beta)\frac{a}{\ap} (1-e^2) \right]^{1/2} \cos(I)
\ee
to compare to Figure \ref{fig:trajectory}.

Figure \ref{fig:velocities} shows a histogram of $u_\infty/v_{\rm p}$ for four runs with Jupiter-like planet, a Sun-like star, and $(\ap/R_\odot,\beta)=(10,0.01), (20,0.01), (10,0.1),(20,0.1)$. The normalizations of each curve are different due to different $f_{\rm pl}$. Each distribution is peaked near $u_\infty/v_{\rm p} = 0.4-0.6$, with a tail extending to lower and higher values. For small $\beta$ this corresponds to $T_\infty = 2.65-2.85$ at the time of impact. As in Figure \ref{fig:trajectory}, these values are much smaller than when the particle enters resonance, and are determined by the decrease in $T$ during resonance as well as the subsequent decrease after the resonance is broken.

Figure \ref{fig:mus} shows the cosine of the angle between the inward radial direction and the dust velocity when it impacts the planet, for the same parameters as used in Figure \ref{fig:velocities}. Aside from the difference in normalization due to different $f_{\rm pl}$ in each run, each distribution is broad but dominated by vertical entry.

\section{ stopping and ablation }
\label{sec:stopping}

The goal of this section is to study the stopping and ablation of the dust accreted by the planet. The distribution of ablated mass with altitude forms the {\it meteoric source} for gas-phase atoms discussed in Sections \ref{sec:sourcedensity} - \ref{sec:lines}. The size distribution of the partially ablated remnant dust particles will be used in Section \ref{sec:remnantdust} to compute dust abundance profiles and continuum opacity.

\subsection{ background }

As dust grains enter the atmosphere at highly supersonic speeds, they are decelerated by collisions with gas particles. The collisions occur in the kinetic regime as the mean free paths of gas particles are much larger than the dust grain radius in the upper atmosphere. For the size ranges of interest here, $s \sim 0.1-100\, \rm \mu m$, two main mechanisms contribute to dust ablation, sputtering, and thermal evaporation. 

In thermal evaporation, frictional heating raises the dust temperature, leading to melting, vaporization, and mass loss from the dust particle. This is most important for larger dust particles that take longer to decelerate. 

Sputtering refers to the process where an incident nucleus penetrates the surface layers of the solid grain, inducing a cascade of collisions among a large number of atoms in the solid, some producing a recoil sufficiently near the surface such that a target atom exits the solid \citep{1990ecpi.book.....J}. Sputtering occurs for all grain sizes, but is most important for small grains. 

For both processes, the ablated gas-phase atoms emerge at high speeds, and will be collisionally ionized. However, since the subsequent settling times of the ions (see Section \ref{sec:sourcedensity} below) are longer than ionization and recombination timescales, they will soon come to the ionization state for that layer in the atmosphere. 

\subsection{ radiative heating }
\label{sec:radheat}

While the dust particle is orbiting the star, the stellar radiation field sets the dust temperature. However, as it approaches the planet and is accreted, it also encounters the planet's radiation field, which has comparable flux to that of the star when near the planet.
The equilibrium temperature of the planet's optically thick lower atmosphere is
\be
T_{\rm eq} & =& T_{\rm s} \left( \frac{R_{\rm s}}{2\ap} \right)^{1/2} \simeq 1300\, \rm K\, \left( \frac{T_{\rm s}}{5800\, \rm K} \right)
\left( \frac{10\, R_{s}}{\ap} \right)^{1/2},
\ee
which can approach the dust grain sublimation temperature for close-in orbits. Here $T_{\rm s}$ is the stellar effective temperature.
As grains enter the atmosphere, they may be heated by both the stellar and planetary radiation fields, as well as frictional heating with the atmosphere. 

After stopping, dust grain temperatures in the upper atmosphere are set by a balance of radiative heating and cooling, and are decoupled from the gas temperature. This may allow grains to survive even when the gas temperature is well above the melting temperature of the dust, as for H II regions in the interstellar medium. For gas temperature $T\sim T_{\rm eq}$, number density $n=P/k_{\rm B}T$ and thermal velocity $v_{\rm th} = (8k_{\rm B}T/\pi \mu m_{\rm u})^{1/2}$, the heating rate due to collisions with the gas is $n v_{\rm th} \pi s^2 2k_{\rm B}(T-T_{\rm d})$ \citep{2011piim.book.....D}. The large-grain radiative heating rate due to a blackbody radiation field at temperature $T_{\rm eq}$ is $4\pi s^2 \sigma T_{\rm eq}^4$. Collisions with gas atoms are then only important \citep{1997ApJ...490..368C, 2011piim.book.....D} at pressure levels deeper than 
\be
P_{\rm d, th} & \simeq & \frac{2\sigma T_{\rm eq}^4}{v_{\rm th}} \simeq 2\, \rm mbar \left( \frac{T_{\rm eq}}{10^3\, \rm K} \right)^{7/2} 
\ee
well below the stopping layer. At $P < P_{\rm d, th}$, the dust temperature is set by the stellar and planetary radiation, while below this layer the dust temperature is regulated to be near the gas temperature.

To compute the radiative heating of the grain, the stellar radiation field is taken to be a blackbody with temperature $T_{\rm s}$ and solid angle on the sky $\Omega = \pi (R_{\rm s}/\ap)^2$. The optically thick lower atmosphere of the planet is assumed to produce radiation with temperature $T_{\rm atm} \simeq T_{\rm eq}$ and solid angle $\Omega = 2\pi$. The gas in the upper atmosphere is optically thin to the bulk of the radiation field. For dust temperature $T_{\rm d}$, the net radiative heating rate is then
\be
\dot{E}_{\rm rad} & = & \pi s^2 \sigma \left[ 4 T_{\rm eq}^4  \varepsilon(s,T_{\rm s}) \Theta(\rm day)
\right. \nonumber \\ & + & \left. 2 T_{\rm eq}^4\varepsilon(s,T_{\rm eq})  - 4 T_{\rm d}^4 \varepsilon(s,T_{\rm d}) \right]
\ee
where the step function $\Theta(\rm day)$ is 1 on the dayside and 0 on the nightside. In the numerical calculations, the stellar irradiation will be included for simplicity, $\Theta(\rm day)=1$, as if all the dust was entering on the dayside of the planet. 
These changes in numerical factors will affect the critical dust sizes for which  emissivity is inefficient and also for which the temperature may be raised above the melting temperature by frictional heating. 

The dust grain absorption efficiency for a blackbody radiation field of temperature $T$  is taken to be $\varepsilon(s,T) = {\rm min} \left[ 1, (s/0.1\, \mu \rm m)(T/10^3\, \rm K)^2\right]$, where the numerical coefficient in the second term can vary by 30\% depending on composition \citep{2011piim.book.....D}. This formula interpolates between the case of large dust for which $\varepsilon=1$, and small dust with inefficient emission and absorption with $\varepsilon \propto s$. The critical dust size below which the efficiency is less than unity is $s_{\rm cr}(T)=0.1\, \rm \mu m\, (T/10^3\, \rm K)^{-2}$. Large grains for which $s > s_{\rm cr}(T_{\rm eq}) $ will have dayside temperature $T_{\rm d} = (3/2)^{1/4}T_{\rm eq} \simeq 1.1\, T_{\rm eq}$ independent of grain size. Small grains with $s< s_{\rm cr}(T_{\rm s})$ can be {\it superheated} with dayside temperature $T_{\rm d} \simeq 1.6\, T_{\rm eq} \left( a_{\rm p}/ 10R_{\rm s}  \right)^{1/6} $, again independent of grain size \footnote{ Assuming that they are not so small that heating by collisions with gas particles is dominant over radiative heating and cooling.} . Medium-size grains have a temperature dependent on size, interpolating between these two regimes. 

\subsection{ stopping calculations }

Numerical solutions are computed here using the equations from \citet{2004ACP.....4..627P}, which are used to model stopping in the Earth's atmosphere, as well as alterations described below. While this model only includes thermal evaporation, sputtering will be included here as well (see also \citealt{2008ACP.....8.7015V} for a more comprehensive treatment of both processes). The equations for the altitude $z$, vertical velocity $u_z$ and horizontal velocity $u_x$ are
\be
\frac{dz}{dt} & = & u_z 
\\
\frac{du_z}{dt} & = & - g - \frac{3\Gamma }{4} \left( \frac{\rho}{\rho_{\rm d}} \right) \left( \frac{ u u_z}{s} \right)
\\
\frac{du_x}{dt} & = &  - \frac{3\Gamma }{4} \left( \frac{\rho}{\rho_{\rm d}} \right) \left( \frac{ u u_x}{s} \right).
\ee
The stopping layer is idealized as being geometrically thin with constant gravity $g$, as well as isothermal with gas temperature $T=3000\, \rm K$ and mean molecular weight $\mu=1.23$, appropriate for a mixture of atomic H and He at a pressure level near $P=1\, \mu \rm bar$ \citep{2017ApJ...851..150H}. Here 
$\rho(z)=\rho_0 \exp(-z/H)$ is the atmospheric density, $\rho_0$ is the density at the  $1\, \mu \rm bar$ level and $H=k_{\rm B}T/\mu m_p g$ is the scale height, $\rho_{\rm d}=2\, \rm g\, cm^{-3}$ is the assumed dust grain density, $u = \sqrt{u_x^2 + u_z^2}$ and $\Gamma = 1.2$ is the drag coefficient. Below the melting temperature, or above the melting temperature but for net negative heating, the equations for the dust temperature $T_{\rm d}$ and radius $s$ are
\be
\frac{dT_{\rm d}}{dt} & = & \frac{3}{4\rho_{\rm d} C_{\rm d} s} \left[ \frac{\Lambda}{2} \rho u^3 
\right. \nonumber \\ & + & \left.  4 \sigma T_{\rm eq}^4 \varepsilon(s,T_{\rm s})  + 2 \sigma T_{\rm eq}^4 \varepsilon(s,T_{\rm eq})- 4 \sigma T_{\rm d}^4 \varepsilon(s,T_{\rm d}) \right]
\\
\frac{ds}{dt} & = & - \frac{u \bar{m}_{\rm t}}{2\rho_{\rm d}} \sum_i n_i Y_i
\label{eq:dsdt1}
\ee
while above the melting temperature and for net positive heating they become
\be
\frac{dT_{\rm d}}{dt} & = & 0
\\
\frac{ds}{dt} & = & \frac{1}{ \rho_{\rm d}L} \left[ 
 \frac{\Lambda}{2} \rho u^3 
\right. \nonumber \\ & + & \left.  4 \sigma T_{\rm eq}^4 \varepsilon(s,T_{\rm s})  + 2 \sigma T_{\rm eq}^4 \varepsilon(s,T_{\rm eq})- 4 \sigma T_{\rm d}^4 \varepsilon(s,T_{\rm d})
 \right]
 \nonumber \\ & - & 
 \frac{u \bar{m}_{\rm t}}{2\rho_{\rm d}} \sum_i n_i Y_i.
\label{eq:dsdt2}
\ee
 Here $\Lambda = 0.5$ is the heat transfer coefficient, $C_{\rm d} = 10^7\, \rm erg\, g^{-1}\, K^{-1}$ is the specific heat, $\sigma$ is the Stefan-Boltzmann constant, $L = 3 \times 10^{10}\, \rm erg\, g^{-1}\, K^{-1}$ is the latent heat, and the melting temperature is taken to be $T_{\rm melt} = 1800\, \rm K$. The choices for the coefficients are discussed in \citet{2004ACP.....4..627P} and \citet{2008ACP.....8.7015V}. We include a prescription for the emissivity of small dust, which is more important for the high temperatures near the star than for the Earth.

In the frame of the dust grain, atmospheric atoms of mass $m$ approach the grain with energies $E = 50\, \rm eV (m/m_{\rm u}) (u/100\, \rm km\, s^{-1})^2$, where $m_{\rm u}$ is the atomic mass unit. The sputtering rate is proportional to the number of impacts with energy $E$ above the threshold $E_{\rm thr}$ required for sputtering, which is typically much larger than the lattice energy. While H atoms are the most abundant atmospheric particles, their  energies $E$ are significantly lower, and  He and O impactors will also be included as they are more robustly above the sputtering threshold energies. 

The low-energy approximation for sputtering yields from \citet{1979ApJ...231...77D} is used (see their Section 4.b), in which the dust mass-loss rate is computed as
\be
\dot{m}_{\rm d} = 4\pi \rho_{\rm d} s^2 \dot{s}_{\rm sput} = - 2\pi s^2 u \bar{m}_{\rm t} \sum_i n_i Y_i
\ee
and the sputtering yield has the form $Y_i \simeq  {\rm min}(Y_{i, \rm max}, A^\prime_i (E - E_{\rm thr,i})^2) $. Here $A^\prime$ is a normalization constant and the sputtering threshold energy is $E_{\rm thr,i }$. This form for the mass-loss rate has been used to derive the change in radius in Equations \ref{eq:dsdt1} and \ref{eq:dsdt2}.
A MgSiO$_3$ composition is assumed for the dust, for which the average ejected atomic mass from the target is $\bar{m}_{\rm t} = 20\, m_{\rm u}$. The sum is over the number density $n_i$ and sputtering yield $Y_i $ of each of the three impactors, with the formulas given in Equations 31-33 of \citet{1979ApJ...231...77D}, and shown in their Figure 4. The low energy formula increases as $E^2$ well above threshold, and is used until the broad plateau seen in their Figure 4 is reached. A maximum yield $Y_{\rm i, \rm max}$ is then enforced. For the number densities $n_i$ of the impactors, solar abundance ratios are used for H, He, and O \footnote{ If mixing and meteoric deposition are sufficiently weak, and the abundances of both He and O are greatly decreased above a homopause much lower in the atmosphere, the sputtering rate solely due to H may be much smaller, especially for cases with low entry speeds}. 

The threshold speeds for $Y_i$ to be nonzero are 76, 33, and 17 $\rm km\, s^{-1}$ for H, He and O, respectively, and they approach the maximum yield at significantly higher speeds. Hence, H can only sputter at fairly high incident speeds, while O's threshold speed is below the escape speed for the planets considered here. While O has a low threshold speed, it's low abundance (here taken to be solar) gives a maximum yield at high energy of $(n_{\rm O}/n)Y_{\rm O} \la 5 \times 10^{-4}$. The high energy yield of H and He are much higher, at $(n_{\rm H}/n)Y_{\rm H} \sim (n_{\rm He}/n)Y_{\rm He} \sim 3 \times 10^{-3}$. 

Sputtering typically leads to a tail of mass deposition higher in the atmosphere than that for thermal evaporation, since sputtering shuts off after deceleration.
For ablation solely by sputtering, and assuming the yield can be expressed in the power-law form $\sum_i n_i Y_i \equiv n Y_0 (u/u_0)^\alpha$, then for vertical incidence the $ds/dt$ and $du_z/dt$ equations can be combined to yield a solution for the final grain mass $m_{\rm fin}  =  m_{\rm init} \exp(-\gamma)$, where the exponent is
\be
\gamma & = & \frac{2}{\Gamma} \left( \frac{\bar{m}_t}{\mu m_{\rm u}} \right) \frac{Y_0}{\alpha}
\nonumber \\ & \simeq &  0.2 \left( \frac{1.2}{\Gamma} \right) \left( \frac{\bar{m}_t}{20\, m_{\rm u}} \right) \left( \frac{1.2}{\mu} \right) \left( \frac{Y_0}{0.03} \right) \left( \frac{4}{\alpha} \right).
\label{eq:gamma}
\ee
Hence, for $\alpha=4$ above threshold and below the maximum yield, these fiducial parameters imply that each dust grain will undergo mass loss by the amount of $m_{\rm fin}/m_{\rm init} \simeq \exp(-0.2) = 0.8$. This base level of mass loss is roughly what is found for dust which does not undergo thermal evaporation. The exact values can vary depending on if and by how much each impactor is above threshold. For instance, if all species are at such high speeds than the yields are constant ($\alpha=0$), a power-law solution $m_{\rm fin}/m_{\rm init} \simeq (u_{\rm fin}/u_{\rm init})^\gamma $ results, where now $\gamma$ in Equation \ref{eq:gamma} is redefined without the $1/\alpha$ factor and $Y_0$ may be a factor of 2 higher if the contribution from H is included.

Ignoring ablation, deceleration requires the dust particle of mass $m_{\rm d} = 4\pi \rho_{\rm d} s^3/3$ to collide with its own mass in atmospheric particles, $\pi s^2 H \rho \simeq m_{\rm d}$. This allows an estimate of the pressure level for stopping to be
$ \rho_{\rm d} g s \simeq 0.2\, \rm \mu bar \left(\rho_{\rm d}/2\, \rm g\, cm^{-3} \right) \left( g/10^3\, \rm cm\, s^{-2}\right)
\left( s/1\, \rm \mu m \right)$. This estimate is applicable for smaller dust that does not undergo thermal ablation. For larger dust, a better approximation is that most of the mass is deposited where the particle reaches the melting temperature. Equating the frictional heating per unit area, $ \Lambda \rho u^3/2$, to the thermal emission per area, $4 \sigma  T_{\rm d}^4$, gives an estimate for the stopping pressure to be
\be
P_{\rm ablate} & \simeq & \left( \frac{kT}{\mu m_p} \right) \left( \frac{8 \sigma   T_{\rm d}^4}{\Lambda u^3} \right)
\simeq 3\, \rm \mu bar\,  \left( \frac{T}{3000\, \rm K} \right)
\nonumber \\ & \times &  \left( \frac{1.23}{\mu} \right) \left( \frac{T_{\rm d}}{1800\, \rm K} \right)^4 \left( \frac{ 100\, \rm km\, s^{-1}}{u} \right)^3,
\label{eq:Pablate}
\ee
in  agreement with the numerical results. Note that this estimate is independent of the dust size.

The stopping pressure depends strongly on both the melting temperature and the entry speed. The value $T_{\rm melt}=1800\, \rm K$ is adopted here, motivated by the more detailed study in \citep{2008ACP.....8.7015V}. Models in which elements can have different melting temperatures have also been used to better model the deposition of Na and K for the Earth's atmosphere \citep{2008ACP.....8.7015V}. With $T_{\rm melt}=1200\, \rm K$ for Na, they will be deposited at a factor of 10 lower pressure. Similarly the higher melting temperatures for Ca and Al minerals would lead to deposition at slightly larger pressures. 

Such sublimation could also occur  before the dust accretes onto the planet for orbits close enough to the star. The case of Na and K are particularly interesting. It is expected that planets interior to the sublimation zone for Na and K would not accrete these elements as they would be lost from the dust particles before they are accreted onto the planet. If dust accretion can affect the transmission spectra of exoplanets, it may be expected to see a difference for planets interior to the sublimation zone as compared to exterior planets.

\subsection{ results }

\begin{figure*}[htb!]
\begin{center}
\includegraphics[width=5in, height=5in, trim = {0 2.0in 0 0.5in}, clip]{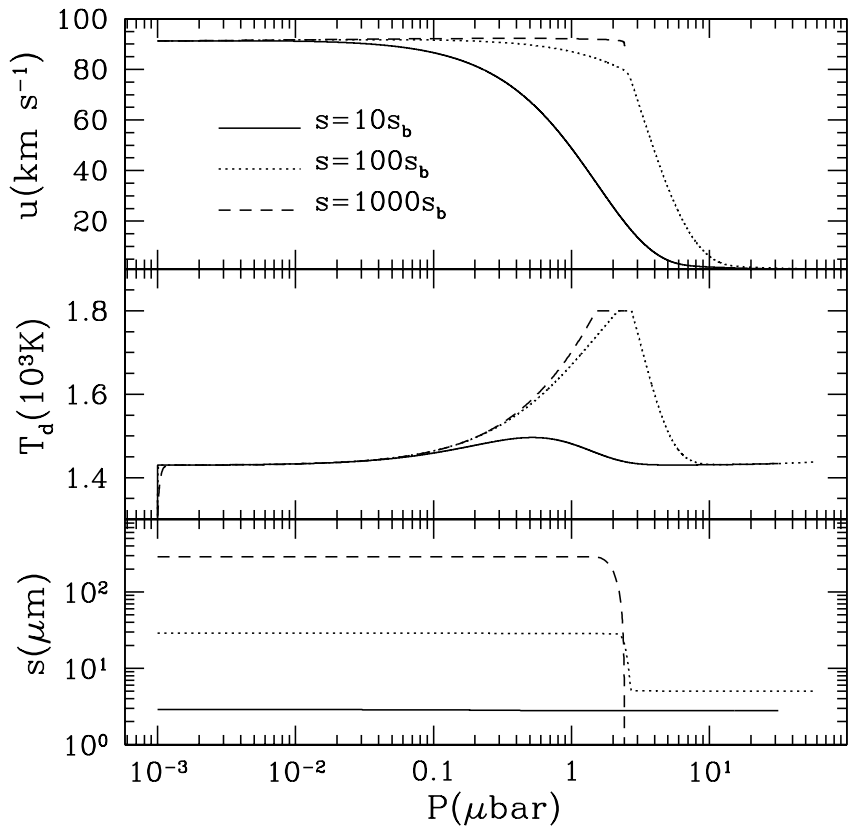} 
\caption{ Individual dust trajectories for sizes $s=10s_b$ (solid) lines, $100s_b$ (dotted lines) and $1000s_b$ (dashed lines) for vertical entry into a medium-size planet at initial speed $u=90\, \rm km\, s^{-1}$. Here the blowout radius is $s_b=0.29\, \mu \rm m$. The three panels show dust speed (top), temperature (middle), and radius (bottom) versus atmospheric pressure as the height coordinate.
\label{fig:traj_fid}
}
\end{center}
\end{figure*}

Figure \ref{fig:traj_fid} shows trajectories for three different size dust particles. The parameters used are a medium-size planet, a Sun-like star, an orbital radius of $\ap=10\, R_\odot$, and an initial speed $u=90\, \rm km\, s^{-1}$ at vertical incidence. The equilibrium temperature is $T_{\rm eq}=1300\, \rm K$, and the expected dust temperature set by radiation is $T_{\rm d, eq} \simeq 1.5^{1/4} T_{\rm eq} =1440\, \rm K$. The three dust sizes are $s \simeq 3, 30$, and $300\, \mu m$.

The $s=3\, \mu \rm m$ dust (solid line) rapidly adjusts to $T_{\rm d, eq}$ and frictional heating leads to only a small temperature increase. It decelerates by pressure levels $P \sim 1\, \mu \rm bar$ and suffers only minor mass loss due to sputtering. The $s=30\, \mu \rm m$ dust (dotted line) must travel deeper into the atmosphere before it begins to decelerate. Frictional heating causes significant decrease in dust size over a fraction of a scale height. Rapid deceleration then leads to a decrease in frictional heating and temperature, and it cools back to $T_{\rm d, eq}$. The $s=300\, \mu \rm m$ dust (dashed line) also undergoes strong frictional heating, but does not decelerate in time to avoid catastrophic mass loss. The trajectory was stopped when the dust reached a size $s=0.01\, \mu \rm m$. 

While this example used a fixed entry speed, Equation \ref{eq:Pablate} implies that a range of speeds will lead to a range of stopping altitudes and distribution of mass deposited with altitude. Since large dust undergoes more mass loss, grain size distributions with mass dominated by large grains may deposit more mass as gas-phase atoms, while small grain size distributions may retain more remnant grains that have undergone little ablation. Lastly, planets closer to the star will have $T_{\rm d, eq}$ closer to $T_{\rm melt}$ and the smaller required heating to melt implies mass deposition higher in the atmosphere as compared to planets further from the star.

\begin{figure*}[htb!]
\epsscale{0.9}
\vspace*{-0.5in}
\plotone{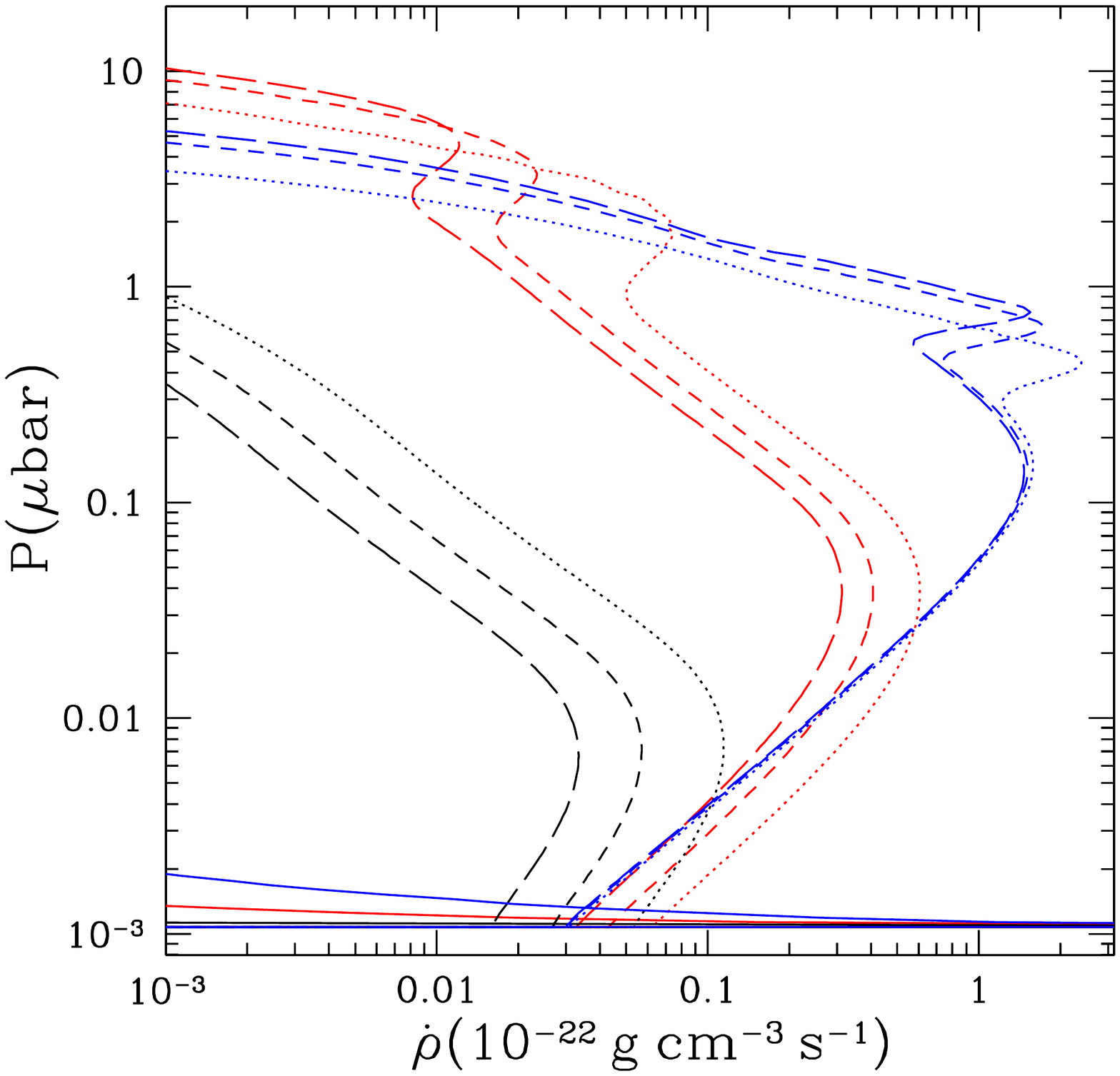}  
\vspace*{-1.3in}
\caption{ Deposition rate of ablated dust mass per volume as a function of pressure for the G85 size distribution. The black, red, and blue lines correspond to small, medium, and large planets. The solid lines that form spikes near $P\sim 1\, \rm nbar$ are for close-in planets at $\ap=6\, R_\odot$. The dotted, short-dashed, and long-dashed lines are then for $\ap/R_\odot=10, 14$, and $18$. The assumed accretion rate into the atmosphere is $f_{\rm pl} \dot{M}_{\rm d} = 10^8\, \rm g\, s^{-1}$.
\label{fig:rhodot_grun}
}
\end{figure*}

\begin{figure*}[htb!]
\epsscale{0.9}
\vspace*{-0.5in}
\plotone{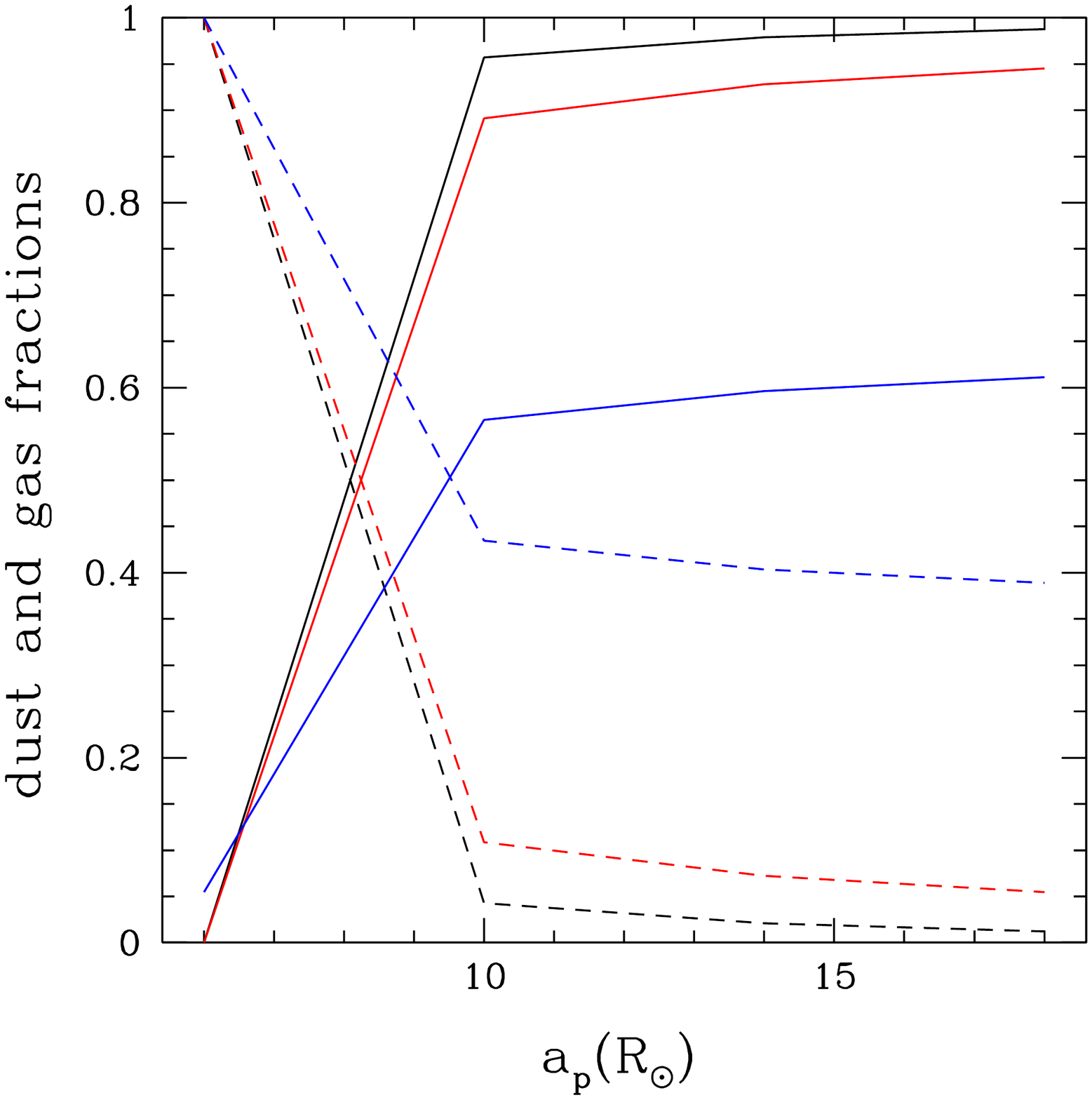}  
\vspace*{-1.3in}
\caption{ Fraction of decelerated dust mass in gas-phase atoms and remnant dust particles. The G85 dust distribution has been used, with a cutoff at $\beta=1$. Solid lines are for dust and dashed for gas. The black, red, and blue lines represent small, medium, and large planets.
\label{fig:ablate_efficiency_grun}
}
\end{figure*}

Further progress requires a choice of grain size distribution that depends on the size distribution and number density at the source, PR drag and radiation pressure, collisional agglomeration, and erosion, encounters with planets, and the stellar wind environment near the star \citep{1985Icar...62..244G, 2000A&A...362.1158I, 2005A&A...433.1007W, 2007A&A...472..169T, 2008ACP.....8.7015V, 2009Icar..201..395K, 2010ApJ...713..816N, 2011ApJ...743..129N, 2012ChSRv..41.6507P, 2014A&A...571A..51V, 2015GeoRL..42.6518C, 2016ApJ...816...50R}. 
 Two extreme limits are considered here. \citet{1985Icar...62..244G}, G85 from here on, gave a model for the solar system dust size distribution, including both PR drag and grain-grain collisions. Due to the relatively low grain density inside 1au, the collision rate is small for this model and it has both mass and area dominated by larger grains $s \sim 100\, \rm \mu m$. We augment their result by imposing a cutoff at the blowout radius. By contrast, the calculations in \citet{2014A&A...571A..51V} are for a much higher grain density in which erosive collisions lead to a distribution in which number, mass, and area are dominated by grains with $\beta \sim 0.5$ near the blowout size.

Motivated by Figure \ref{fig:velocities}, the distribution of $u_\infty/\vp$ is chosen to be a Gaussian with mean $0.5$ and standard deviation $0.1$, and the escape speed contribution is then included as $u = \sqrt{ u_\infty^2 + 2G\Mp/\Rp}$. The angle distribution is chosen to be $P(\mu)d\mu = 2\mu d\mu$ to represent  the curves in Figure \ref{fig:mus}. 

Figures \ref{fig:rhodot_grun} and \ref{fig:ablate_efficiency_grun} show the results for dust ablation for the G85 size distribution.
 For comparison, the G85 distribution is shown in Figure \ref{fig:size_mass_med_grun} as the dashed line.

Figure \ref{fig:rhodot_grun}  shows the ablated mass per volume per time, $\dot{\rho}$, as a function of pressure. Three different planet sizes are shown as well as four different orbital separations. For all planet masses, at orbital separation $\ap=6\, R_\odot$ the equilibrium temperature is high enough that only a small amount of frictional heating leads to thermal evaporation immediately near the top of the atmosphere, which is set to be at $P=1\, \rm nbar$. For the other three orbital separations, the mass is distributed much lower in the atmosphere. All three planet masses show a peak higher up due to sputtering, while the $\Mp=1$ and $10\, M_{\rm J}$ runs show a secondary peak lower in the atmosphere due to thermal evaporation. The fraction of the incident dust mass that is ablated is larger for the high mass planet as compared to the two lower mass cases.

Figure \ref{fig:ablate_efficiency_grun} shows the fraction ($\epsilon_{\rm abl}$) of incident dust mass which is ablated as gas-phase atoms and which remains as remnant dust particles ($\epsilon_{\rm rem}=1-\epsilon_{\rm abl}$). 
Again, the large speeds at small $\ap$ imply all the dust is ablated. For larger separations, the escape speed has a significant effect, with the massive planet having a much larger amount of ablation. Beyond $\ap \sim 10\, R_\odot$ the escape speed contribution dominates over the orbital relative velocity.

\begin{figure*}[htb!]
\epsscale{0.9}
\vspace*{-0.5in}
\plotone{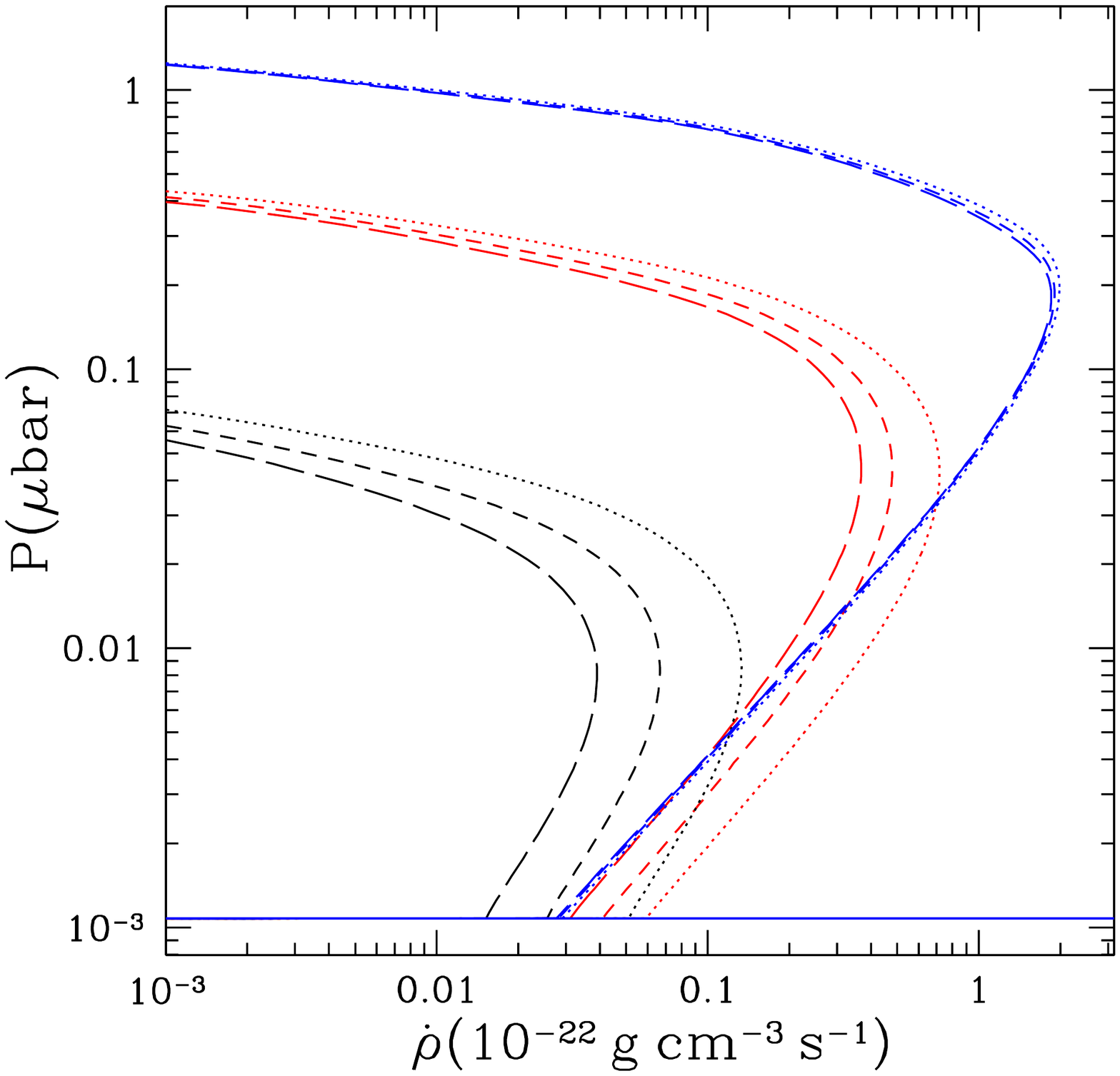}  
\vspace*{-1.3in}
\caption{ Same as Figure \ref{fig:rhodot_grun} but for all dust with the same initial size, set by $\beta=0.5$. 
\label{fig:rhodot_beta0p5}
}
\end{figure*}

Figure \ref{fig:rhodot_beta0p5} is the same as Figure \ref{fig:rhodot_grun}, but for a delta-function size distribution with all initial grains with $\beta=0.5$, or $s=0.57\, \mu \rm m$. The contribution from sputtering is still present in this case, but the deeper thermal ablation peaks are absent as the grains are able to decelerate before rising to the melting temperature. The fraction of mass which is ablated and remaining in the remnant grains is similar to that found for the G85 distribution in Figure \ref{fig:ablate_efficiency_grun}.

The scenario studied here, of a distant dust source and nearly circular orbits, will produce small collision speeds. The collision speeds for highly eccentric orbits will tend to be larger. A dust particle with $a \gg \ap$ and $a(1-e) \ll \ap$ will have relative speed $u_\infty \simeq 3^{1/2} \vp$ due to radial motion near the escape speed. A dust particle with $a \gg \ap$ but $a(1-e) \simeq \ap$ will have 
$u_\infty \simeq \vp \left( \sqrt{2(1-\beta)} \pm 1 \right)$ for prograde and retrograde dust orbits. This formula gives $12-72\, \rm km\, s^{-1}$ for the Earth. Significantly higher dust speeds will result in more of the dust mass being ablated into gas-phase atoms than the cases considered here.

\section{ gas-phase atoms }
\label{sec:sourcedensity}

In this section the abundance profile of a trace species deposited into the atmosphere by dust accretion, and slowly drifting downward due to gravitational settling, will be considered. It will be shown that a constant mixing ratio results (see e.g., \citealt{1980JAtS...37.1342H}), and that the resulting abundances are comparable to solar composition for a dust accretion rate comparable to $\dot{M}_{\rm d, iss}$. The effects of vertical mixing, molecular diffusion, and upward motion of the background species due to atmospheric escape will also be included in Section \ref{sec:diffusion}.

Let the trace species have mass $m$, number density $n(z)$, vertical speed $u(z)$ and flux $F(z)=n(z)u(z)$. Similarly, the background species, which we will consider to be H$_2$, has mass $m_{\rm b}$, density $n_{\rm b}(z)$ velocity $u_{\rm b}(z)$ and flux $F_{\rm b}(z)=n_{\rm b}(z) u_{\rm b}(z)$. The gas is assumed isothermal with temperature $T$. The collision frequency of the trace with the background will be written as $\nu = \langle \sigma v \rangle n_{\rm b}$. 

In the hard-sphere model, the transport collision rate coefficient between a trace particle and the background particles is  \citep{1987tpaa.book.....C} 
\be
\langle \sigma v \rangle & =& \frac{8\sqrt{2}}{3\sqrt{\pi}}\, \left(\frac{k_{\rm B}T m_b}{(m_b+m)m}   \right)^{1/2} \pi \left(R_{\rm coll} + R_{\rm coll,b}\right)^2
\nonumber \\ & \simeq & 1.7 \times 10^{-9}\, \rm cm^3\, s^{-1}\, \left( \frac{T}{10^3\, \rm K} \right)^{1/2}\, 
\nonumber \\ & \times & 
\left( \frac{m_b m_u}{m(m+m_b)} \right)^{1/2}\, \left( \frac{R_{\rm coll}+R_{\rm coll,b} }{3 \AA} \right)^2.
\ee
Here $m_b$ is the mass of the background species and $R_{\rm coll, b}$ and $R_{\rm coll}$ are the hard-sphere radii for the background and trace. Atomic radii are only measured for a few species, as discussed in \citet{2007P&SS...55.1414G} and \citet{1973aero.book.....B}. Following \citet{1973aero.book.....B}, from here on we will assume a radius $R_{\rm coll, t}=R_{\rm coll, b}=1.5 \times 10^{-8}\, \rm cm$ for all species. 

A heavy atom accelerated downward by gravity $g$, and colliding with the lighter background atoms with timescale $\nu^{-1}$ will have a downward drift velocity of
\be
u & \simeq & - \frac{g}{\nu}
\nonumber \\ & \simeq & 
- 0.22 \, \rm cm\, s^{-1} 
 \left( \frac{P}{1\mu \rm bar} \right)^{-1} \left( \frac{T}{10^3\, \rm K} \right)^{1/2}\, 
\nonumber \\ & \times & 
\left( \frac{g}{g_{\rm J}} \right) \left( \frac{m(m+m_{\rm b})}{m_{\rm b}m_{\rm u}} \right)^{1/2},
\ee
where the gravity of Jupiter is $g_{\rm J}=2700\, \rm cm\, s^{-2}$.
The settling time over a scale height is
\be
t_{\rm settle} & = & \frac{H\nu}{g} = 1.4 \times 10^8\, \rm s\, 
 \left( \frac{P}{1\mu \rm bar} \right) \left( \frac{T}{10^3\, \rm K} \right)^{1/2}\, 
\nonumber \\ & \times & 
\left( \frac{g}{g_{\rm J}} \right)^{-2} \left( \frac{m_u^3}{m_b m(m+m_b)} \right)^{1/2}.
\ee
This timescale is much longer than photoionization and recombination times, and the atoms quickly come into local ionization equilibrium \footnote{ Here the atoms are treated as neutral. Including the upward charge-separation electric field will effectively decrease the downward gravity, and using ion-neutral cross sections will give larger collision rates than the neutral-neutral rates considered here. Hence, the results underestimate the densities of ionized species. Neutral atoms are considered for simplicity.}.

The settling speed can be used to find the mixing ratio 
\be
\frac{n}{n_{\rm b}} & = & \frac{F}{u n_{\rm b}} = - \frac{F \nu}{ g n_{\rm b} } = - \frac{ F \langle \sigma v \rangle }{ g }
\label{eq:vmr_source}
\ee
where $F<0$ for downward motion. Since the background density has canceled out of the right-hand side, the result will vary slowly with altitude.

The particle flux below the source can be written
\be
F & = & - f_{\rm pl} \epsilon_{\rm abl} f \frac{\dot{M}_{\rm d}}{4\pi R_{\rm p}^2 \bar{m}}
\nonumber \\ & \simeq & 
- 6.5 \times 10^9\, \rm cm^{-2}\, s^{-1}\, f  \left( \frac{ f_{\rm pl} \epsilon_{\rm abl} \dot{M}_{\rm d} }{ 10^8\, \rm g\, s^{-1}} \right)
\left( \frac{ R_{\rm p} }{ R_{\rm J} } \right)^{-2}. 
\ee
Here $ \dot{M}_{\rm d}$ is the accretion rate of the dust population as it moves toward the star under PR drag, $f$ is the abundance by number of this element in the dust particles, and $\bar{m} \simeq 15\, m_{\rm u}$ is the mean mass of the elements making up the dust particles.

\begin{deluxetable}{cccc}
\tablenum{1}
\tablecaption{Abundances by Number \label{tab:abundances}}
\tablewidth{0pt}
\tablehead{
\colhead{Element} & \colhead{Solar} & \colhead{Chondritic}  & \colhead{Dust Layer}}
\startdata
         H        &      9.256e-01        &      3.144e-01        &               \\
        He        &      7.343e-02        &      3.454e-08        &      6.329e-11         \\
         C        &      2.272e-04        &      4.417e-02        &      2.897e-05         \\
         N        &      6.259e-05        &      3.165e-03        &      1.790e-06         \\
         O        &      4.535e-04        &      4.319e-01        &      2.147e-04         \\
        Ne        &      6.894e-05        &      1.347e-10        &      5.342e-14         \\
        Na        &      1.846e-06        &      3.286e-03        &      1.148e-06         \\
        Mg        &      3.274e-05        &      5.947e-02        &      1.966e-05         \\
        Al        &      2.699e-06        &      4.751e-03        &      1.418e-06         \\
        Si        &      3.210e-05        &      5.719e-02        &      1.641e-05         \\
         S        &      1.428e-05        &      2.544e-02        &      6.407e-06         \\
        Ar        &      3.290e-06        &      5.502e-10        &      1.116e-13         \\
        Ca        &      2.018e-06        &      3.413e-03        &      6.897e-07         \\
        Cr        &      4.127e-07        &      7.508e-04        &      1.173e-07         \\
        Fe        &      2.690e-05        &      4.936e-02        &      7.184e-06         \\
        Ni        &      1.534e-06        &      2.733e-03        &      3.787e-07         \\
        \enddata
\tablecomments{Columns 2 and 3 are the solar photospheric and CI chondrite abundances  taken from \citet{2003ApJ...591.1220L}. All elements with abundances 1\% that of Si or larger are shown. The fourth column shows the mixing ratio below the meteoric dust layer from Equation \ref{eq:big_result_vmr} for $f_{\rm pl} \epsilon_{\rm abl} \dot{M}_{\rm d}  = 10^8\, \rm g\, s^{-1} $,  $M_{\rm p} = M_{\rm J}$ and $T=1000\, \rm K$. }
\end{deluxetable}

\begin{figure*}[htb!]
\epsscale{1.0}
\plotone{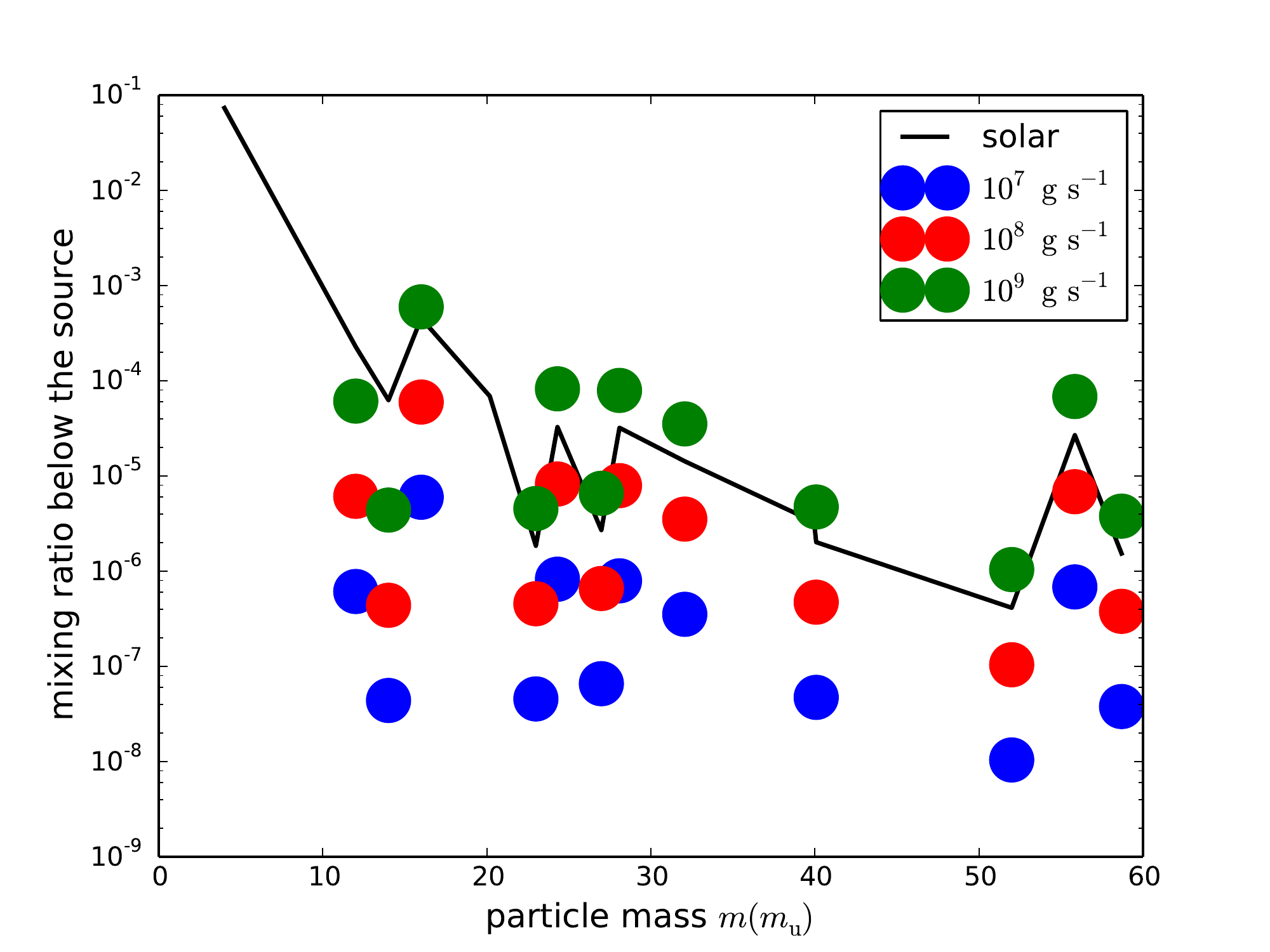}  
\caption{ Gas-phase mixing ratios below the source (dots) as compared to solar abundances (line). The three different dots are for 
 accretion rates $f_{\rm pl} \epsilon_{\rm abl} \dot{M}_{\rm d}[\rm g\, s^{-1}]  = 10^{7,8,9}  $.
 \label{fig:abundances}
}
\end{figure*}

The results are now assembled to compute the mixing ratio below the meteoric source using Equation \ref{eq:vmr_source}. The result is
\be
\frac{n}{n_{\rm b}} & = & 
4.1 \times 10^{-3} \rm cm^{-3}\, f\, \left(   \frac{ f_{\rm pl} \epsilon_{\rm abl}  \dot{M}_{\rm d} }{ 10^8\, \rm g\, s^{-1}} \right)
\nonumber \\ & \times & 
\left( \frac{ M_{\rm p} }{ M_{\rm J} } \right)^{-1}
\left( \frac{T}{10^3\, \rm K} \right)^{1/2}\, 
\left( \frac{m_{\rm b} m_{\rm u}}{m(m+m_{\rm b})} \right)^{1/2}.
\label{eq:big_result_vmr}
\ee
The relative number of different elements scales as $f/m$ for $m \gg m_{\rm b}$, and favor small planets due to the smaller gravity.

The mixing ratios below the meteoric source from Equation \ref{eq:big_result_vmr} are shown in Table \ref{tab:abundances} and Figure \ref{fig:abundances}.  If $f_{\rm pl} \epsilon_{\rm abl}   \dot{M}_{\rm d} \simeq 10^8-10^9\, \rm g\, s^{-1} $ the abundances of C, O, Na, Mg, Al, Si, S, Ca, Cr, Fe, and Ni are comparable to their solar abundance. Given the previous successes in modeling transmission spectra for several lines using solar abundances (e.g. \citealt{2017ApJ...851..150H}), the meteoric dust model provides a natural explanation for these abundances. However, some elements are notably underabundant, the main example being He, which has tiny abundance in meteoritic dust. 

Given the rough agreement between the abundances due to the meteoric source and solar abundances, and the difficulty in measuring absolute abundances for a number of elements, it may be difficult to tell the difference between heavy elements mixed up from below and the effect of dust accretion.

\section{ mixing in the upper atmosphere  }
\label{sec:Kzz}

A crucial consideration for high altitude absorbers is the size of the vertical mixing coefficient $K$ (short for $K_{\rm zz}$). Opposite to the case of heavy elements mixed up from below, where mixing must be strong to loft them to high altitudes, mixing must be sufficiently {\it weak} for the meteoric source to dominate and set a constant mixing ratio \footnote{ Unless  the dust accretion rate is so high that the implied abundances due to the meteoric source were well in excess of the deep, well-mixed atmosphere.}. Turbulent mixing, as parameterized by $K$, arises naturally in a convective atmosphere, where $K \sim u_{\rm eddy} \ell_{\rm eddy}$ is related to the size ($\ell_{\rm eddy}$) and speed ($u_{\rm eddy}$) of the energy-bearing eddies. However, for strongly irradiated extrasolar planets, the entire atmosphere at $P \la 10^2-10^3 \, \rm bar$ may be stably stratified (e.g. \citealt{2006ApJ...650..394A}) and mixing is instead due to winds driven by day-night temperature differences. 

An analytic theory for $K$ due to winds in stably stratified atmospheres has been developed by  \citet{2018ApJ...866....1Z}, and a detailed comparison to global circulation models including tracer particles was carried out by \citet{2019ApJ...881..152K}. The latter paper includes radiation transfer in the two-stream and two-frequency band (optical and infrared) approximation, with the upper boundary near the infrared photosphere at $P \sim 1\, \rm mbar$ and the lower boundary well below the optical photosphere at $P \sim 0.1\, \rm bar$. A drag force is included near the lower boundary, to represent friction with the interior planet where winds are negligible. Rayleigh drag is also included with constant drag timescale throughout the simulation domain. Excluding Rayleigh drag, and for long chemical equilibration times, their numerical results show turbulent mixing as large as $K \sim 0.01\, c_s H_{\rm b} \sim 10^{10}\, \rm cm^2\, s^{-1}$ for the case of strongly irradiated atmospheres with high wind speeds, where $c_s \sim 1\, \rm km\, s^{-1}$ is the sound speed. Since the wind speeds in this simulation increased upward, $K$ showed a strong vertical increase. When Rayleigh drag is included, wind speeds can be strongly suppressed for large drag, thereby decreasing $K$. 

A key question is how to extend these results for $K$ to lower pressures in the $P \sim \rm mbar-\mu bar$ range appropriate for transit observations. 

Since vertical ballistic motion of a fluid element at speed $\sim c_{\rm s}$ can only result in 1 scale height of  displacement, the deposition of the stellar optical continuum flux at the optical photosphere, and subsequent transport of heat up to the infrared photosphere by radiative diffusion and fluid motion, may not be able to coherently drive fluid motions in the  $P \sim \rm mbar-\mu bar$ region, well into the optically thin portion of the atmosphere (along a vertical path) \footnote{ Upward traveling {\it thermal tide} waves excited in the lower atmosphere may also give mixing in the upper atmosphere \citep{1970attg.book.....C, 1971ASSL...25..122L, 2010ApJ...714....1A}. The focus here is instead on atmospheric circulation, as that has been investigate in more detail.}. Rather,  the local imbalance of optically thin heating and cooling, which taps into only a small fraction of the total stellar radiation, may drive the motions in the upper atmosphere. Simulations including appropriate input physics (optically thin, non-LTE, magnetic field effects, etc.) are required to address this question.

A second effect that becomes of primary importance in the upper atmosphere is the {\it Lorentz drag}. Effectively, due to the finite magnetic diffusivity, gas moving perpendicular to the magnetic field will experience a drag force attempting to damp out that motion \footnote{ For a Jupiter-size magnetic field, magnetic pressure dominates gas pressure for $P \la 1\, \rm \mu bar$ and the gas will be incapable of inducing large perturbations to the field in the high conductivity limit.}. This is analogous to the friction with the deep planetary interior. The {\it weather-layer}, with strong winds relative to the mean planetary rotation, should be confined between these upper and lower layers where gas is forced to corotate.

To briefly review the origin of this force, stellar heating-driven flows at speed $u$ attempt to advect the magnetic field ($B$), creating currents ($J$). Ohmic diffusion \footnote{ The Hall and ambipolar diffusion effects will also become important at sufficiently low density, but are ignored here for simplicity.} attempts to dissipate these currents with a diffusion coefficient $\eta$. When advection balances diffusion, currents of size $J \sim (c/4\pi) (uB/\eta)$ give rise to a drag force per volume $JB/c \sim uB^2/(4\pi \eta)$ and {\it Lorentz drag} timescale $t_{\rm drag} = 4\pi \rho \eta / B^2$. When the drag time becomes smaller than the characteristic flow timescale $\Rp/u \sim (10^{10}\rm cm/ \rm 1\, km\, s^{-1}) \sim 1\, \rm day$, the drag forces can greatly decrease the flow speeds  \citep{2018ApJ...866....1Z} and hence the mixing coefficient $K$.

In particular the Lorentz drag timescale becomes  short at lower pressures,  due to the decrease in both density $\rho$ and also diffusivity  $\eta$. The magnetic diffusivity is given by $\eta \simeq 230\, \rm cm^2\, s^{-1} \sqrt{T/K}\, x_e^{-1}$ \citep{2010ApJ...719.1421P}, where $x_e$ is the mixing ratio of electrons. While it is appropriate to determine $x_e$ using collisional ionization of Na and K deep in the atmosphere, above the $P \sim 1\, \rm mbar$ level photoionization dominates \citep{Fortney2003, 2014ApJ...796...15L, 2017ApJ...851..150H, 2017ApJ...847...32L}. If a solar abundance of Na were fully ionized, the resulting $x_e \simeq 4.3 \times 10^{-6}$ would give a diffusivity $\eta = 2 \times 10^9\, \rm cm^2\, s^{-1}\, \sqrt{T/10^3\, \rm K}$ and a drag timescale $t_{\rm drag}=0.04\, \rm s\, \left(P/1\, \rm \mu bar \right) \left( T/10^3\, \rm K \right)^{-1/2} \left( B/4.3\, \rm G \right)^{-2}$, where Jupiter's equatorial magnetic field has been used. In a detailed calculation for HD 189733b, \citet{2017ApJ...847...32L} find $x_e \sim 10^{-8}$ ($10^{-6}$) at $P=1\, \rm mbar$ ($1\, \rm \mu bar$), giving the range $t_{\rm drag} = 10^4 (10^{-1})\, \rm s$. Even the longer drag time $t_{\rm drag}=10^4\, \rm s$ was found to greatly decrease the flow speeds and $K$ by \citet{2019ApJ...881..152K}.

To estimate the upper atmosphere flow speeds and mixing coefficient, we follow the one-zone model of  \citet{2017ApJ...836...73Z}, now applied locally to the upper atmosphere instead of the lower atmosphere. Using dimensional analysis on the horizontal Euler equation gives
\be
\frac{u^2}{\Rp} + \Omega u & \sim & \frac{c_s^2}{\Rp} - \frac{u}{t_{\rm drag}}
\ee
where the terms represent fluid acceleration, the Coriolis force, the day-night pressure gradient force, and the drag force. Here we have assumed large day-night temperature difference and ignored the factors of order unity in the more detailed formula in  \citet{2017ApJ...836...73Z}. For weak drag and Coriolis terms, the flow speed is $u \sim c_{\rm s}$. However, when drag forces decrease the flow speed, the two terms on the left-hand side become negligible and a balance of the two terms on the right-hand side gives
\be
u & \sim & c_{\rm s} \left( \frac{ t_{\rm drag}}{t_{\rm adv}} \right),
\label{eq:ustrongdrag}
\ee
which is smaller than $c_{\rm s}$ when $t_{\rm drag}$ is smaller than the fiducial (drag-free) horizontal advection timescale $ t_{\rm adv} \equiv \Rp/c_{\rm s} \sim 1\, \rm day$. As a numerical example, for the range $t_{\rm drag} = 10^4 - 10^{-1}\, \rm s$, and advection time is $t_{\rm adv}=10^5\, \rm s$, the flow speed would be reduced a factor of $t_{\rm drag}/t_{\rm adv}=10^{-1}-10^{-6}$ over the range $P=\rm mbar - \mu bar$.

Clearly, in the upper atmosphere of planets receiving strong ionizing stellar flux and which creates high ionization levels, the Lorentz drag timescale may be orders of magnitude smaller than $t_{\rm adv}$, and significantly decrease the flow speeds. \citet{2019ApJ...881..152K} find a day-night morphology for the flow when drag is strong, as opposed to an equatorial zonal wind when drag is weak.

The characteristic vertical speed $w \sim u (H_{\rm b}/\Rp)$ gives rise to a vertical mixing coefficient \citep{2018ApJ...866....1Z, 2019ApJ...881..152K}
\be
K & \simeq & \frac{\left( \frac{uH_{\rm b}}{\Rp} \right)^2}{t_{\rm chem}^{-1} + \frac{u}{\Rp}},
\ee
where $t_{\rm chem}$ is the  timescale for source terms to significantly change the tracer density due to e.g. photoionization and recombination times for atoms, and growth times for dust. Hence, $K \propto u^2 (u)$ for short (long) $t_{\rm chem}$. For long $t_{\rm chem}$ and weak drag, $u \simeq c_{\rm s}$ gives $K \simeq c_{\rm s}H_{\rm b} \left( H_{\rm b} /\Rp \right)$. This is the expected value for the lower atmosphere when drag is weak. In the strong drag case, plugging in the flow speed from Equation \ref{eq:ustrongdrag} gives
\be
K & \simeq &c_{\rm s} H_{\rm b} \left( \frac{H_{\rm b}}{\Rp} \right)  \left( \frac{ t_{\rm drag}}{t_{\rm adv}} \right),
\ee
which is smaller by a factor of $t_{\rm drag}/t_{\rm adv}$ than the weak drag estimate. If the horizontal flow timescale $t_{\rm adv}(t_{\rm drag}/t_{\rm adv})^{-1}$ becomes longer than the chemical timescale, the expression becomes
\be
K & \simeq & c_{\rm s} H_{\rm b} \left( \frac{H_{\rm b}}{\Rp} \right) 
 \left( \frac{ t_{\rm chem}}{t_{\rm adv} }\right)  \left(\frac{ t_{\rm drag}}{t_{\rm adv}} \right)^2,
\ee
which is now smaller by both the  factors $t_{\rm chem}/t_{\rm adv} \ll 1$ and $\left( t_{\rm drag}/t_{\rm adv} \right)^2 \ll 1$. For the fiducial range $t_{\rm drag}/t_{\rm adv}=10^{-1}-10^{-6}$ over the range $P=\rm mbar - \mu bar$, this would decrease $K$ by $10^{-1}-10^{-6}$ for the linear scaling and at least $10^{-2}-10^{-12}$ for the quadratic scaling.

While a more detailed calculation is warranted, it's clear that Lorentz drag is an important effect over the pressure range $P \sim \rm mbar-\mu bar$ and that most likely $K$ is much smaller than the fiducial value $c_{\rm s} H_{\rm b}^2/\Rp  \sim 10^{10}\, \rm cm^2\, s^{-1}$ applicable in the lower atmosphere. If $K$ is indeed reduced by many orders of magnitude, then the homopause at $D  = K$ may occur near near the base of the upper atmosphere, where Lorentz drag first begins to suppress the day-night flows. In this case, since heavy elements may not be mixed up from below, the meteoric source may be required for heavy elements to be present in the upper atmosphere probed by optical and infrared transmission spectra.

\section{ diffusion, mixing, and escape }
\label{sec:diffusion}

In Section \ref{sec:sourcedensity}, only the effect of gravitational settling was included. Here vertical mixing, molecular diffusion, and upward drag due to atmospheric escape of the background species are included as well. To goal is to understand the conditions under which the meteoric dust abundance estimates from Section \ref{sec:sourcedensity} are applicable. 

For a heavy trace species, the abundance profiles can be  affected by their much smaller scale height. Let $a=m/m_{\rm b}$ be the mass ratio of the heavy trace particle to the background particle. The scale height of the background is $H_{\rm b} = k_{\rm B}T/(m_{\rm b} g)$ and that of the trace is $H = H_{\rm b}/a$. A second important consideration is the size of the molecular diffusion coefficient $D$ as compared to the  vertical mixing coefficient $K$. Typically $D$ will increase upward faster than $K$ and above the homopause altitude 
$z_{\rm h}$, at which  $D(z_{\rm h})=K(z_{\rm h})$, the atmosphere will no longer be well mixed and diffusion effects will become important.
 
The vertical momentum equation of the trace is
\be
0 & \simeq & - k_{\rm B}T \frac{dn}{dz} - m n g + m n \nu \left( u_{\rm b} - u \right)
\label{eq:diff1}
\ee
which can be rewritten as 
\be
D \left( \frac{d n}{dz} +  \frac{n}{H} \right) & =& n \left( \frac{ F_{\rm b} }{n_{\rm b}} -  \frac{ F } { n } \right)
\ee
where the diffusion coefficient is $D = k_{\rm B}T/(m\nu)$.
 For the case of zero fluxes, $F_{\rm b}=F=0$, the solution is called {\it diffusive equilibrium}
\be
\frac{n(z)}{n_{\rm b}(z)} & =& \frac{n(z_0)}{n_{\rm b}(z_0)} e^{-(a-1)(z-z_0)/H_{\rm b}},
\ee
where the background density profile $n_{\rm b}(z) = n_{\rm b}(z_0) \exp(-(z-z_0)/H_{\rm b})$. Species  heavier than the background ($a \gg 1$) will decrease upward much faster than the background and their mixing ratios will become small. Hence, in diffusive equilibrium, the optical depth of the trace will decrease upward strongly and the transit radius will be limited to be near the homopause.

The transition between the lower, well-mixed layer and the upper layer where diffusion is important can be evaluated by adding in the phenomenological mixing term
\be
D \left( \frac{d n}{dz} +  \frac{n}{H} \right) + K \left( \frac{d n}{dz} +  \frac{n}{H_{\rm b}} \right) & =& n \left( \frac{ F_{\rm b} }{n_{\rm b}} -  \frac{ F } { n } \right)
\label{eq:diff2}
\ee
where $K$ is the vertical mixing  coefficient. The term with $K$ tries to keep the scale height of the trace to be the same as that of the background.
Following \citet{1987tpaa.book.....C}, the effects of a meteoric source layer and drag forces from atmospheric escape can be included through solution of the two equations for $n(z)$ and $F(z)$
\be
\frac{dn}{dz} & = & - \frac{1}{D+K} \left[ \left(  \frac{D}{H} + \frac{K}{H_{\rm b}} - \frac{ F_{\rm b} }{n_{\rm b}}  \right) n - F \right]
\\
\frac{dF}{dz} & = & \frac{F_{\rm s}  }{ \sqrt{2\pi} \sigma_{\rm s} } e^{-(z-z_{\rm s})^2/2\sigma_{\rm s}^2}.
\ee
While the detailed profiles of $dF/dz=f\dot{\rho}/m$ from Section \ref{sec:stopping} could be used as the source here, for simplicity, the atomic source has been given by a Gaussian profile with total flux $F_{\rm s}>0$, centered at altitude $z_{\rm s}$, and with width $\sigma_{\rm s}$.
The density of the background will be referenced to the homopause as $n_{\rm b}(z) = n_{\rm b}(z_{\rm h}) \exp(-Z)$ where $Z = (z-z_{\rm h})/H_{\rm b}$. Assuming a constant $K$ everywhere, the diffusion coefficient can be written $D(z) = K \exp(Z)$. The boundary conditions are the mixing ratio $\xi \equiv n(z_{\rm 0})/n_{\rm b}(z_0)$ at a reference altitude $z_{\rm 0}$ deep in the well-mixed layer, and $F(z_0) = F_{\rm esc} - F_{\rm s}$, where $F_{\rm esc} $ is only nonzero when blowoff is occurring for large $F_{\rm b}$, as discussed below. This choice for the flux gives $F = F_{\rm esc}$ at the top of the atmosphere and the source gives a downward flux $-F_{\rm s}$.

\begin{figure*}[htb!]
\epsscale{1.0}
\plottwo{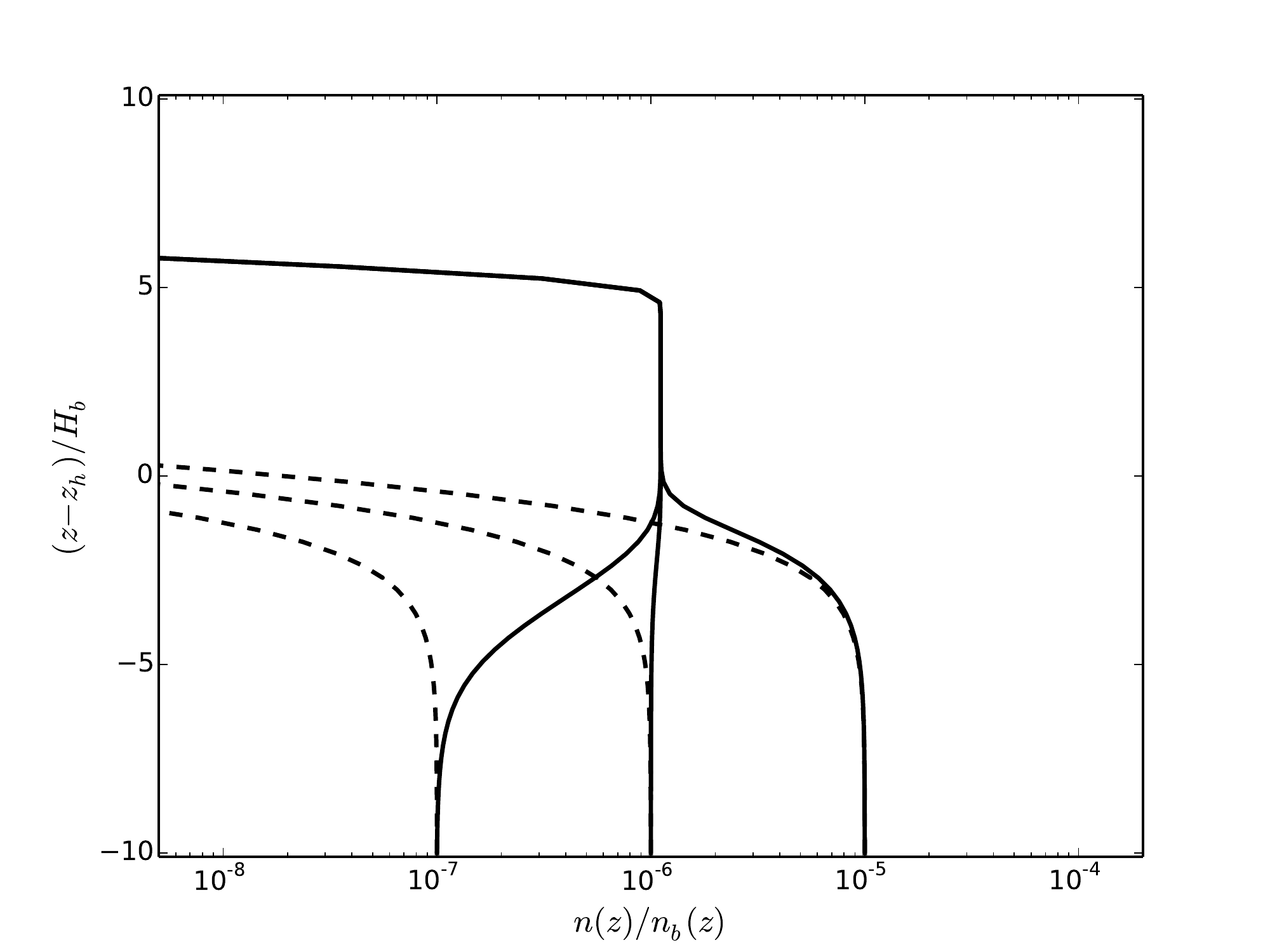}{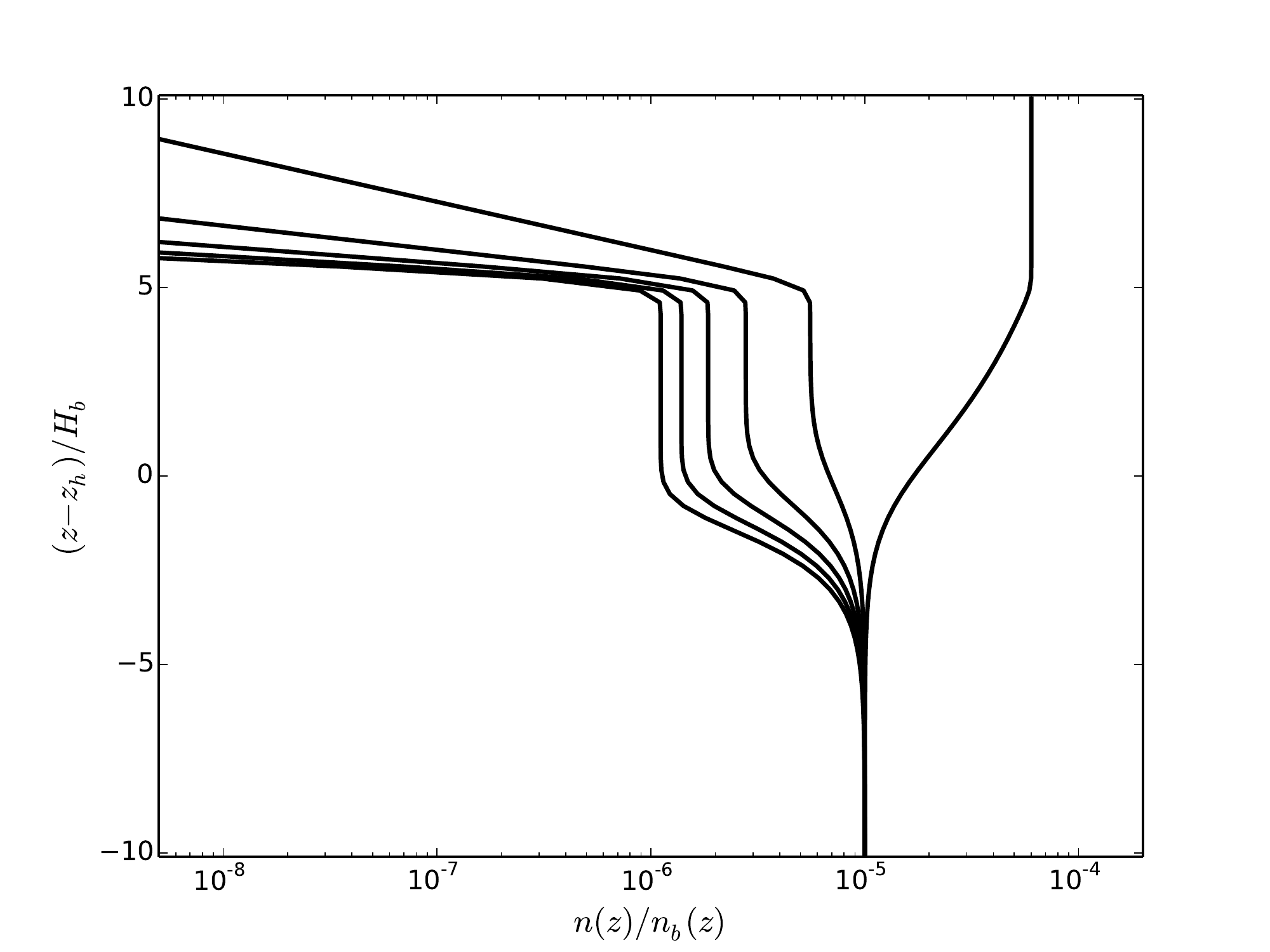} 
\caption{ (Left panel) Mixing ratio $n(z)/n_{\rm b}(z)$ as a function of altitude for the case with background flux $F_{\rm b}=0$. The meteoric source is centered 5 scale heights above the homopause. The source flux $F_{\rm s}$ and turbulent mixing parameter $K$ are chosen so that the mixing ratio below the source is $\simeq 10^{-6}$. The three solid lines are for different mixing ratio $\xi = 10^{-7,-6,-5}$ deep in the well-mixed layer. The dashed lines are the diffusive equilibrium solution for no meteoric source. The particle mass is taken to be $a=m/m_{\rm b}=10$. (Right panel) In this case the same source parameters are used, and a single $\xi = 10^{-5}$ is chosen. The lines show different upward flux of the background particle $F_{\rm b}/F_{\rm b,cr} = 0, 0.2, 0.4, 0.6, 0.8$, and $1.0$ relative to the critical flux for blowoff.
\label{fig:escape_no_blowoff}}
\end{figure*}

The left panel of Figure \ref{fig:escape_no_blowoff} shows examples of abundances profiles including a meteoric source 5 scale heights above the homopause, and for three different abundances deep in the atmosphere. The mixing ratio due to the meteoric source is close to the $\xi=10^{-6}$ line and so it looks almost straight. However, when the deep and meteoric mixing ratios do not match, there is a region a few scale heights thick over which the ratio transitions. For this case of a meteoric source well above the homopause, the solutions including the source (solid lines) retain large mixing ratio while the diffusive equilibrium solutions (dashed lines) decay sharply above the homopause for this case of $a=10$.

The right panel of Figure \ref{fig:escape_no_blowoff} uses the same source parameters and a deep mixing ratio $\xi=10^{-5}$. The different lines represent different escape flux $F_{\rm b}$ of the background. For small $F_{\rm b}$, the solution agrees with the rightmost solid line in the left panel, including the diffusive equilibrium above the source, giving rise to rapid falloff. As $F_{\rm b}$ increases, the mixing ratio below the source increases and the upward decrease above the source is slower. For the value $F_{\rm b} = F_{\rm b, cr} \equiv K n_{\rm b}(z_{\rm h}) (a-1)/H_{\rm b}$ the abundance actually increases below the source layer and is constant above.

\begin{figure*}[htb!]
\epsscale{0.8}
\plotone{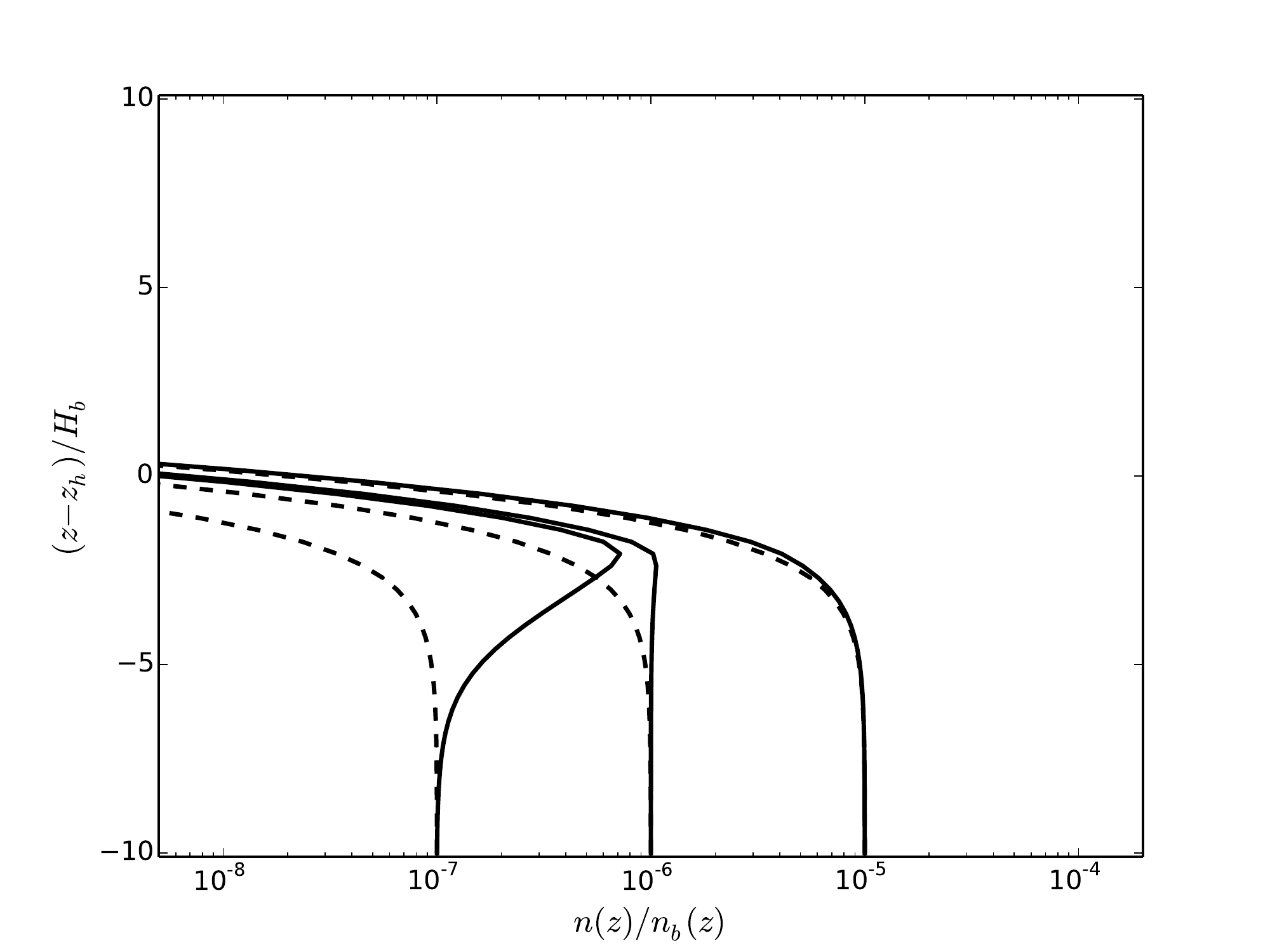}
\caption{ The same as the left pane of Figure \ref{fig:escape_no_blowoff} but with source position 2 scale heights below the homopause.
 \label{fig:abundances_low_source}}
\end{figure*}

Figure \ref{fig:abundances_low_source} shows the case of a dust source two scale heights below the homopause. For the case of meteoric source layer mixing ratio smaller or comparable to the deep mixing ratio the effect of the dust source is small. However, when the dust source mixing ratio is larger than that of the deep atmosphere the abundances can have a significant impact on the abundance.

These numerical results can be recovered with an analytic solution below the meteoric source \citep{1987tpaa.book.....C}
\be
\frac{n(z)}{n_{\rm b}(z)} & = & \Xi +  \left( \xi - \Xi  \right) \left( \frac{1 + e^{Z_0} }{ 1 + e^Z } \right)^{a_{\rm eff} - 1}
\label{eq:mixrat_analytic}
\ee
where the effective particle mass is
\be
a_{\rm eff} & = & a - \frac{ F_{\rm b} H_{\rm b} }{ K n_{\rm b}(z_{\rm h}) },
\ee
$Z_0 = (z_0-z_{\rm h})/H_{\rm b}$ and the mixing ratio below the meteoric source is
\be
\Xi & = & - \frac{ F H_{\rm b}} {K (a_{\rm eff}-1) n_{\rm b}(z_{\rm h}) }.
\ee
This solution interpolates between the value deep in the well-mixed layer, $\xi$, and the value between the homopause and the meteoric source, $\Xi$. Above the meteoric source, the solution again decays as the diffusive equilibrium form
\be
\frac{n(z)}{n_{\rm b}(z)} & \simeq & \Xi e^{-(a_{\rm eff} - 1)Z},
\ee
but now with the slower decay $a_{\rm eff}$ instead of $a$. When $a_{\rm eff} \leq 1$, the drag force from the upward-moving light background particles overcomes gravity. This occurs for particle masses below a critical mass
\be
m_{\rm cr} & =& m_{\rm b} + \frac{k_{\rm B}T F_{\rm b}}{bg},
\ee
or equivalently, above the critical flux
\be
F_{\rm b, cr} & \equiv &  \frac{ K n_{\rm b}(z_{\rm h}) (a-1)}{H_{\rm b}}.
\ee
Above this critical flux, the particles with mass $m$ may be carried along with the outflow, and their upward flux $F_{\rm esc}$ will be proportional to the background flux $F_{\rm b}$ \citep{1987tpaa.book.....C}.

The critical mass loss rate of the background species to give a blowoff state for the trace is
\be
\dot{M}_{\rm b, crit}  & = & 4\pi \Rp^2 m_{\rm b} F_{\rm b, cr} = \frac{4\pi G\Mp m}{\langle \sigma v \rangle}
\nonumber \\ & = & 2.6 \times 10^{12}\, \rm g\, s^{-1}\, \left( \frac{m}{28m_{\rm b}} \right)^2 \sqrt{ 1 + \frac{m_{\rm b}}{m} } 
\nonumber \\ & \times & \left( \frac{\Mp}{M_{\rm J}} \right) \left( \frac{T}{10^3\, \rm K} \right)^{-1/2} 
 \left( \frac{m_{\rm b}}{m_{\rm u}} \right)^{3/2} \left( \frac{R_{\rm coll} + R_{\rm coll, b}}{3\AA} \right)^{-2}.
\ee
The numerical estimate is appropriate for an atomic hydrogen background and $^{28}$Si atoms and for a medium-size planet. Due to the $m^2$ scaling, the required mass-loss rate to cause a blowoff state ranges from $\sim 10^{11}\, \rm g\, s^{-1}$ for He to $\sim 10^{13}\, \rm g\, s^{-1}$ for Fe. The latter mass-loss rate is large enough to evaporate the entire planet away in $\sim 6$ Gyr. 
Large transit depths of heavy elements require their presence high in the atmosphere, and a blowoff state may be invoked to accomplish this, requiring high mass-loss rates. The alternative suggested here is that the meteoric source may supply heavy elements to the upper atmosphere without the need for such high mass-loss rates.

\section{ transit radius for atomic lines }
\label{sec:lines}

Atomic resonance lines interact with the gas much more strongly than the neighboring continuum, leading to larger transit depth in the neighborhood of the lines. If the $\tau=1$ radius for the continuum is $\Rp$ and that of the line is $\Rp + z(\lambda)$, then this annulus of area $ \simeq 2\pi \Rp z(\lambda)$ gives a contribution to the transit depth of size $\Delta F_\lambda/F_\lambda \simeq 2\Rp z(\lambda)/R_{\rm s}^2$ over that of the continuum. The goal of this section is to give examples of how $z(\lambda)$ depends on dust accretion rate and vertical mixing coefficient $K$. 

For definiteness, the Na D doublet will be used for the line.
Different values of vertical mixing coefficient will be used to show the effect of the homopause. 
The source is taken to be a delta function at $P=10^{-2}\, \rm \mu bar$. Deep in the well-mixed layer the abundance is assumed to be solar with mixing ratio $\xi$ relative to the background species. The abundance below the meteoric source will depend on the ablation rate $\epsilon_{\rm abl}f_{\rm pl}\dot{M}_{\rm d}$. 

Constant temperature is not a good approximation over the entire atmosphere, as it is expected to vary from $T \sim T_{\rm eq}\sim 1000\, \rm K$ near the $P=1\, \rm mbar$ level to $T \sim 3000-10,000\, \rm K$ above the $P=1\, \rm \mu bar$ level due to thermospheric heating. While the wings of the line may have their $\tau=1$ altitude where $T \sim T_{\rm eq}$, wavelengths near line center may have $\tau=1$ where $T \gg T_{\rm eq}$. To take these two cases into account, but still using the isothermal assumption for the atmosphere, two cases will be shown. In the first case, more appropriate for the wings, the temperature is chosen to be $T=1500\, \rm K$ atmosphere with H$_2$ as the background. In the second case, more appropriate for line center, a $T=5000\, \rm K$ atmosphere with H atoms is used as the background. In both cases, the $P=1\, \rm mbar$ level with be chosen as the $z=0$ point, but keep in mind that the scale height is very different in the two cases.

For absorber number density that increases strongly downward, the optical depth can be evaluated as the effective path length 
$\simeq \sqrt{2\pi \Rp H_{\rm b}}$ times the density at the altitude of closest approach, giving an optical depth
\be
\tau(z,\nu) & \simeq &  \sqrt{2\pi \Rp H_{\rm b}} n(z) \sigma(\nu).
\label{eq:tau}
\ee
Here the cross section is 
\be
\sigma(\nu) & = & \frac{\pi e^2}{m_e c}\, f_{\rm os}\, \frac{H(x,a_{\rm v})}{\sqrt{\pi} \Delta}
\ee
where $f_{\rm os}$ is the oscillator strength, $H(x,a_{\rm v})$ is the Voigt function, $x=(\nu-\nu_0)/\Delta$, $a_{\rm v}=\Gamma/(4\pi \Delta)$, and $\Delta = \nu_0 \sqrt{2k_{\rm B}T/mc^2}$. The number density $n(z)$ of the absorber is the product of the  background density and the mixing ratio in Equation \ref{eq:mixrat_analytic}. Plugging into Equation \ref{eq:tau}, root solving can then be used to find the value of $z(\lambda)$ where 
$\tau(z(\lambda),\lambda)=1$.

\begin{figure*}[htb!]
\epsscale{0.9}
\plotone{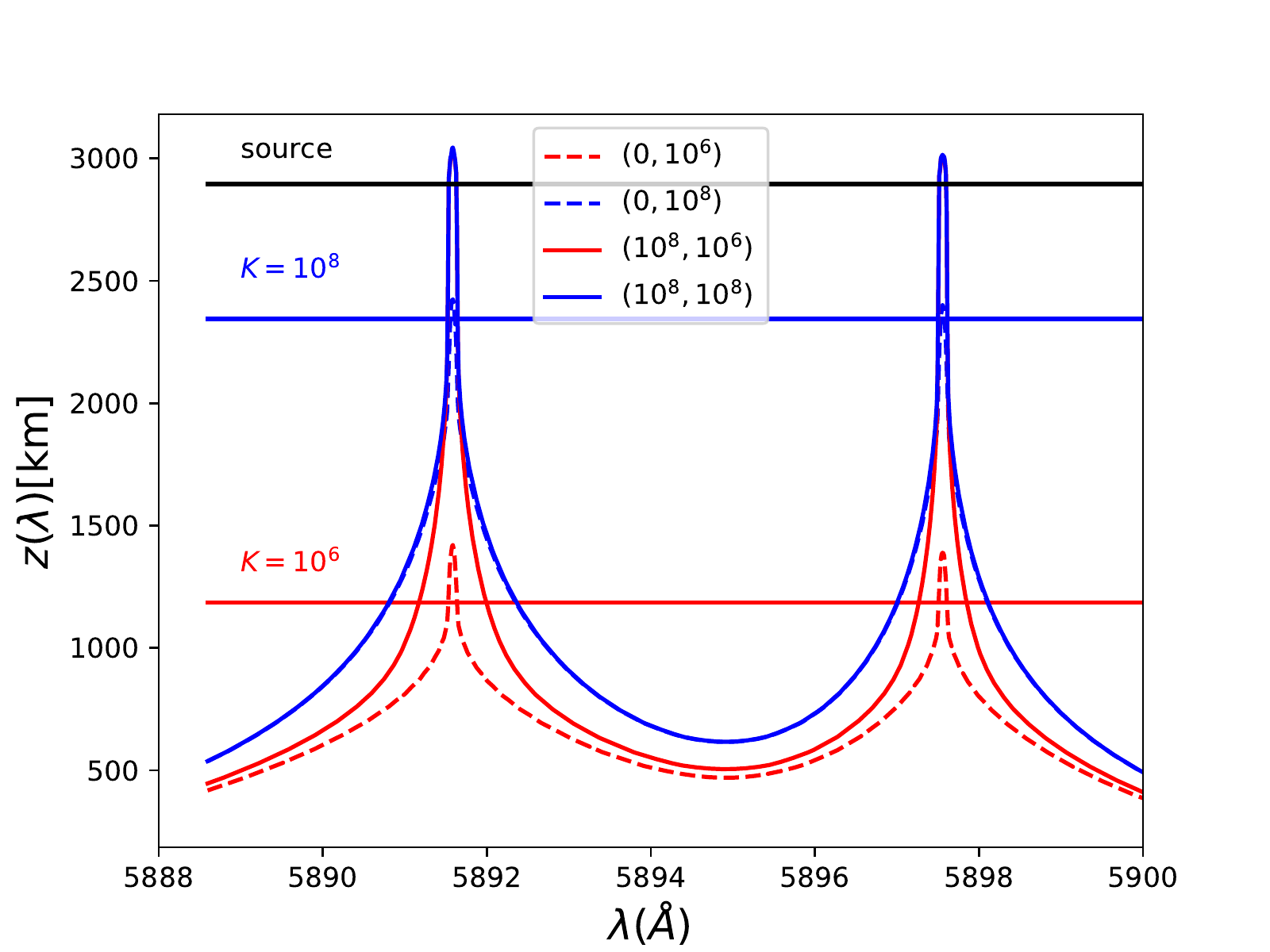}
\caption{Transit radius versus wavelength for the Na doublet for an isothermal atmosphere with $T=1500\, \rm K$ and H$_2$  background. The medium-mass planet is used. The $z=0$ altitude is set at $P=1\, \rm mbar$ and the source is taken to be a delta function at $P=10^{-2}\, \rm \mu bar$, shown as the black line labeled ``source". The four lines represent two different vertical mixing coefficients and two dust accretion rates, and are labeled by $(f_{\rm pl}\epsilon_{\rm abl}\dot{M}_{\rm d}(\rm g\, s^{-1}),K(\rm cm^2\, s^{-1}))$. The red and blue horizontal lines are the homopause altitudes for $K=10^6$ and $10^8\, \rm cm^2\, s^{-1}$, respectively.
\label{fig:transit_radius_sodium_lower}
}
\end{figure*}

Figure \ref{fig:transit_radius_sodium_lower} shows $z(\lambda)$ as a function of wavelength $\lambda$ for a $T=1500\, \rm K$ atmosphere with H$_2$ background for a medium-size planet. 
Four different profiles are shown, labeled by dust accretion rate and vertical mixing coefficient.
When the meteoric source is absent (dashed lines), the density decreases strongly above the homopause since the Na scale height is 11.5 times smaller than the H$_2$ scale height. Even at the center of the line, where the cross section is largest, $z(\lambda)$ extends above the homopause by less than a scale height. Hence, in the absence of dust accretion, the transit radius is limited by the homopause. Only in the case that the homopause extends to very high altitudes, where $\tau < 1$ even with constant mixing ratio $\xi$, is the effect of the meteoric source on the line profile ignorable.

By contrast, the meteoric source, here with $f_{\rm pl} \epsilon_{\rm abl} \dot{M}_{\rm d}=10^8\, \rm g\, s^{-1}$, allows the transit altitude to be set by the source, rather than the homopause, for both cases where the homopause is well below the source. The solid red and blue lines agree near the center of the line, and only begin to disagree below the respective homopauses where the mixing ratio becomes $\xi$ rather than $\eta$.

\begin{figure*}[htb!]
\epsscale{0.9}
\plotone{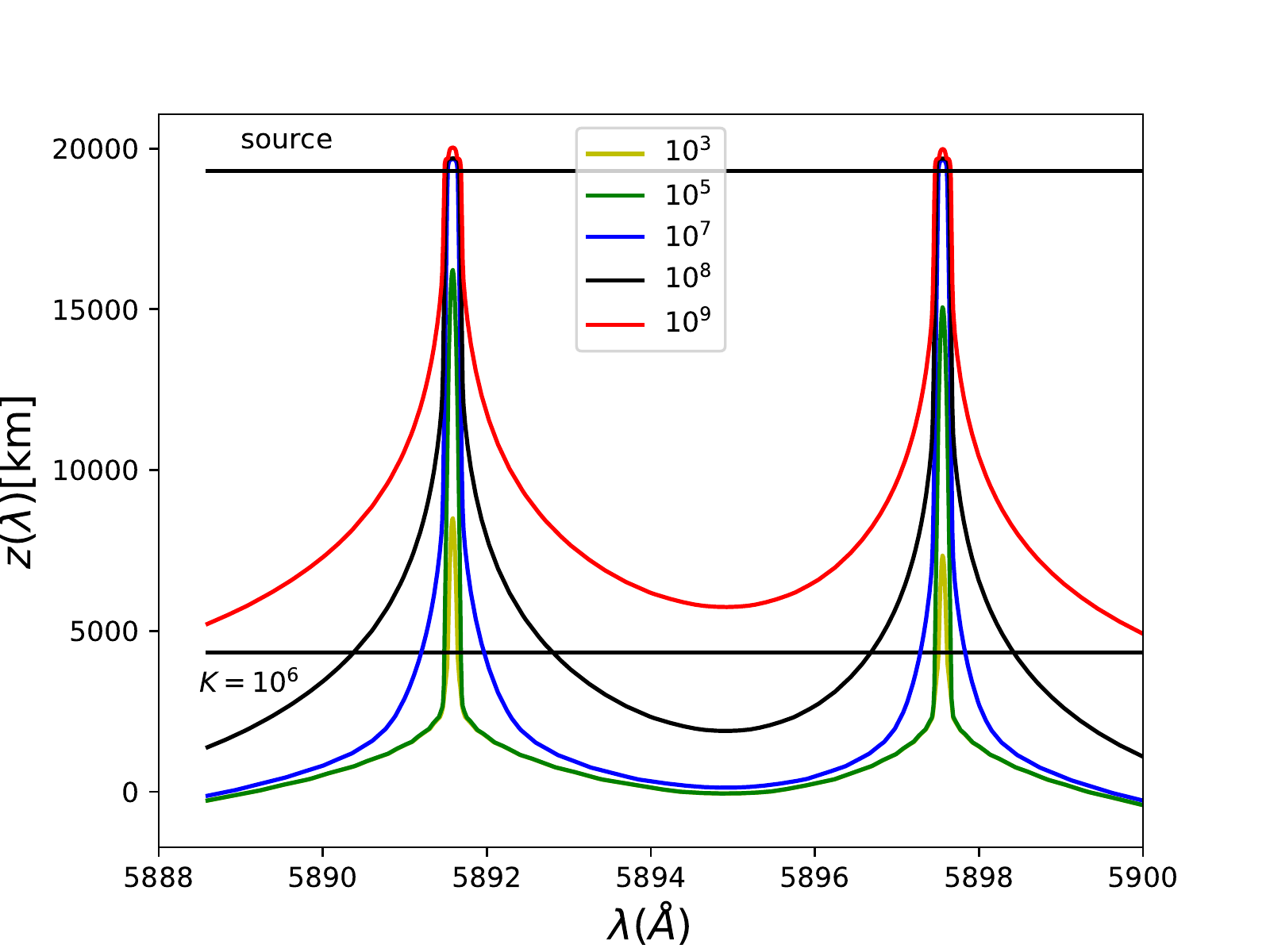}
\caption{Same as Figure \ref{fig:transit_radius_sodium_lower} but now for the hot atmosphere with $T=5000\, \rm K$ at atomic hydrogen background.
The lines are labeled by the value of the accretion rate $f_{\rm pl}\epsilon_{\rm abl}\dot{M}_{\rm d}(\rm g\, s^{-1})$. The mixing coefficient is $K=10^6\, \rm cm^2\, s^{-1}$. 
\label{fig:transit_radius_sodium_upper_6}
}
\end{figure*}

\begin{figure*}[htb!]
\epsscale{0.9}
\plotone{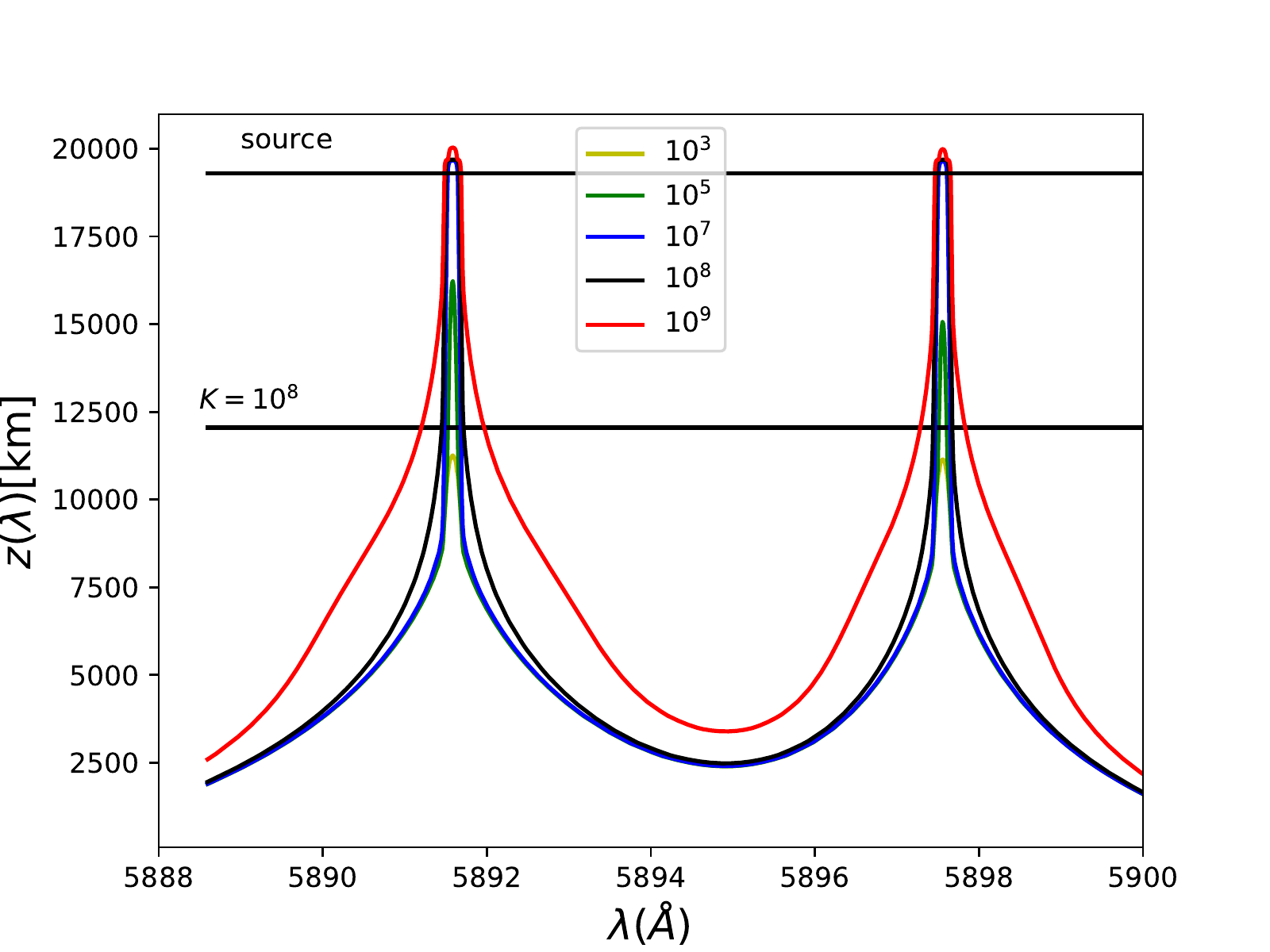}
\caption{Same as Figure \ref{fig:transit_radius_sodium_upper_6} but with larger $K=10^8\, \rm cm^2\, s^{-1}$. 
\label{fig:transit_radius_sodium_upper_8}
}
\end{figure*}

\begin{figure*}[htb!]
\epsscale{0.9}
\plotone{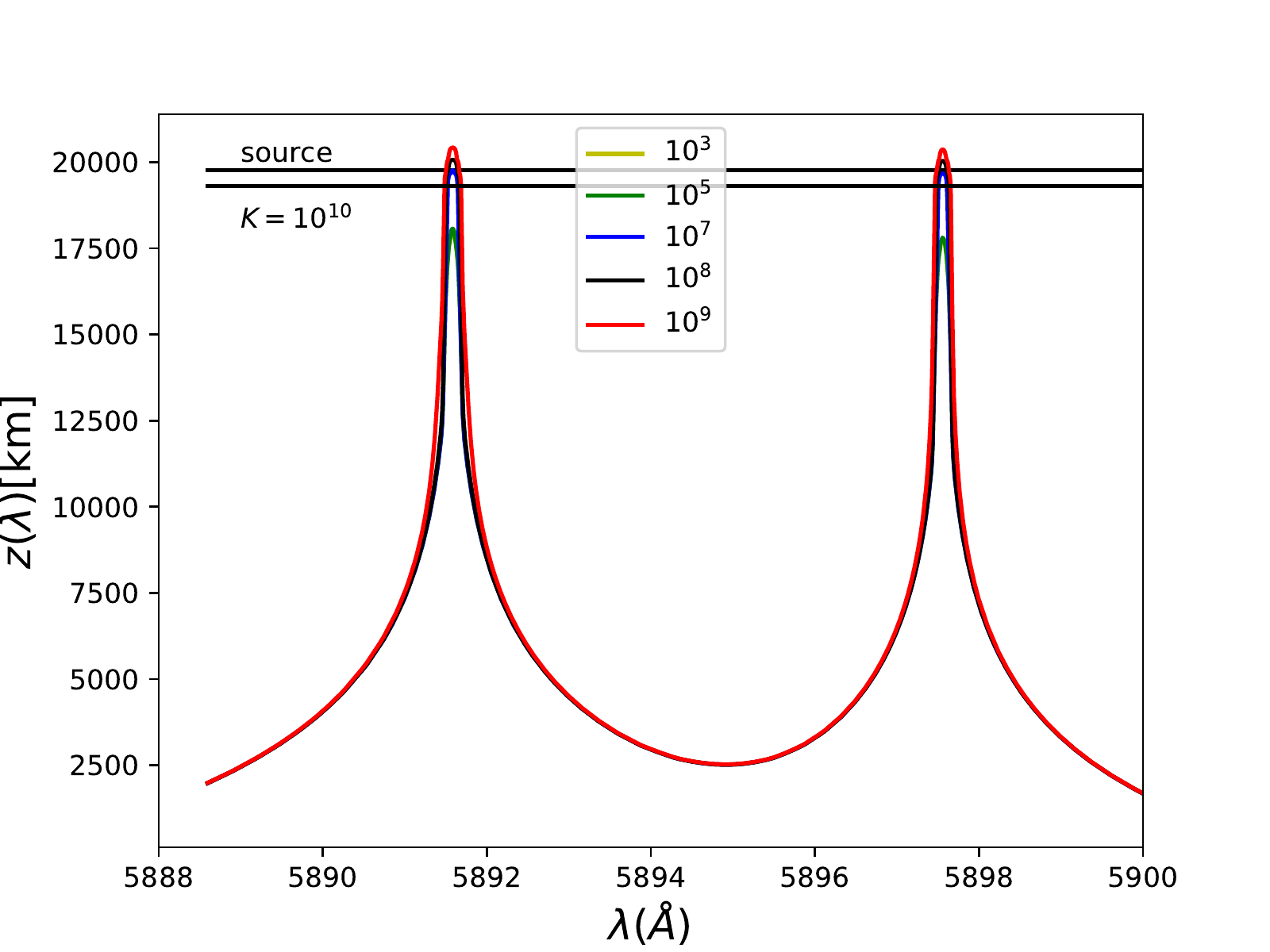}
\caption{Same as Figure \ref{fig:transit_radius_sodium_upper_6} but with larger $K=10^{10}\, \rm cm^2\, s^{-1}$. 
\label{fig:transit_radius_sodium_upper_10}
}
\end{figure*} 

Figures \ref{fig:transit_radius_sodium_upper_6} - \ref{fig:transit_radius_sodium_upper_10} show the transit altitude for the hotter atmosphere with $T=5000\, \rm K$ and H atoms as the background, over a large range of $f_{\rm pl}\epsilon_{\rm abl}\dot{M}_{\rm d}$ in each plot and  three different values of the mixing coefficient $K$. 

For the large mixing ratio $K=10^{10}\, \rm cm^2\, s^{-1}$ used in Figure \ref{fig:transit_radius_sodium_upper_10}, the transit radius is nearly the same for all the accretion rates, since the homopause is near the meteoric source, and the mixing ratio is $\xi$ over nearly the entire atmosphere. By contrast, for the smaller value  $K=10^{6}\, \rm cm^2\, s^{-1}$ used in Figure \ref{fig:transit_radius_sodium_upper_6}, there is a significant difference in the transit altitude even for very small accretion rates. At $ f_{\rm pl} \epsilon_{\rm abl} \dot{M}_{\rm d} \ga 10^2-10^3\, \rm g\, s^{-1}$, the transit altitude of the line center begins to extend above the homopause. By $10^5\, \rm g\, s^{-1}$, well below $\dot{M}_{\rm d, iss}$, the line core is approaching the source altitude. For the $10^7-10^8\, \rm g\, s^{-1}$ cases, the line core extends above the source altitude, and the line wings are also significantly affected. The intermediate case of $K=10^{8}\, \rm cm^2\, s^{-1}$ shown in Figure \ref{fig:transit_radius_sodium_upper_8}, still shows a difference on the line wings for $10^9\, \rm g\, s^{-1}$, and the line core is near the source for $ f_{\rm pl} \epsilon_{\rm abl} \dot{M}_{\rm d} \ga 10^7\, \rm g\, s^{-1}$.

These results shown that it is possible for the meteoric source to  affect the transmission spectrum of strong lines if the vertical mixing is sufficiently small. The importance will be limited if very strong mixing extends all the way up to the meteoric source region.

\section{ remnant dust }
\label{sec:remnantdust}

In this section the abundance and optical depth of the remnant dust particles left over from ablation are considered. 

\subsection{  size distribution }

Let the size distribution of the remnant dust be ${\cal P}(s)$, which is normalized to unity, $\int ds {\cal P}(s) = 1$, and has mean mass $\bar{m}_{\rm d} = \int ds {\cal P}(s) (4\pi/3) \rho_{\rm d} s^3$. Below the meteoric source, there is a  flux of dust particles per size interval $ds$ which is denoted $\Phi(s)$. 
This flux can be written as
\be
\Phi(s) & = & - \frac{f_{\rm pl} \epsilon_{\rm rem} \dot{M}_{\rm d}}{4\pi \Rp^2 \bar{m}_{\rm d}} {\cal P }(s),
\label{eq:Phis}
\ee
where the negative sign denotes a downward flux.

\begin{figure*}[htb!]
\epsscale{0.9}
\plotone{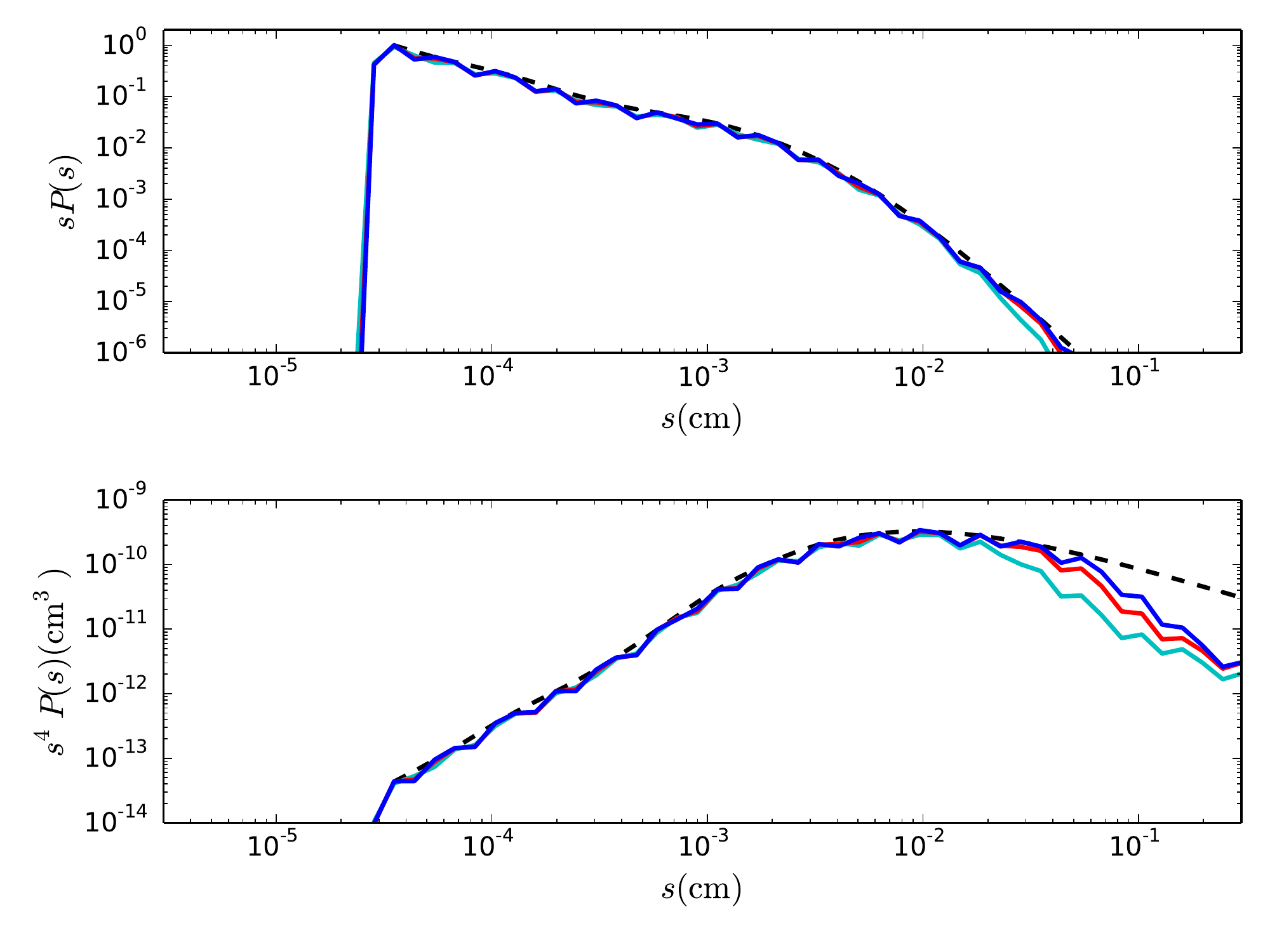}
\caption{ Dust size (upper panel) and mass (lower panel) distributions for the small planet. The black dashed line is the size distribution of the incident dust, assumed to be the G85 distribution with a cutoff at $\beta=1$. The solid cyan, red and blue lines represent the remnant dust distributions for orbital separations $\ap/R_\odot=10, 14$ and $18$.
 \label{fig:size_mass_small_grun}
 }
\end{figure*}

The size distribution for the remnant dust was computed as part of the stopping calculations in Section \ref{sec:stopping}. Figure \ref{fig:size_mass_small_grun} uses the G85 size distribution and parameters for the small planet. Shown are the size and mass distributions for the incident dust before it hits the atmosphere (dashed line) and the remnant dust after it has stopped (solid lines), for three different orbital separations. 
Large millimeter-size dust is depleted by as much as an order of magnitude in mass; however, the smaller sizes are less affected. Ablation is larger for closer-in planets. 

Results are not shown for the $\ap=6\, R_\odot$ case since all remnant dust was in the lowest size bin ($s=0.01\, \mu \rm m$), implying that it would have been further ablated if the calculation had not been stopped. All runs at this orbital separation have such large entry speeds that little mass is left in remnant dust.

\begin{figure*}[htb!]
\epsscale{0.9}
\plotone{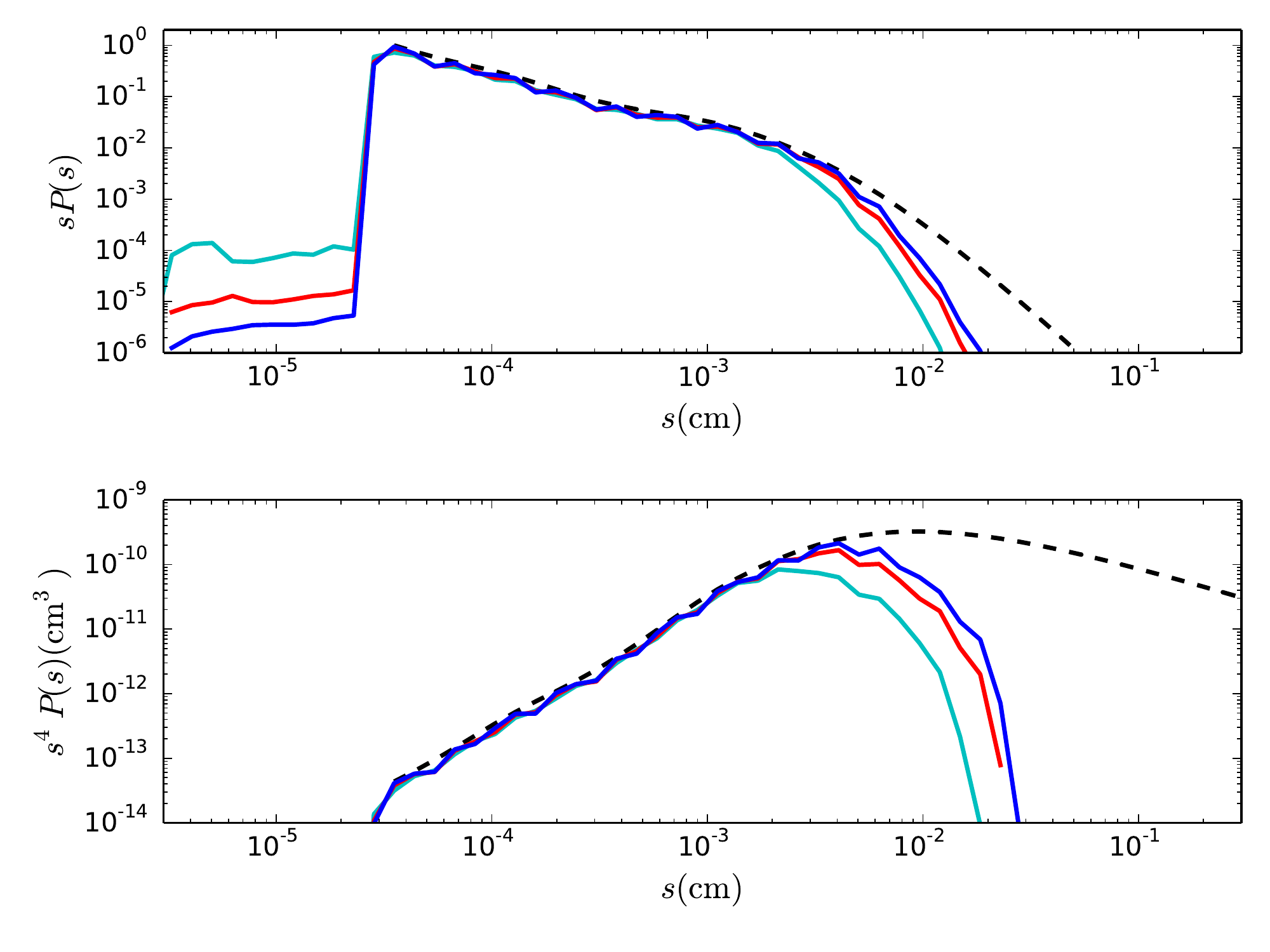}
\caption{ Same as Figure \ref{fig:size_mass_small_grun} but for the medium-size planet.
 \label{fig:size_mass_med_grun}
 }
\end{figure*}

Figure \ref{fig:size_mass_med_grun} shows results for the medium-size planet. Now there is significant ablation at the peak of the mass distribution at all orbital separations shown, due to the higher escape speed. Ablation is larger for closer planets. Also apparent is small remnant dust, below the blowout size at $s =0.3\, \rm \mu m$, created by ablation of larger dust. The amount of small remnant dust is larger for closer orbits.

\begin{figure*}[htb!]
\epsscale{0.9}
\plotone{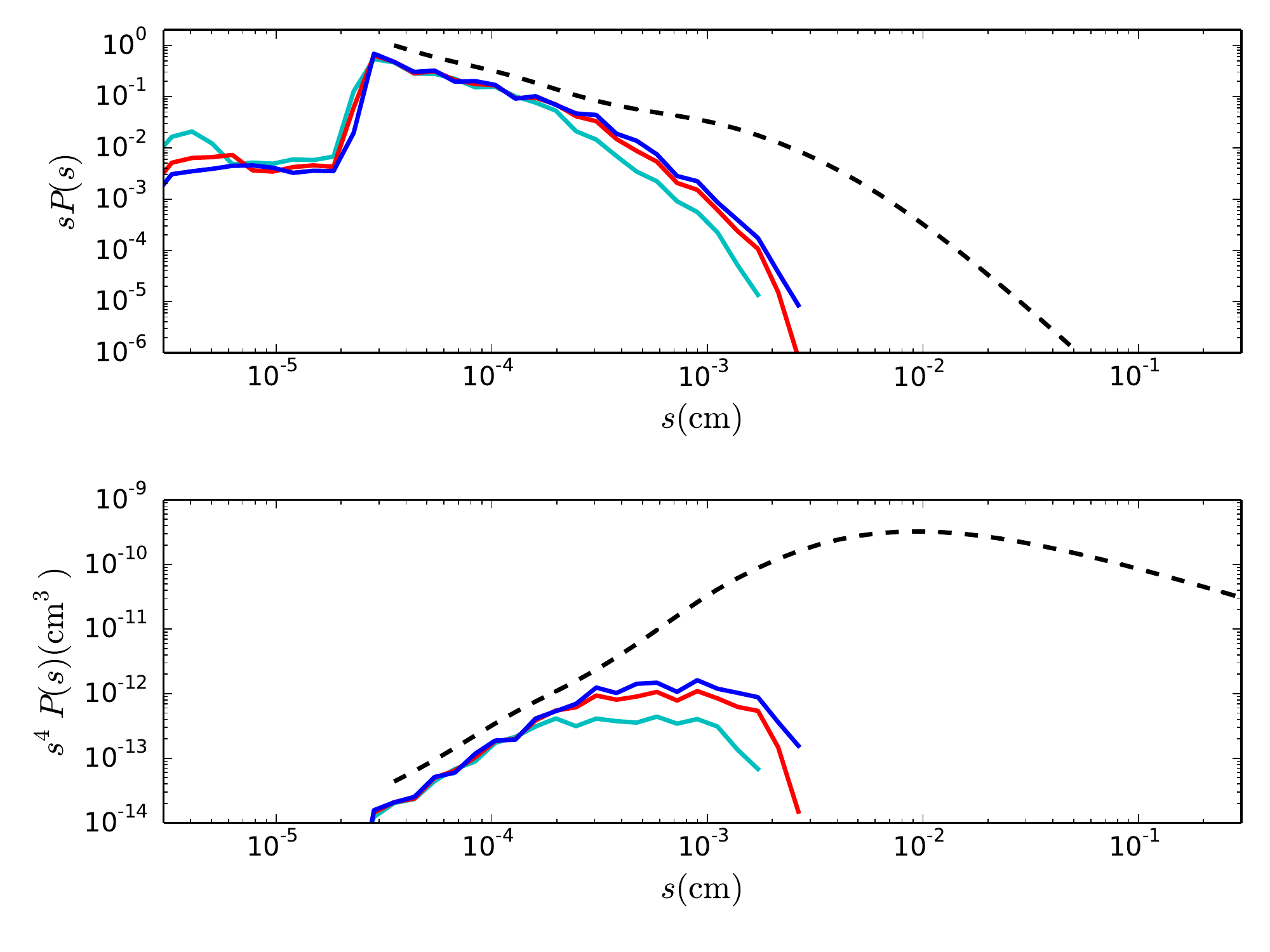}
\caption{ Same as Figure \ref{fig:size_mass_small_grun} but for the large-size planet.
 \label{fig:size_mass_large_grun}
 }
\end{figure*}

Figure \ref{fig:size_mass_large_grun} shows results for the large-size planet. In addition to significant depletion at large masses, micron-size dust is also noticeably depleted. The abundance of remnant dust below the blowout size is now significantly larger, although still smaller than the peak of the initial distribution.

The above cases were for the G85 dust-size model. For the single-dust-size model with $\beta=0.5$, or $s=0.58\, \rm \mu m$, the dust size is small enough that thermal evaporation does not occur and there is only mass loss due to sputtering.

\subsection{ abundance below the source }
\label{sec:dustabundance}

Given the size distribution for the downward flux of dust, the profile of number density is required to compute the optical depth through the atmosphere. When mixing is negligible, the dust density is determined solely by the vertical settling velocity $u$, as in the atomic case discussed in Section \ref{sec:sourcedensity}. The number density of dust per size interval $ds$ is a function of pressure and size, and will be denoted ${\cal N}(P,s)$. It is normalized to the total number density $n(P) = \int ds {\cal N}(P,s)$, and is given by
\be
{\cal N}(P,s) & = &  \frac{\Phi(s)}{u(P,s)}.
\ee
The dust velocity is computed as $u(P,s) = -u_{\rm set}(P,s)$ where the (positive) settling speed is computed as $u_{\rm set} = {\rm max}(u_{\rm ep},u_{\rm st})$, where the larger of the Epstein and Stokes drift speeds is used.
At high altitudes, the gas mean free path $\ell$ is larger than the dust radius $s$, and the Epstein settling velocity applies with value \footnote{ This formula assumes that the downward gravity force is balanced by the upward drag force. However, above the optical and infrared photospheres, where the radiation field is highly anisotropic, both the stellar and planetary radiation fields imply a radiation force on the dust grain. In the case of the upward planetary radiation field, the dust will feel an upward radiative acceleration $\beta_{\rm p} g$ so that the effective gravity of the planet becomes $(1-\beta_p)g$, and the setting speed is reduced by a factor $1-\beta_{\rm p}$. This may  increase the grain density for small grains with $\beta_{\rm p} \la 1$. Further, grains with the appropriate size and composition to give $\beta_{\rm p} \ga 1$ would experience a net upward {\it radiative levitation} force. While these effects may be important, they are ignored here for simplicity.} 
\be
u_{\rm ep}(P,s) & = & \left( \frac{\rho_{\rm d}}{\rho} \right) \left( \frac{gs}{v_{\rm th}} \right),
\ee
where $\rho = \mu m_u P/(k_{\rm B} T)$ is the mass density, $\mu = 2.3$ is the mean molecular weight in the lower atmosphere discussed here.
For a delta-function size distribution ${\cal P}(s')=\delta(s'-s)$, the number density is
\be
n(P) & = & \left( \frac{f_{\rm pl} \epsilon_{\rm rem} \dot{M}_{\rm d}}{4\pi \Rp^2 \bar{m}_{\rm d}} \right) \left( \frac{\rho v_{\rm th}}{\rho_{\rm d} g s} \right).
\ee
Since the dust density is proportional to the gas density, the mixing ratio is fairly constant with altitude. This result can be re-expressed as the  ratio of dust and gas densities
\be
f_{\rm d} & \equiv & \frac{\bar{m}_{\rm d} n(P)}{\rho}  =   \frac{f_{\rm pl} \epsilon_{\rm rem} \dot{M}_{\rm d} v_{\rm th}}{4\pi G\Mp \rho_{\rm d} s} 
\nonumber \\ & \simeq & 9.4 \times 10^{-8}\, \left( \frac{  f_{\rm pl} \epsilon_{\rm rem} \dot{M}_{\rm d} }{ 10^8\, \rm g\, s^{-1}} \right) 
\left( \frac{T}{10^3\, \rm K} \frac{2.3}{\mu} \right)^{1/2} 
\nonumber \\ & \times & \left( \frac{\Mp}{M_{\rm J}} \right)^{-1} \left( \frac{\rho_{\rm d}}{2\, \rm g\, cm^{-3}} \right)^{-1} \left( \frac{s}{1\, \mu m} \right)^{-1}.
\label{eq:fd}
\ee
This estimate shows that for dust accretion rates near the solar system value, and micronsize dust, the dust is settling sufficiently fast that its mass fraction is small compared to the solar metallicity value $0.01$. 

Along a vertical path, the optical depth due to remnant dust settling in the atmosphere is $\tau_{\rm vert} \sim f_{\rm d} (P/g) \pi s^2/\bar{m}_d 
\simeq 2.9\times 10^3 \, f_{\rm d} (P/1\, \rm mbar) (g/g_{\rm J})^{-1} (s/1\, \rm \mu m)^{-1}$. Hence, at the $P \sim 1\, \rm mbar$ level a mass fraction $f_{\rm d} \ga 3.4 \times 10^{-4}$ is needed for vertical optical depth unity, which would require dust accretion rates far larger than $\dot{M}_{\rm d, iss}$. 

While the Epstein formula applies, the main contribution to the optical depth along a straight-line path through the atmosphere during transit occurs at the point of closest approach, where the effective path length for an isothermal atmosphere is $\sqrt{2\pi H_{\rm b} \Rp}$. The resultant extinction optical depth along a path with maximum pressure $P$  is
\be
\tau(P) & \simeq & \sqrt{2\pi H_{\rm b} \Rp} n(P) 2\pi s^2 \nonumber \\ & = & 
 \frac{3}{2\pi} \left( \frac{G\Mp}{\Rp} \right)^{-3/2}  \left( \epsilon_{\rm rem}  f_{\rm pl} \dot{M}_{\rm d} \right) \left( \frac{P}{(\rho_{\rm d} s)^2} \right)
   \nonumber \\ & \simeq & 
  0.18\, \left( \frac{P}{1\, \rm mbar} \right) \left( \frac{  \epsilon_{\rm rem}  f_{\rm pl} \dot{M}_{\rm d} }{ 10^8\, \rm g\, s^{-1} } \right)
  \left( 2\beta \right)^2 Q_{\rm pr}^{-2} \nonumber \\ & \times & 
  \left( \frac{\Mp}{M_{\rm J}} \right)^{-3/2} \left( \frac{\Rp}{R_{\rm J}} \right)^{3/2} \left( \frac{L_{\rm s}}{L_\odot} \right)^{-2}
  \left( \frac{M_{\rm s}}{M_\odot} \right)^2.
  \label{eq:taudust}
\ee
In the last step the dust size was expressed in terms of $\beta$ using Equation \ref{eq:beta}. An extinction cross section $2\pi s^2$ has been used, which assumes that $2\pi s \ga \lambda$.

Equation \ref{eq:taudust} shows that the optical depth during transit may be order unity at $\sim \rm mbar$ pressures. By using $\beta=0.5$, the dust mass is efficiently used to give a large area to block starlight. Grain size distributions with most of the mass at sizes much larger than a micron will give smaller optical depths by comparison. For the small, medium and large planets, the ratio of $(G\Mp/\Rp)^{-3/2}$ is $20:1:0.1$, which favors low-mass planets. This is offset partially by the  decreased fraction $f_{\rm pl}$ accreted onto smaller mass and radius planets, but reinforced by the larger $\epsilon_{\rm rem}$ for small planets. Lastly, for all other parameters fixed, smaller stellar mass may lead to significantly higher optical depth, at constant $\beta$, as the blowout size is smaller. For instance, the value of $\tau$ for a K star such as HD , with $M_{\rm s}=0.85\, M_\odot$ and $L_{\rm s}=0.33\, L_\odot$, will be larger by a factor $(0.84/0.33)^2=6.6$.

At lower altitudes, the higher gas density may lead to $\ell \la s$, in which case Stokes drag applies with
\be
u_{\rm st} & = & \frac{2}{9} \left( \frac{\rho_{\rm d} g s^2}{\eta_{\rm dyn}} \right),
\ee
where $\eta_{\rm dyn} \simeq 1.3 \times 10^{-4}\, \rm g\, cm^{-1}\, s^{-1}\, \left( T/10^3\, \rm K \right)^{1/2}$ is the dynamic viscosity for a H$_2$ dominated atmosphere using the hard sphere cross section. In this case the settling speed varies slowly with altitude, giving rise to a roughly constant number density and hence decreasing mixing ratio with depth. The pressure level at which the settling speed switches from the Epstein to Stokes limit is
\be
P_{\rm sw}(s) & = & 0.7\, \rm bar\, \left( \frac{1\, \rm \mu m}{s} \right) \left( \frac{T}{10^3\, \rm K} \right).
\ee
The Epstein formula applies for small dust down to deep in the atmosphere, while larger dust switches to the Stokes formula higher in the atmosphere.

\subsection{ dust distribution including settling and vertical mixing }
\label{sec:dustmixing}

In Section \ref{sec:dustabundance}, the dust was assumed to settle below the source. When vertical mixing is included, the dust abundance may be significantly altered below the dust homopause where $K \ga u_{\rm set} H_{\rm b}$. The dust abundance will be determined from solutions of the equation
\be
\frac{d{\cal N}}{dz} + {\cal N} \left( \frac{1}{H_{\rm b}} + \frac{u_{\rm set}}{K} \right) & = & - \frac{\Phi}{K}.
\label{eq:dusteqn}
\ee
with constants $K$, $H_{\rm b}$ and $\Phi(s)$. The settling speed $u_{\rm set}$ interpolates between the Stokes and Epstein limits.

Equation \ref{eq:dusteqn} is numerically integrated for each dust size $s$ from a base pressure of $P=10\, \rm bar$ up to $P=1\, \rm \mu bar$. The dust flux is taken from Equation \ref{eq:Phis}, and self-consistently includes the incident size distribution of the remnant dust, G85 or single-size, as well as $\epsilon_{\rm rem}$ for the given planet size and distance from the star. The accretion rate onto the planet is taken to be  $f_{\rm pl} \dot{M}_{\rm d}=10^8\, \rm g\, s^{-1}$ in all results shown. Since Equation \ref{eq:dusteqn} is linear, the optical depth can be scaled up or down for different accretion rates. 

The boundary condition ${\cal N}=0$ is used at the base. This represents destruction of dust in deeper layers where the temperature goes above the dust melting temperature. With this boundary condition, the effect of mixing is to {\it decrease} the dust density below the $K = u_{\rm set} H_{\rm b}$ layer, as dust is mixed down to the base where it is destroyed. This boundary condition neglects dust production through upward mixing and condensation, and focuses solely on meteoric source and the remnant dust settling down from above.

In order to include the reduced extinction cross section when $2\pi s < \lambda$, the cross section used is $\sigma(s,\lambda) = 2\pi s^2\,
 {\rm min}(1,(2\pi s/\lambda)^4)$.  This will lead to decreasing $\tau$ when the bulk of the dust is smaller than the wavelength.

\begin{figure*}[htb!]
\epsscale{0.9}
\plotone{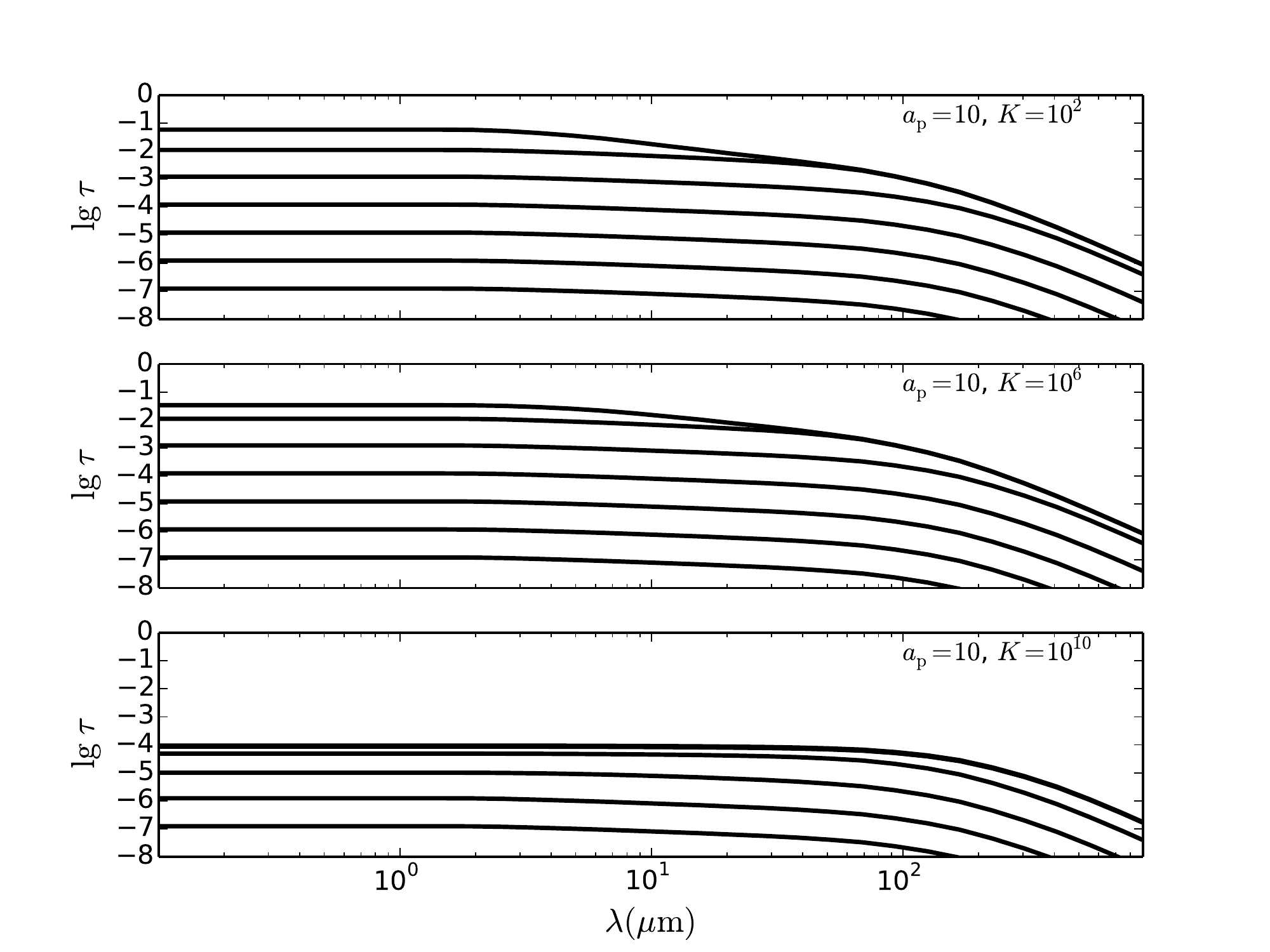}
\caption{ Optical depth versus wavelength for the medium-size planet and the G85 dust size distribution. Each line represents a pressure level $\lg(P/\rm bar)=-6,...,0$ from bottom to top. Each panel is labeled by orbital separation $\ap/R_\odot$ and turbulent diffusion coefficient $K(\rm cm^2\, s^{-1})$.
 \label{fig:tau_med_grun}
 }
\end{figure*}

Figure \ref{fig:tau_med_grun} shows optical depth versus wavelength, with the different lines representing pressure levels from $P=1\, \mu \rm bar - 1\, \rm bar$. The G85 dust size distribution is used, as well as a medium-planet size. The distance from the star is $\ap = 10\, R_\odot$. The three panels show results for mixing coefficients $K=10^2$, $10^6$ and $10^{10}\, \rm cm^2\, s^{-1}$.

In the top panel of Figure \ref{fig:tau_med_grun}, the small values of $K$ lead to mixing being unimportant even for small dust, and optical depth increases proportional to pressure as in Equation \ref{eq:taudust}. The maximum optical depth achieved is $\tau \sim 0.1$ at the $P = 1\, \rm bar$ level. In the middle and lower panels, increased mixing decreases the dust abundance at deeper levels in the atmosphere leading to a smaller maximum optical depth.

\begin{figure*}[htb!]
\epsscale{0.9}
\plotone{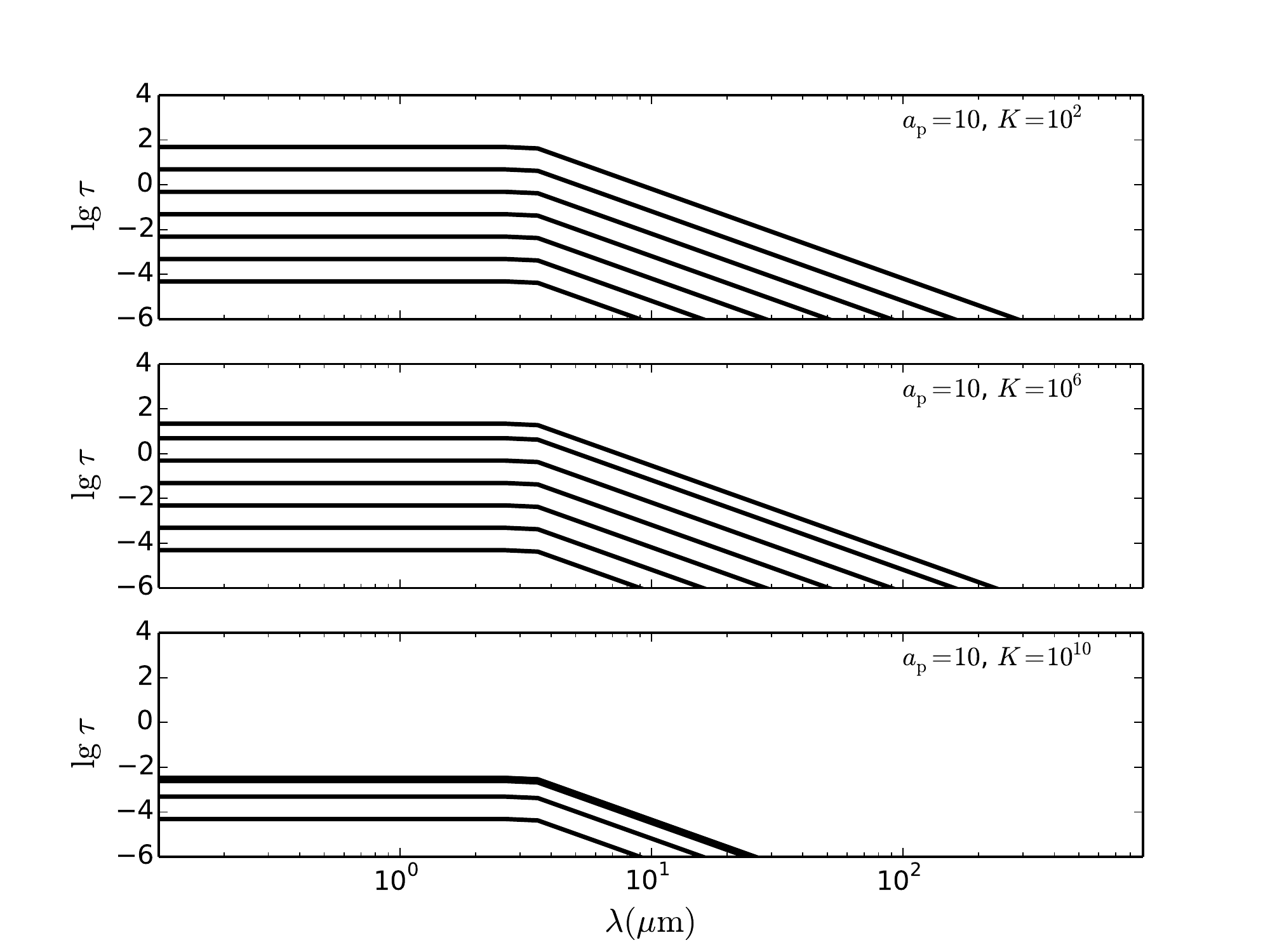}
\caption{ The same as Figure \ref{fig:tau_med_grun} but for the single-size distribution with $\beta=0.5$.  \label{fig:tau_med_beta0p5}
 }
\end{figure*}

Figure \ref{fig:tau_med_beta0p5} shows the results for the single-size distribution with $\beta=0.5$. The change from small to large dust compared to the wavelength is now more apparent. The key difference is in the maximum optical depth. As the same dust accretion rate is now in smaller dust, the number density is higher, leading to significantly larger  $\tau \sim 10^2$ at the 1 bar pressure level, for the case of small mixing. Again for larger mixing the maximum optical depth achieved is much smaller at the deeper levels.

\section{ discussion }
\label{sec:discussion}

The calculations presented in this paper motivate that accretion of interplanetary dust may occur at much larger rates for close-in, massive exoplanets than for the Earth, and that this {\it meteoric source} of heavy elements in the upper atmosphere could in principle contribute significantly to atmospheric opacity. One would like to be able to estimate the contribution from both the deep atmosphere and the meteoric source to the opacity for any specific planet. There are, however, two main sources of uncertainty that prevent firm conclusions about the importance of dust accretion at this time. 

The first uncertainty concerns mixing. As discussed in Section \ref{sec:Kzz}, the level of mixing must be sufficiently {\it weak} in order for the meteoric source to be important, while for strong mixing the meteoric source is overwhelmed by the reservoir of heavy elements deep in the planet. Given that the region of interest is not undergoing thermal convection with the associated turbulent mixing, mixing occurs by the atmospheric circulation patterns and estimates of $K$ are more complicated and uncertain (e.g. \citealt{2018ApJ...866....1Z}). More accurate calculations appropriate for the region of interest likely require resistive MHD simulations including a conductivity profile with dayside photoionization in the upper atmosphere, and are beyond the scope of this paper.

The second uncertainty is observational. The accretion rate onto the planet has been written as a product of three factors, with $f_{\rm pl} \epsilon_{\rm abl} \dot{M}_{\rm d}$ for gas and  $f_{\rm pl} \epsilon_{\rm rem} \dot{M}_{\rm d}$ for remnant dust. For distant debris disk sources and for dust densities low enough that dust collisions are infrequent on the inspiral time, the efficiencies of accretion onto the planet $f_{\rm pl}$ and amount ablated into gas or remaining as dust can be estimated for any particular planetary system. The dust size distribution and the accretion rate of dust toward the star, $\dot{M}_{\rm d}$, are however poorly constrained. One might have hoped that emission from the outer debris disk could be observed, and hence $\dot{M}_{\rm d}$ near the star estimated. However, \citet{2021A&A...651A..45A} find no strong correlation between the presence of a near infrared excess for the hot dust and a mid or far infrared excess for an outer debris disk. Hot dust excesses are not necessarily accompanied by warm or cold dust excesses. This raises the issue of the origin of the hot dust. Presumably the debris disk cannot be in the inner solar system as its collisional lifetime is then much shorter than the age of the star \citep{2007ApJ...658..569W}. But if the debris disk is very far from the star, a mechanism must deliver the dust to the small radii where it is observed (e.g. \citealt{2012A&A...548A.104B}), and the dust level in the outer disk must remain small enough to remain undetected.

How many zodis of dust are required to affect the transmission spectrum for gas-phase elements? If heavy element abundances near solar are required, then Equation \ref{eq:big_result_vmr}, Table \ref{tab:abundances} and Figure \ref{fig:abundances} show that $f_{\rm pl} \epsilon_{\rm abl} \dot{M}_{\rm d} \sim 10^8\, \rm g\, s^{-1}$ is required for a medium-size planet. Orbital radius $a=8\, R_\odot$ and dust with $\beta=0.1$ would have $f_{\rm pl} \epsilon_{\rm abl} \simeq 0.1$. Hence the required dust accretion rate to achieve solar abundance would be $\dot{M}_{\rm d} = 10^9\, \rm g\, s^{-1} \sim 100\, \dot{M}_{\rm d, iss}$ (100 zodi's). While significantly more than what is found in our own solar system, it would likely be undetectable as a {\it H-} or {\it K-}band excess by an order of magnitude.

Implicit in the above argument is that dust density and thermal emission is proportional to accretion rate $\dot{M}_{\rm d}$. If the collision time is shorter than the PR inspiral time, then destructive collisions may then limit $\dot{M}_{\rm d} \la {\rm a\, few} \times 10^8\, \rm g\, s^{-1}$ for a solar type star and grains near the blowout size \citep{2005A&A...433.1007W,2014A&A...571A..51V}. If true, then this maximum accretion rate may limit the abundances due to accreted dust to be subsolar by a factor of 10 or more. Such small accretion rates would also likely not deliver dust to near the sublimation zone at a high enough rate to explain the observed near-infrared excesses \citep{2021A&A...651A..45A}.

While there are examples of systems with both a near-infrared excess and also a close-in planet (e.g. HD 4113 and HD 20794), we are not aware of any that are transiting. Hence, at present there is no system where transmission spectroscopy could be used to see if dust excess directly translates into an affect on the line or continuum spectra.

\citet{2018MNRAS.480.5560B} calculated the fraction of dust hitting the planet and showed that it can be order unity for massive planets near the star. The results presented in Section \ref{sec:rebound} confirm this central result. As \citet{2018MNRAS.480.5560B} did not include a finite size star, significant differences are expected for planets in close orbits as the dust pericenter may lower to the point that it hits the  sublimation zone around the star, giving large values of $f_{\rm st}$. In addition to the orbit integrations, modeling of each scattering event by the \"{O}pik approximation (see the Appendix) shows qualitative and rough quantitative agreement with the orbit integrations, and is a much faster method. Lastly, the present study also computed the impact speed and angle as the dust collides with the planet's atmosphere.

While the orbit simulations in Section \ref{sec:rebound} included the radiation pressure and PR drag due to the star's light, the contribution from the planet's light was neglected for simplicity. This is a good approximation beyond a few planetary radii away from the planet, which is adequate for most of the dust particle's evolution. But since the outgoing flux from the planet is comparable to the incoming  flux from the star, when very near the planet, its flux should be included. Effectively the planet will have a mass $\Mp (1-\beta_{\rm p})$, where $\beta_{\rm p}$ is the ratio of radiation pressure to gravitational force for the planet, and the results from a lower-planet mass model could be used as a first approximation. To model this well, the position-dependent scattered optical light and thermal infrared emission on the dayside and nightside should be included, as well as the difference in $\beta$ for the different wavelengths of the stellar and planetary flux. This is beyond the scope of the present paper.

Likewise, for grains in the upper atmosphere either due to dust accretion or upward mixing, the slower settling speeds implied by radiation pressure $\beta_{\rm p} \la 1$, or even radiative levitation for small grains with $\beta_{\rm p} \ga 1$, has been ignored. This issue deserves further study.

\citet{2017ApJ...847...32L} considered stopping and ablation of dust in exoplanet atmospheres using the equations from \citet{1992Icar...99..368M}. The latter study included thermal evaporation but not sputtering (see Section \ref{sec:stopping}), which was found here to be a significant source of ablation for the large impact speeds implied for massive planets and close orbits. Sputtering gives rise to a distribution of ablated dust mass extending to lower pressures than the $P \sim 1\, \mu \rm  bar$ found for thermal evaporation.

Another key aspect of \citet{2017ApJ...847...32L} was to include radiative heating and cooling for grains. Their numerical results show that grain temperatures are close to the gas temperature at $P \ga 10\, \rm mbar$ and are determined by the radiation field at lower pressures, where differences are found between the temperatures of large and small grains in that study. The analytic description in Section \ref{sec:radheat} uses the radiative absorption efficiency for blackbody radiation from \citet{2011piim.book.....D} to show the effect of {\it super-heating} of small grains. When the grain size is smaller than $\lambda/2\pi$ for its emitted thermal radiation ($s=0.1\, \mu \rm m$ for $T_{\rm d}=1000\, \rm K$), the dust temperature rises, as compared to $T_{\rm eq}$, due to inefficient cooling. This rise continues until the dust becomes so small ($s=30\, \AA$ for $T_{\rm s}=5800\, \rm K$) that absorption of stellar radiation becomes inefficient as well. 

Although not discussed in detail here, the increased temperature for small grains may lead to significantly increased detachment rates, due to the exponential dependence on temperature, as compared to the case where gas and grain temperatures are equal \citep{1981ASSL...88..317D, 2008MNRAS.384..165L}. The resulting smaller nucleation rates, which would occur at low pressures when the grain temperature is set by radiation, may make nucleation at such low pressures more difficult as compared to higher gas pressure regions where the grains are not superheated.

While the focus here has been on the effect of dust on the transmission spectrum, it may also affect the dayside emission spectrum. If the vertical optical depth of grains at $P \la 1\, \rm mbar$ becomes order unity, they may mask the expected dayside molecular absorption features (e.g. H$_2$O rovibrational transitions, \citealt{2016Natur.529...59S}). Further, if this dust is superheated, its emission temperature may be significantly different than would occur from gas at the that pressure level, similar to the mid-infrared spectrum in the protoplanetary disk case \citep{1997ApJ...490..368C}.

This paper has focused on the gas-phase atoms and remnant dust directly produced by the incident dust particles, and has ignored subsequent molecule formation and dust growth. However, the gas-phase atoms and remnant dust may augment heavy elements mixed up from below by providing seeds for heterogeneous nucleation, in addition to those produced by homogeneous nucleation \citep{2018A&A...614A.126L, 2018ApJ...860...18P}, as well as heavy-element monomers that may stick to the seeds.
A parameter study for heterogeneous growth of dust grains through an assumed downward flux of seeds at the upper boundary
was carried out in \citet{2018ApJ...855...86G}.

In addition, as described \citet{2017ApJ...847...32L}, the gas-phase atoms may form molecules that then undergo photolysis  to generate {\it soot} layers up near $\sim \mu \rm bar$ levels in the atmosphere. The mass fluxes reported there were in the range of $10^{-14}-10^{-12}\, \rm g\, cm^{-2}\, s^{-1}$, which are comparable to the accretion rates $f_{\rm pl} \epsilon_{\rm abl} f \dot{M}_{\rm d}/(4\pi \Rp^2) = 10^{-13}\, \rm g\, cm^{-2}\, s^{-1} (f_{\rm pl} \epsilon_{\rm abl} f \dot{M}_{\rm d}/10^8\, \rm g\, s^{-1}) (R_{\rm J}/\Rp)^2$ from dust accretion at rates $\ga \dot{M}_{\rm d, iss}$. The results are similar as these dust accretion rates give rise to nearly solar composition abundances of species contained in the dust, while in \citet{2017ApJ...847...32L} they assume solar abundances deep in the atmosphere and strong vertical mixing.

\section{ summary }
\label{sec:summary}

Observed near-infrared excesses around nearby stars imply large amounts of hot dust orbiting near the sublimation radius around the star. The goal of this paper was to understand if a close-in, massive planet could accrete a significant fraction of the hot dust, and whether or not this could affect atmospheric abundances of gas-phase atoms or  dust enough to be detected through the transmission spectrum. The calculations and results are as follows.

\begin{enumerate}

\item Three-body simulations of the star, planet, and dust particle including PR drag were carried out to compute the fraction of inspiraling dust that is accreted by the planet. Under the assumption that the dust orbits have circularized by the time they reach the planet, and that the dust and planet orbits are nearly coplanar, it is found that Jupiter-mass planets near the star may accrete a significant fraction, up to 10-50\%, of the dust for favorable parameters.
\item Calculations for the stopping and ablation of the dust particles in the planet's atmosphere were carried out to determine the distribution of deposited gas-phase atoms and remnant dust particles with altitude. Both thermal evaporation and sputtering are included, and two extreme dust size distributions were examined. It is found that the large impact speeds for planets close to the star or for massive planets tend to cause dust to be efficiently ablated into gas-phase atoms, while small planets further from the star tend to lead to less ablation and more mass in remnant dust. The dust-stopping layer is at low pressures $P \la \mu \rm bar$,  where the gas temperature may be above the dust melting temperature. However, the dust temperature in these regions is set by radiation, and is regulated to be near the equilibrium temperature for large dust.
\item The abundance of gas-phase, neutral atoms below the meteoric source is computed assuming that the particles descend at their terminal velocity. For dust accretion rates roughly 10-100 larger than that in our inner solar system, the heavy element abundances in the upper atmosphere can approach the solar value. This is a central result of the paper. 
\item It is argued that the increase of electrical conductivity and Lorentz drag with altitude may lead to significantly smaller flow speeds and mixing in the upper atmosphere. 
\item For atomic resonance lines, the dependence of the transit radius on the level of mixing and dust accretion rate is exhibited for a range of each parameter. In the absence of the meteoric source, the transit radius decreases as the level of mixing decreases. Including the meteoric source, even for no mixing, the transit radius can approach the meteoric source altitude even for dust levels of order one zodi. As the dust accretion rate increases, the transit radius is increased further from line center. These results motivate that, for sufficiently large dust accretion rate and sufficiently weak mixing, that large transit radii may reflect the altitude of the dust source.
\item The continuum opacity of the remnant dust particles is maximized when the size distribution is peaked in near the blowout radius and mixing is weak. Optical depths of order 0.1 at mbar pressure levels may then be achieved for dust accretion rates a factor of 100 higher than in our solar system. 

\end{enumerate}

\acknowledgments

The authors thank Ilse Cleeves, Nick Ballering, Jonathan Ramsey, Jonathan Tan, Shane Davis, Craig Sarazin, and the Living, Breathing Planet group for useful discussions. This work was supported by the Living Breathing Planet grant funded by the NASA Nexus for Exoplanet System Science under grant NNX15AE05G. Support was also provided by the NASA ATP grant 80NSSC18K0696, ``Exoplanetary MHD Outflows Driven by EUV Heating, Lyman alpha Radiation Forces and Stellar Tides."

\software{Rebound version 2.20.5 \citep{2012A&A...537A.128R} }
          
\appendix

\section{ the \"{O}pik approximation }
\label{sec:opik}

The direct simulation of the orbits in Section \ref{sec:rebound} involved time-consuming calculations lasting days to weeks. The \"{O}pik approximation  \citep{1976iecr.book.....O} allows a speedup by assuming that the particle orbits on a Keplerian orbit around the star when outside the planet's Hill sphere, and a Keplerian orbit around the planet when inside the Hill sphere. This assumption  is  appropriate after the resonance is broken and Hill sphere encounters begin to occur. The orbital motion may be bypassed and the trajectory receives velocity kicks during Hill sphere crossings. We augment this method by allowing PR drag to affect the orbit between Hill sphere encounters, allowing the Jacobi constant to evolve in time. While it is common to use the Fokker-Planck approximation and sum over many encounters (e.g. \citealt{ 1980A&A....85...77Y, 1987AJ.....94.1330D,1993ASPC...36..335T, 2004ApJ...600..804M, 2008CeMDA.100....1B}), here each encounter is simulated as the number of encounters before ejection or destruction may be small.

Ejection of a particle to infinity occurs through a {\it gravity assist} trajectory where the particle is deflected so that the final direction is nearly parallel to that of the planet. Ejections are strongly constrained by the relative speed and direction of the dust as it enters the Hill sphere. Positive orbital energy after the encounter requires $ (\vec{u}^\prime + \vec{v}_{\rm p})^2 \geq 2G\Ms(1-\beta)/\ap$, where $\vec{u}^\prime = \vec{u} + \Delta \vec{u}$ is the relative velocity after the encounter, $\vec{u}$ is the relative velocity before, and the kick is $\Delta \vec{u}$. Keplerian motion implies that $u^2 = (\vec{u} + \Delta \vec{u})^2$, i.e. the relative speed is constant but the direction is rotated. The minimum relative speed required to eject occurs when $\vec{u}^\prime$ and $\vec{v}_{\rm p}$ are parallel, and is $u_{\rm min}^\prime = v_{\rm p} \left( \sqrt{2(1-\beta)} - 1 \right)$. Hence there is a finite range of speeds for an initially bound particle to reach infinity through a gravity assist. This translates into a finite range of $T_\infty$. If particles emerged from resonance outside this range of $T_\infty$ where ejections could occur, their only options would be physical collisions with the star and planet.

Monte Carlo simulations of  \"{O}pik's method are performed as follows. An initial $a$, $e$ and $I$ are chosen to represent the post-resonance orbit. The dust orbital period is then $P(a) = 2\pi \sqrt{a^3/G\Ms(1-\beta)}$. The probability of encountering the planet's Hill sphere per dust orbit is taken to be $\epsilon_{\rm Hill}=\pi r_{\rm Hill}^2/(2\pi \ap r_{\rm Hill})$ \citep{2004ApJ...600..804M} \footnote{More detailed discussions of the collision probability for two Keplerian orbits (not in resonance) are given in  \citet{1976iecr.book.....O, 2017AJ....153..235J}.}. The time to the next encounter is chosen to be from an exponential distribution with mean time $P(a)/\epsilon_{\rm Hill}$. PR drag acts over this time. The relative velocity vector $\vec{u}$ entering the Hill sphere can be constructed from the orbital parameters up to the signs of the radial and vertical component, which are chosen randomly. The impact parameter $\vec{b} \perp \vec{u}$ is chosen uniformly in area inside the Hill radius and uniformly in angle. If the minimum separation with the planet $r_{\rm min}(b)=(b^2/b_0)/(1+\sqrt{1+(b/b_0)^2})$ is smaller than $\Rp$ then a physical collision has occurred and the simulation ends. If there is no physical collision, then in the planet's reference frame, the velocity kick given to the dust particle is
\be
\Delta \vec{u} & = & -  \frac{2}{ 1 + (b/b_0)^2} \left( \vec{u} + \frac{u}{b_0} \vec{b} \right),
\ee
where the impact parameter for $90^\circ$ scattering is $b_0 = G\Mp/u^2$. 
The first term is the parallel kick, slowing the particle down relative to its incoming trajectory, and the second term is the perpendicular kick along the impact parameter $\vec{b}$. This velocity kick rotates $\vec{u}$ while keeping the same speed before and after the collision. Given the post-encounter velocity vector $\vec{u} + \Delta \vec{u}$, it is checked if the particle is ejected to infinity, and if so the simulation is stopped. If the particle is not ejected, the new values of $a$, $e$ and $\cos(I)$ are reconstructed from the position, which is assumed not to have changed, and the post-encounter velocity. 
After each PR step and each kick, it is checked if $a(1-e) < \Rsub$, and if so the particle has hit the dust sublimation radius and the simulation is stopped. Further it is checked if $a(1+e)<\ap$, so that no further encounters with the Hill sphere will occur, and the simulation is stopped as the particle will eventually hit the star due to PR drag.

With the focus on orbits close to the star, only a small number of encounters may occur before the dust fate is sealed, and hence each encounter is simulated. For more distant orbits for which many encounters happen before ejection or collision, the Fokker-Planck approximation may be more efficient \citep{1980A&A....85...77Y, 1987AJ.....94.1330D,1993ASPC...36..335T, 2008CeMDA.100....1B}. In the present case though, even when {\it cometary diffusion} of $a$ and $e$ occurs, diffusion toward ejection must compete with diffusion toward hitting the star, which would present an additional boundary condition that is complicated to enforce. Simulating each Hill sphere encounter is efficient for the cases considered here. 

\begin{figure*}[htb!]
\epsscale{0.9}
\plotone{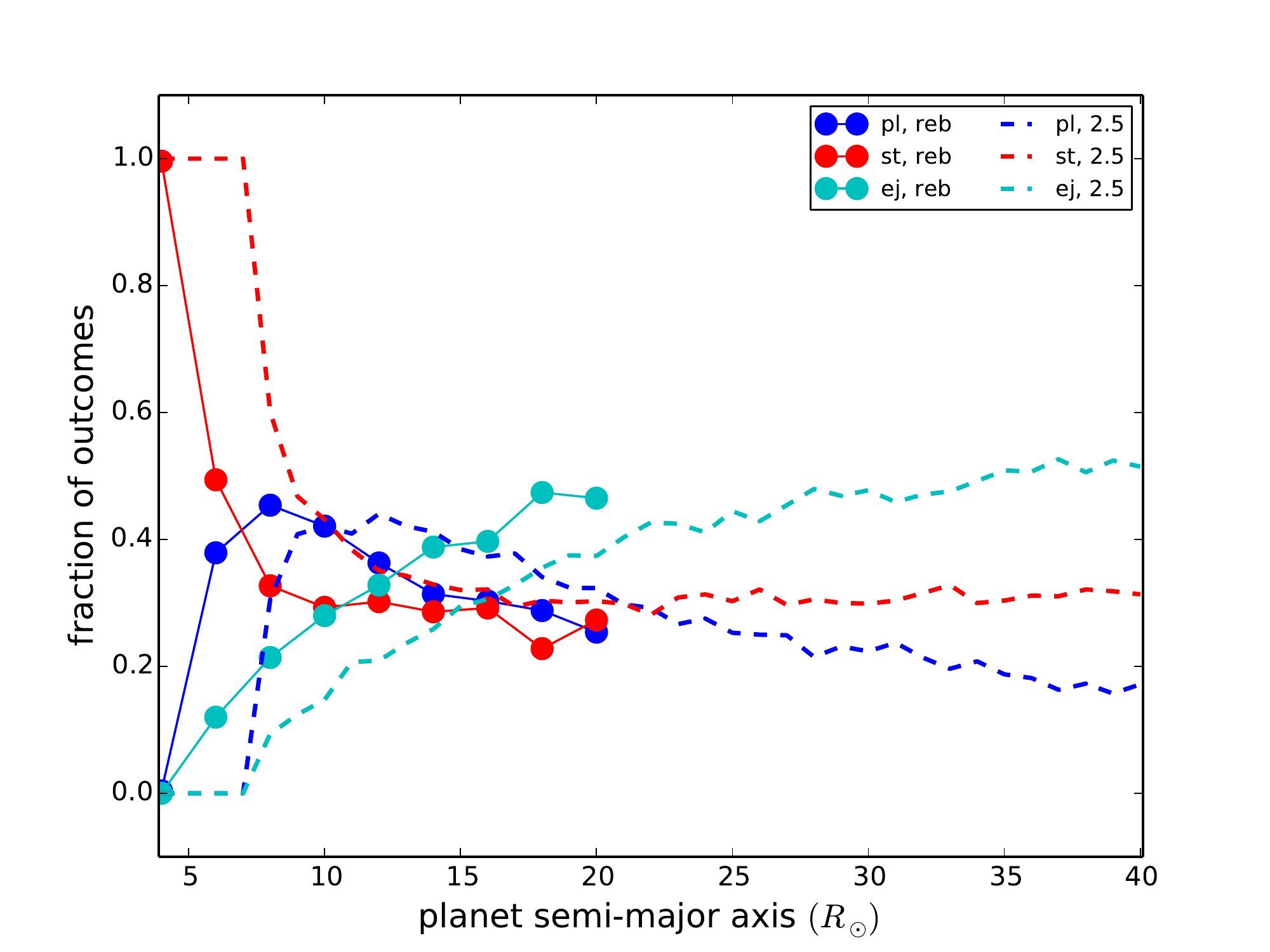} 
\caption{ Outcome fractions computed using {\it Rebound} (dots and solid lines) and \"{O}pik's method for $\beta=0.01$. An initial value $u_\infty/v_{\rm p}= 2.5$ was used for \"{O}pik's method. The dots connected by solid lines show the Rebound simulations from Figure \ref{fig:medfractions}. The dashed lines used \"{O}pik's method.
\label{fig:opik}}
\end{figure*}

Figure \ref{fig:opik} shows the results of 1000 Monte Carlo simulations for $\beta=0.01$ for comparison to the solid lines in Figure \ref{fig:medfractions}. A single  initial value $u_\infty/v_{\rm p}= 2.5$ was used for \"{O}pik's method. The range of $\ap$ has been extended out to $40\, R_\odot$ to see trends more clearly. There is both qualitative and rough quantitative agreement with the full numerical simulations, at the $\sim 10\%$ level for this particular $u_\infty$. While not shown here, other values of $u_\infty=2.4-2.8$ are qualitatively similar and show agreement to 10-20\%. The true answer, as shown in Figure \ref{fig:velocities}, is that particle leave the resonance with a range of $u_\infty$ and so an average should be performed over $u_\infty$ to try to obtain a more precise answer. That is beyond the scope of the present paper, where rough quantitative agreement is sufficient.

\bibliography{ref}{}
\bibliographystyle{aasjournal}

\end{document}

%% file: mydef.tex
\newcommand{\be}{\begin{eqnarray}}
\newcommand{\ee}{\end{eqnarray}}
\newcommand{\beq}{\begin{equation}}
\newcommand{\eeq}{\end{equation}}

\def\la{\mathbin{\lower 3pt\hbox
      {$\rlap{\raise 5pt\hbox{$\char'074$}}\mathchar"7218$}}}
\def\ga{\mathbin{\lower 3pt\hbox
      {$\rlap{\raise 5pt\hbox{$\char'076$}}\mathchar"7218$}}} 

\renewcommand{\vec}[1]{\mathbf{#1}}

